\documentclass[review]{elsarticle}

\usepackage{lineno}
\usepackage[toc]{appendix}
\modulolinenumbers[5]
\usepackage{comment}

\usepackage{graphicx}
\usepackage{amsmath,amssymb}
\usepackage{amsthm}
\usepackage{xcolor}
\usepackage{mathrsfs}
\usepackage{paralist}
\usepackage{multirow}
\usepackage[T1]{fontenc}
\usepackage[utf8]{inputenc}
\usepackage[english]{babel}
\DeclareMathAlphabet\mathbfcal{OMS}{cmsy}{b}{n}
\usepackage{caption}
\usepackage[labelformat=simple]{subcaption}

\usepackage{bbm}
\usepackage[normalem]{ulem}
\usepackage[percent]{overpic}
\usepackage{color}
\usepackage{tikz}
\usetikzlibrary{positioning}
\usepackage{yhmath}
\usepackage{pict2e}

\usepackage{hyperref}
\hypersetup{
    colorlinks=true,     
    linkcolor=red,       
    citecolor=green,     
    filecolor=magenta,   
    urlcolor=blue        
}

\newcommand{\ie}{\emph{i.e.}} 
\newcommand{\eg}{\emph{e.g.}} 
\newcommand{\Rey}{\mathrm{Re}} 
\newcommand{\Ma}{\mathrm{Ma}} 

\newcommand{\f}[1]{\overline{#1}} 
\newcommand{\ff}[1]{\widetilde{#1}} 

\newcommand{\del}[1]{}

\usepackage[a4paper,bindingoffset=0.0in,%
            left=1.0in,right=1.0in,top=1.in,bottom=1.in,%
            footskip=.4in]{geometry}

\begin{document}

\begin{frontmatter}

\title{A data-driven study on Implicit LES using a spectral difference method}

\author[mymainaddress]{Nicola Clinco}
\author[mymainaddress]{Niccol{\`o} Tonicello\corref{mycorrespondingauthor}}
\cortext[mycorrespondingauthor]{Corresponding author}
\ead{ntonicel@sissa.it}
\author[mymainaddress]{Gianluigi Rozza}

\address[mymainaddress]{Mathematics Area, International School of Advanced Studies, SISSA, Via Bonomea, 265, 34136 Trieste, Italy}

\begin{abstract}
In this paper, we introduce a data-driven filter to analyze the relationship between Implicit Large-Eddy Simulations (ILES) and Direct Numerical Simulations (DNS) in the context of the Spectral Difference method. The proposed filter is constructed from a linear combination of sharp-modal filters where the weights are given by a convolutional neural network trained to replicate ILES results from filtered DNS data. In order to preserve the compactness of the discretization, the filter is local in time and acts at the elementary cell level. The neural network is trained on the data generated from the Taylor-Green Vortex test-case at $\Rey=1600$. In order to mitigate the temporal effects and highlight the influence of the spatial discretization, the Implicit Large-Eddy Simulations are periodically restarted from DNS data for different time windows. Smaller time windows result in higher cross-correlations between ILES and the filtered DNS snapshots using the data-driven filters.
The modal decay of the filter for the smallest time window considered aligns with classical eigenanalysis, showing better energy conservation for higher orders of approximation. Similarly, an analysis of the filter's kernel in the Fourier space confirms that higher polynomial orders are less dissipative compared to lower orders. As large time windows are considered, the trained filter encounters difficulties in representing the data due to significant non-linear effects. Additionally, the impact of the data-driven filter on the resolved kinetic energy has been assessed through the evaluation of the sub-grid production term which results in both direct and inverse cascades with the former being more likely on average. The presence of backscatter suggests that Implicit Large-Eddy Simulations based on Discontinuous Spectral Element Methods might be equipped with an intrinsic mechanisms to transfer energy in both directions with a predominance of direct kinetic energy cascade.
\end{abstract}


\end{frontmatter}

\section{Introduction}\label{sec:introduction}
Computational Fluid Dynamics (CFD) is essential for turbulence modeling as it provides an efficient framework for solving complex equations governing turbulent flows, bypassing possibly expensive experimental facilities. Turbulence is inherently chaotic, with eddies, swirls, and vortices spanning a wide range of scales, making its dynamics extremely difficult to accurately predict using simple analytical models. Over the past 50 years, significant advancements in computational resources have enhanced CFD's capability to simulate increasingly complex turbulent flows. 
Based on the level of spatial resolution and modeling complexity, different approaches for simulating turbulent flows are commonly known in the literature. Direct Numerical Simulation (DNS) describes accurately all the eddies in a flow field as it solves the entire range of the spatial and temporal scales involved in the flow. However, performing a DNS for practical applications is still beyond reach even for nowadays high-performing machines. On the other side of the spectrum, the Reynolds Average Navier-Stokes (RANS) approach models turbulent flows by decomposing the flow field into an average value and a fluctuation component. The objective of RANS is consequently predicting the average value only, reducing the computational burden by avoiding expensive unsteady simulations and exploiting symmetries in the domain. Thus, many industries rely on RANS as an efficient way to deal with turbulent flows as it is computationally cheap. 
However, it can lack in accuracy for more complex applications, as the scale resolving capabilities of the approach is quite limited.

One way to keep the computational cost affordable without sacrificing accuracy is the Large-Eddy Simulation (LES) approach \cite{Pierre_Sagaut_2001_INC,eric_garnier_2009,PIOMELLI1999335}. This methodology models turbulent flows by directly simulating the large-scale eddies, while modeling the effects of the small scales by a so-called Sub-Grid Scale (SGS) model.
Approaches such as DNS and LES can significantly benefit from highly resolving spatial discretizations, as they get closer to the asymptotic convergence regime. The increasing computational power experienced in the last few decades has generated a strong interest in the development of high order numerical methods. In particular, Spectral Element Methods (SEMs) have garnered significant attention due to their geometrical flexibility and promising computational efficiency. These includes the Discontinuous Galerkin (DG)~\cite{hesthaven:book,cockburn:98,cockburn:98b}, Flux Reconstruction (FR)~\cite{huynh2007flux} and Spectral Difference (SD)~\cite{kopriva1996conservative,liu:06a,wang:07} methods. All of these have shown promising results in a large variety of fluid dynamics problems such as compressible flows~\cite{cuong2022large,TONICELLOramp,Tonicello_Lodato_Vervisch_2022,chapelier2024comparison,KURZ2025109388}, combustion~\cite{MarchalSDcombustion,lv2014discontinuous,lv2023recent} and multiphase flows~\cite{manzanero2020entropy,ntoukas2022entropy,TONICELLO2024112983,OrlandoDG,pandare2023design}.

Within the LES formalism, two main approaches are known in the literature: the first one is called Explicit Large-Eddy Simulation (ELES) where explicit analytical functions of the resolved scales are used to model unclosed sub-grid scale terms \cite{Pierre_Sagaut_2001_INC,eric_garnier_2009}. The second approach is known as Implicit Large-Eddy simulation (ILES) \cite{Grinstein_Margolin_Rider_2007} where the numerical scheme itself acts as an implicit SGS model. Often, explicit SGS models are developed using an approach that prioritizes physical principles over considerations on the numerical method employed.
Consequently, classical explicit SGS models (\ie, Smagorinsky~\cite{smagorinsky1963}, WALE~\cite{nicoudWALE}, Vreman~\cite{VremanSGS04} models) yield significantly different outcomes when applied to various spatial discretizations. Therefore, it is not always recommended to build explicit models without considering the effect of the numerical discretization, which can ultimately influence the overall dynamics of the large scale motions. One possible approach is to design numerical schemes so  that physical quantities of interest such as kinetic energy are extactly preserved on the discrete level ~\cite{morinishi1998fully,subbareddy2009fully} and leave the turbulence modeling to explicit models. Another possibility is to tune the numerical scheme in order to act as an implicit SGS model itself or to formulate SGS models which adapt depending on the underlying discretization.
     
One of the most common tool for investigating the inherent numerical characteristics of spatial discretizations in advection-dominated flows is the spectral eigenalysis \cite{moura2015linear,moura2019spatial,mengaldo2018spatial,mengaldo2018spatial2}. This technique is straightforward to implement and computationally efficient. Despite its simplicity, it provides valuable insights into the numerical footprint of the scheme, such as numerical dispersion and diffusion. 
One of the advantages of the eigenanalysis approach is the capability of deriving analytical kernel functions that characterize the dispersion and diffusion properties of a numerical scheme. In this context, numerical dissipation acts as a damping mechanism for the Fourier coefficients across different spatial frequencies. As a result, the dissipation curves can be viewed as a filter that retains low frequencies, corresponding to resolved scales, while eliminating unresolved scales. These findings suggest that certain numerical schemes are well-suited for Implicit Large-Eddy Simulations~\cite{Grinstein_Margolin_Rider_2007,Fernandez17}.
However, due to its intrinsically linear nature, this approach can often miss in predicting the complex non-linear dynamics of more realistic flows. Establishing a direct connection between the numerical discretization of the linear advection equation and complex three-dimensional simulations such as the Navier-Stokes equations can be a quite challenging task. It is clear that characterizing the spatial discretization and thus the filter implied by the numerical scheme requires more complex techniques in these scenarios.
    
Recently, with the widespread use of machine learning, data-driven strategies have been proposed within the LES framework \cite{BECK_19,BeckReviewML,ZHOU2019104319,Wang18dd,Iaccarino23LES1,VANGASTELEN2024113003,Sirignano1,Beck23_RL0,SARGHINI200397,GamaharaANDAttori}. 
Sarghini et al. \cite{SARGHINI200397} developed a scale-similarity model learned from data, Maulik et al. \cite{Maulik_San_2017} presented a data-driven approach to learn the deconvolution of flow variables without assuming a functional form for the filter and Benjamin \& Iaccarino  \cite{Iaccarino23LES1} applied an artificial neural network to improve the accuracy of LES for channel flow by building a map from the flow features to the SGS terms.
Despite these promising results, building a model based entirely on data trained completely in an offline fashion (\ie, by simply using pre-computed filtered DNS data), results in models which are relatively accurate in \emph{a-priori} setting, but they are unstable in \emph{a-posteriori} computations \cite{BECK_19}. Since the discretization error can be of similar magnitude with respect to the modeled subgrid-scale term \cite{GHOSAL199524,KRAVCHENKO1997310,CHOW2003366}, the resulting dynamical system arising from the LES discretization becomes entirely uncorrelated from one of the filtered DNS \cite{BECK_19}, leading to a mismatch between \emph{a-priori} and \emph{a-posteriori} analyses. On the other side, data-driven SGS models developed on the specific discretization in an online fashion (\ie, by running in parallel LES and DNS) demonstrated to be stable also during \emph{a-posteriori} computations \cite{Sirignano1,Sirignano2,sanderse2024,Kochkov21}. In this case, the model is built directly on the discretized PDEs, and thus, it takes into account the accumulation error. However, this method is expensive, as it requires a fully differentiable solver. As a result, using this technique for high-order spatial discretizations such as Spectral Element Method can be impractical due to both computational and storage demands. 

In all the works previously mentioned, machine learning techniques were employed to learn the unclosed SGS term from data without any consideration on the implicit filter induced by the space discretization.
In this paper, instead, we explore the use of machine learning techniques to derive the implicit filter directly from data.
Such approach can provide useful insights in the numerical footprint of the scheme for under-resolved turbulent flows, extending classical linear approaches such as temporal/spatial linear eigenanalysis and improving the current understanding of non-linear effects on numerical dissipation.
Furthermore, a more faithful characterisation of the implicit filter of numerical schemes can lead to the development of \emph{a-posteriori} SGS models which are ``numerics-aware''. For example, the proposed approach could be easily implemented within scale-similarity models or dynamic procedures in classical SGS models (\eg, Smagorisnky, Vreman, WALE). Also, designing filters which are aware of the underlying numerical scheme used in \emph{a-posteriori} computations can mitigate the instability issues related to offline data-driven SGS models introduced earlier in this section.

This paper will be organized as follows. Section \ref{sec:problemdef} introduces the LES formalism and the concept of numerical filter associated to the spatial discretization, as well as the proposed approach to formulate a data-driven filter within the spectral difference framework. In Section \ref{sec:CNN} the specific architecture employed in this work is outlined along with all the details regarding its hyper-parameters. In Section \ref{sec:setup} we introduce the numerical set-up in terms of test case used herein (the Taylor-Green vortex) and the training data preparation. In particular, we will present a strategy to pre-process the data in such a way the influence of the spatial discretization is isolated from accumulation errors associated with time integration. In Section \ref{sec:results} we present a series of analyses on the numerical characteristics of the proposed data-driven filter both in terms of accuracy and interpretability of its predictions. The presented analyses are twofold as they are focused on both strictly numerical properties of the filter and on its influence on the physics of turbulent flows. Among these, the smoothing properties of the data-driven filter are assessed through kinetic energy spectra, averaged modal coefficient in the space of Legendre polynomials and inner-element spectral signature in Fourier space, as well as through the evaluation of the SGS dissipation term in the resolved kinetic energy balance. Finally, in Section \ref{sec:conclusion} we outline the key conclusions and perspectives of this work.
\section{Problem definition: characterization of the implicit filter of the numerical scheme}\label{sec:problemdef}
In this section we present the Large-Eddy Simulation paradigm and the concept of implicit filter induced by the numerical discretization. Once these concepts are properly introduced, we will proceed by presenting the approach proposed in this work to construct a data-driven filter that is representative of the spatial discretization. 
\subsection{The compressible Navier-Stokes equations}\label{sec:NSE}
Large-Eddy Simulations are powerful techniques for simulating turbulent flows by directly resolving the large-scale eddies while modeling the effects of smaller scales. The starting point for LES is the set of governing equations for compressible flows, which consists in the balance of mass, momentum and total energy of the fluid. Namely, the compressible Navier-Stokes equations can be written as:
\begin{align}
    & \frac{\partial \rho}{\partial t} + \frac{\partial}{\partial x_j} \left( \rho u_j \right)= 0,  \label{eq:mass} \\
    &\frac{\partial}{\partial t} \left( \rho u_i \right) + \frac{\partial}{\partial x_j} \left( \mathbf{\rho} u_i  u_j + p\delta_{ij} \right) = \frac{\partial}{\partial x_j} \left( \Sigma_{ij} \right),\label{eq:momentum} \\
    & \frac{\partial}{\partial t}\left( \rho E \right ) + \frac{\partial}{\partial x_j} \left [ \left( \mathbf{\rho}E +p  \right) u_j \right] = -\frac{\partial q_j}{\partial x_j} + \frac{\partial}{\partial x_j} \left( \Sigma_{ij} u_i\right ),\label{eq:TotalEnergy}
\end{align}
where $\rho$ is the fluid density, $u_i$ is the $i$-th velocity component, $p$ is the pressure, $E$ is the total energy per unit mass, $\Sigma_{ij}$ is the viscous stress tensor and  $q_j$ represents the heat flux. Furthermore, $\delta_{ij}$ indicates the Kronecker-delta.
System \eqref{eq:mass}-\eqref{eq:TotalEnergy} is closed by a thermodynamic equation of state for the pressure which, based on the assumption that air behaves like an ideal gas, is given by:
\begin{equation}
    p  = \rho \left(E-\frac{u_i u_i}{2} \right ) \left( \gamma -1\right), \label{eq:state}
\end{equation}
where $\gamma = c_p/c_v$ is the specific heat ratio.

In LES, the governing equations are spatially filtered to separate the large-scale motions, which are resolved, from the small-scale motions, which are modeled. Specifically, for compressible turbulent flows, the Favre-filtering technique is applied to equations \eqref{eq:mass}-\eqref{eq:TotalEnergy}. For a general function $\phi$, the Favre-filtered quantity is defined as $\ff{(\phi)} = \f{\rho \phi}/{\f{\rho}}$, where the operator $\overline{(\cdot)}$ denotes the spatial filtering operation which is commonly assumed to be a convolution operation with a kernel function $G$:
\begin{equation}
    \f{\phi}(\mathbf{x},t) = G(\mathbf{x};\Delta) \star \phi(\mathbf{x},t) = \int G(\mathbf{x}-\mathbf{x'};\Delta) \phi(\mathbf{x'}) d \mathbf{x'} \label{eq:spatial_filtering}.
\end{equation}
The kernel function $G$ is influenced by the cutoff length scale $\Delta$, which defines the threshold separating resolved and under-resolved scales.
Applying the Favre-filtering to these equations results in the so-called filtered equations:
\begin{align}
    & \frac{\partial \f{\rho}}{\partial t} + \frac{\partial}{\partial x_j} \left( \f{\rho} \ff{u_j} \right)= 0,  \label{eq:fmass} \\
    &\frac{\partial}{\partial t} \left( \f{\rho} \ff{u_i} \right) + \frac{\partial}{\partial x_j} \left( \f{\rho} \ff{u_i}  \ff{u_j} + \f{p} \delta_{ij} \right) = \frac{\partial}{\partial x_j} \left( \f{\Sigma_{ij}} \right) -\frac{\partial \tau^{\mathrm{SGS}}_{ij}}{\partial x_j}, \label{eq:fmomentum} \\
    & \frac{\partial}{\partial t}\left( \f{\rho} \ff{E} \right ) + \frac{\partial}{\partial x_j} \left [ \left( \f{\rho}\ff{E} + \f{p} \right) \ff{u_j} \right] = -\frac{\partial \f{q_j}}{\partial x_j} + \frac{\partial}{\partial x_j} \left( \f{\Sigma_{ij} u_i}\right ) -\frac{\partial q_j^{\mathrm{SGS}}}{\partial x_j}, \label{eq:fTotalEnergy}
\end{align}
where $\tau^{\mathrm{SGS}}_{ij} = \f{\rho u_i u_j}-\f{\rho} \ff{u_i} \ff{u_j}$ represents the subgrid scale tensor and $q^{\mathrm{SGS}}_{j} = \f{\rho E u_j+p u_j} - (\f{\rho} \ff{E} \ff{u_j} +\f{p}\ff{u_j})$ accounts for the subgrid scale energy flux. Note that we have assumed that the filtering operation commutes with spatial derivatives.
Further terms arise from the non-linearity of the viscous term when non-linear dependence of the viscosity with respect to the temperature (such as Sutherland law) is used. However, these terms are often negligible and $\tau^{\mathrm{SGS}}_{ij}$ and $q^{\mathrm{SGS}}_{j}$ are usually the main focus of modeling efforts in compressible LES~\cite{Vreman95Priori}.
Two different approaches are commonly employed in Large-Eddy Simulations. The first is known as ELES, where an explicit functional form of the large scale motion is used to model the unknown SGS terms. In contrast, the second approach, known as ILES, relies on the numerical dissipation inherent in the discretization scheme to act as an effective SGS model. Indeed, in this case, it is assumed that the numerical scheme itself provides sufficient numerical dissipation at the grid scale so that no additional explicit terms are needed in the equations \cite{Grinstein_Margolin_Rider_2007,Pierre_Sagaut_2001_INC,eric_garnier_2009}. In both the approaches, however, deriving the analytical functional form of the convolution kernel in equation \eqref{eq:spatial_filtering} is not an easy task. In the context of the ELES approach, the implicit filter kernel is shaped by both the numerical scheme and the explicit SGS model. Instead, the ILES approach defines the implicit filter solely through the spatial discretization's effect. It is clear that in both methodologies, the resolution of LES equations induces a filtering effect by truncating small scales that the grid cannot resolve. 

The implicit filter kernel induced by the numerical scheme can only be analytically derived in the simplified case of the linear advection equation, allowing us to assess the dissipation and dispersion characteristics of the numerical scheme \cite{mengaldo2018spatial,moura2015linear,lele1992compact,bogey2004family,van2008stability}. 
However, extending these analyses to three dimensional simulations of the Navier-Stokes equations is still an open problem.
In the classical linear eigenanalysis, each Fourier mode evolves independently, and thus one can analyze the evolution of each mode in isolation. In contrast, in the non-linear regime, such as in the Navier-Stokes equations, modes interact by transferring energy to one another. The resulting coupling between the modes leads to a complex dynamics which cannot be captured by traditional linear methods.

It is important to point out that each numerical scheme has its own characteristic numerical dissipation and dispersion curves arising from linear eigenanalysis. In this work, we will focus on the spectral difference scheme (SD), which is a specific variant of the Spectral Element Method~\cite{kopriva1996conservative,liu:06a,wang:07}.
\subsection{Construction of a data-driven filter for the Spectral difference scheme}
According to the classical LES formalism, the resolution of the filtered equations should aim at matching the filtered DNS data. The first approach of simply resolving the LES equations and obtaining a ``filtered'' solution is known as \emph{a-posteriori} approach whereas explicitly filtering DNS data is known as \emph{a-priori} analysis. Properly matching \emph{a-priori} and \emph{a-posteriori} analyses has been a long standing challenging task in turbulence modeling~\cite{de3fe897e5744c62a7c59501cb081d2a,vreman1996large,park2005toward}.

Let us define the ``ideal filter'' $\boldsymbol{\mathcal{F}} (\cdot)$ as the operator that exactly links the DNS solution $U^{\mathrm{DNS}}$ to the ILES solution $U^\mathrm{ILES}$ as follows:
\begin{equation}
    {U}^{\mathrm{ILES}}= \boldsymbol{\mathcal{F}} U^{\mathrm{DNS}}. \label{eqn:perfect_filter}
\end{equation}
The exact form of this operator is unknown as it arises implicitly from the resolution of the filtered compressible Navier-Stokes equations. Consequently, trying to link LES and DNS as a simple filtering operation might be problematic.
First, non-local interactions in turbulence, both in time and space, raise doubts about the exact region where the operator $\boldsymbol{\mathcal{F}} (\cdot)$ is applied~\cite{Pierre_Sagaut_2001_INC,eric_garnier_2009,PIOMELLI1999335}. 
Secondly, since we are connecting two discrete quantities (\ie, defined on a grid), the operator must be tailored to the particular discretization applied in both time and space for both DNS and LES \cite{GHOSAL1996187,KravchenkoMoin}. Moreover, the discretization itself influences the scale resolving capabilities of the method and thus non-local interactions.
Consequently, the operator must take into account both the physical and numerical aspects of the problem at hand.
In light of this challenge, it is common practice in the design of SGS models to simply consider the \emph{a-priori} filtered data by a predefined and often arbitrary filter as the ideal target for \emph{a-posteriori} computations~\cite{Pierre_Sagaut_2001_INC,eric_garnier_2009}.

As discussed, in order to build an operator which is as general as possible, non-locality in space and time should be included. However, spectral element methods for compressible flows are usually local in both time (explicit time schemes) and space (compactness of DSEMs). Consequently, we restrict our modeling methodology to operators which are local in space and time. Furthermore, notice that this choice is also motivated by the possible use of such filter in \emph{a-posteriori} computations (\eg, dynamics procedures and/or scale-similarity SGS models).


Note that the present methodology, although formulated for the spectral difference scheme, can be easily generalised to other DSEMs such as Discontinuous Galerkin and Flux Reconstruction as these methods are similar to the spectral difference scheme in terms of compactness.

In the present work, we model the filtering operation \eqref{eqn:perfect_filter} as a combination of $N$ classical filters $G_1,...,G_N$ as follows:
\begin{equation}
    U^{\mathrm{ILES}}  = \mathcal{M}(G_1,G_2,...,G_n) U^{\mathrm{DNS}}, \label{eqn:compfilterBASE}
\end{equation}
where $\mathcal{M}(\cdot)$ indicates the explicit dependence of the model on classical filters.
In order to smooth out small scale fluctuations, different choices of classically used filters are possible (\eg, Gaussian, box, Exponential), acting either in physical~\cite{lesieur2005large,Pierre_Sagaut_2001_INC} or modal~\cite{hesthaven:book} space. 
Among these, the modal filters in the space of Legendre polynomials are particularly suitable to DSEMs as they are relatively simple to implement and they provide a fast way to attenuate the high frequency components in the solution. For the sake of clarity, we present the formulation in 1D. The Extension to higher dimensions is straightforward by tensor-product.
In the SD scheme the solution can be represented as an expansion of orthogonal polynomials in the reference element $[-1,1]$ as follows:
\begin{equation}
    u(x) = \sum_{n=1}^{N}\widehat{u}_n P_n(x), \label{eqn:umodal}
\end{equation}
where $\widehat{u}_n$ are the modal coefficients of the solution, $ \{P_n(x)\}_{n=1,...,N}$ are the Legendre basis functions and $N=\mathrm{p}+1$ is the order of accuracy with $\mathrm{p}$ being the polynomial order.
Modal filters aim to selectively damp high-frequency components (high-order modes) by multiplying the modal coefficients by a filter function $\sigma_n$:
\begin{equation}
    u(x) = \sum_{n=1}^{N}\sigma_n \widehat{u}_n P_n(x).\label{eqn:ufilteredmodal}
\end{equation}
Typically, the filter function $\sigma_n$ is designed to attenuate high order modes more strongly than lower order modes.
In this work, we propose as an approximation of the ideal filter operator \eqref{eqn:perfect_filter} a linear combination of sharp cutoff filters $G_1,...,G_{N}$ in the modal space as:
\begin{equation}
   U^{\mathrm{ILES}}(x) = \sum_{j=1}^{N} \theta^{\mathrm{ANN}}_j G_j \left[ U^{\mathrm{DNS}} \right ]= \sum_{j=1}^{N}\theta^{\mathrm{ANN}}_j\sum_{n=1}^{N}\Gamma_{jn} \widehat{u}_n^{\mathrm{DNS}} P_n(x),\label{eqn:LinearCombSharpModal}
\end{equation}
where the weights of the sharp filters $\theta_j^{\mathrm{ANN}}$ are provided as output of an artificial neural network and $\Gamma_{jn}$ is defined as:
\[ \Gamma_{jn} = 
    \begin{cases} 
     1 & \text{if}\;\;\; n \leq j, \\
     0 & \text{if}\;\;\; n > j. \label{eqn:GAMMA_LINEAR_COMB}
     \end{cases}
\]
This choice is motivated by the fact that considering a single sharp modal filter set on the expected cutoff length scale (as it is commonly done) might be a limited description of the filtering operation implied by the numerical scheme, which could be acting on a wider range of scales.

The connection between equations \eqref{eqn:ufilteredmodal} and \eqref{eqn:LinearCombSharpModal} can be explicitly written by inverting the summations in the latter expression as:
\begin{equation}
    U^{\mathrm{ILES}}(x) = \sum_{n=1}^{N} \overbrace{\left( \sum_{j=1}^{N}\theta^{\mathrm{\mathrm{ANN}}}_j \Gamma_{jn} \right)}^{\displaystyle{\sigma_n}} \widehat{u}_n^{\mathrm{DNS}} P_n(x).  \label{eqn:LinearCombSharpModal1}
\end{equation}
Note that the modal coefficients are linear combinations of the outputs of the artificial neural network.
\section{Convolutional neural network}\label{sec:CNN}
The relationship between the flow features of a fully resolved field ($\mathrm{DNS}$) and the weights associated to each sharp filter can be approximated by an artificial neural network \cite{LeCunDeep,Haykin_NN}. Previous works demonstrated the superiority of convolutional neural networks with respect to feedforward neural networks due to their ability to efficiently handle high-dimensional spatial data and capture local patterns  \cite{Krizhevsky2012ImageNetCW,LeCunCNN}, which are crucial for extracting complex flow features.
Therefore, in this work we chose a convolutional neural network to predict the coefficients $\theta^{\mathrm{ANN}}$ used in the data-driven filter introduced in the previous section. The structure of the architecture is reported in Table~\ref{tab:CNN_architecture}. Notice that the number of points inside each element is equal to $N^{3}$.
The first inner layer is a convolutional layer that uses a kernel of size 3, stride equal to 1 and padding equal to 1. The number of channels in output is set to 1 in order to maintain the complexity and the training cost as low as possible. The output of the convolutional layer is flattened in order to be used for fully connected layers.  Between the hidden layers, the hyperbolic tangent activation function is selected due to its smoothness and effectiveness in mitigating the vanishing gradient problem. No activation function is used after the output layer, but the outputs are normalized such that their sum is equal to $1$. This normalization is required for obtaining a convex combination of filters. The learning rate was fixed to $0.008$.
The parameters related to the CNN are summarised in Table~\ref{tab:lr_table}.
\begin{table}[!ht]
\begin{center}
\scalebox{0.8}{
\begin{tabular}{|c|c|c|c|} 
 \hline
 Layer & Layer name & Input shape & Output shape  \\ [0.5ex] 
 \hline
 Input & Input & $(N,N,N)$ & $(N,N,N)$ \\ 
 \hline
 Convolutional & Conv3D & $(1,N,N,N)$ & $(1,N,N,N)$ \\
 \hline
 Fully-connected & FCC1 &  $N^3$ & $400$ \\
 \hline
 Fully-connected & FCC2 &  400 & 400 \\
 \hline
 Output & FCC3 &  400 & $N$ \\
 \hline
\end{tabular}
}
\end{center}
\vspace{-0.5cm}
\caption{Details of the neural network architecture. $N$ indicates the order of accuracy.}
\label{tab:CNN_architecture}
\end{table}
\begin{table}[!ht]
\begin{center}
\scalebox{0.8}{
\begin{tabular}{|c|c|c|c|c|c|c|} 
 \hline
  Layer name & kernel size & padding & stride & activation function & learning rate & Epochs \\ [0.5ex] 
 \hline
 Conv3D & $3\times 3 \times 3$ & 1  & 1 & Tanh & $8\times 10^{-3}$ & $60$\\ 
\hline
\end{tabular}
}
\end{center}
\vspace{-0.5cm}
\caption{Details of the convolutional layer in the network alongside the activation function, the learning rate and the number of epochs.}
\label{tab:lr_table}
\end{table}

\begin{figure}[h!]
\centering
\includegraphics[width=.99\textwidth]{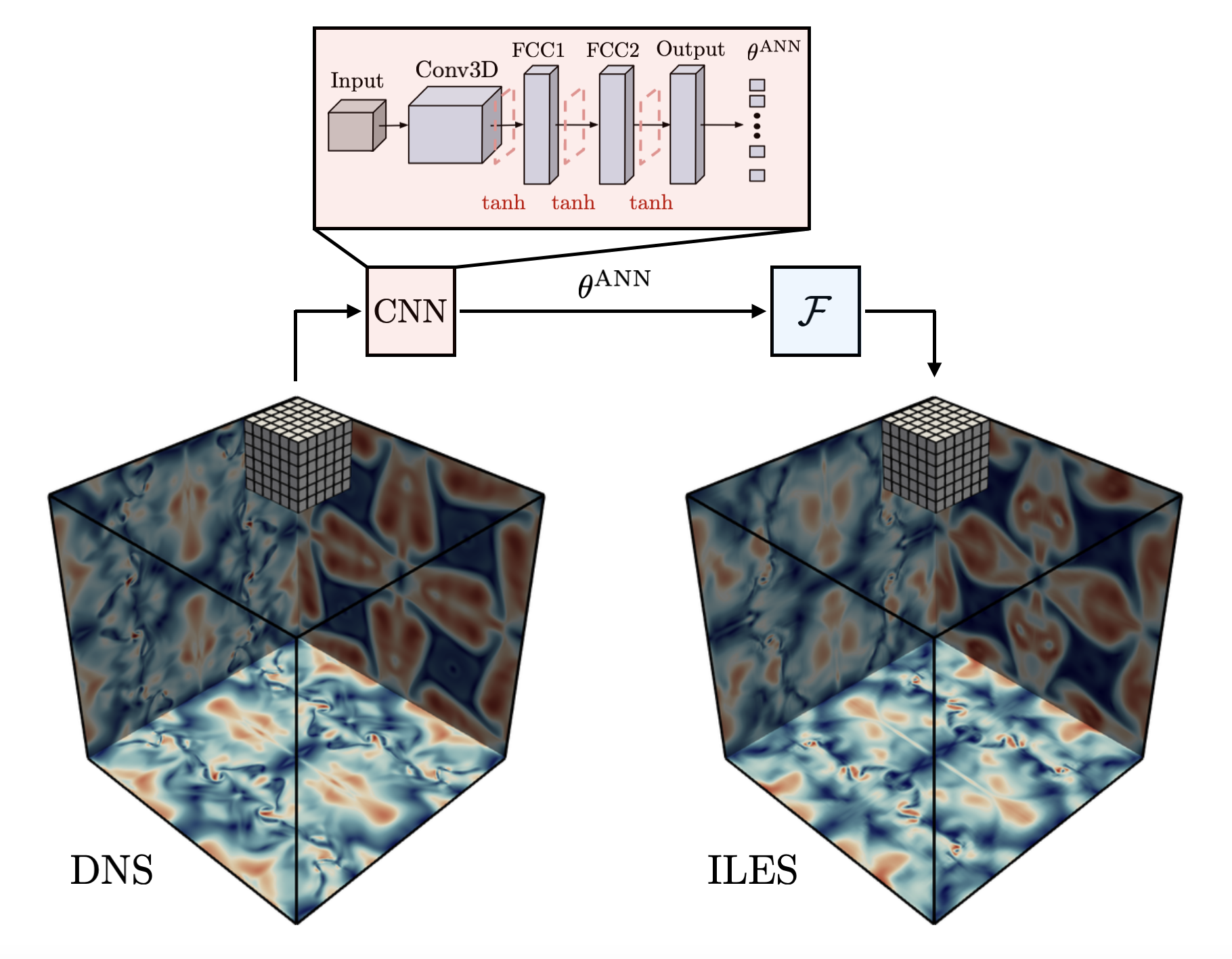}
\caption{A schematic view of the architecture. The inputs and outputs layer are shown for a batch size equal to 1 for clarity. On the edges of the computational domain slices of the velocity magnituded are depicted for DNS (left) and ILES (right) at $t^{*}=15$ for the Taylor-Green vortex test case. Each minibatch consists of the pointwise values of the momentum inside each discontinuous element as depicted in the left portion of the figure. The spectral element shown is significantly enlarged for clarity and ease of visualization.}
\label{fig:ANN_schematic}
\end{figure}

A schematic view of the architecture is shown in Figure~\ref{fig:ANN_schematic} for clarity.
In the proposed architecture, the inputs of the neural network are represented by the pointwise values of the momentum stored at the $N^{3}$ solution points inside each computational cell of the domain:
\begin{equation}
    \mathbf{X^{in}} = \{ (\rho u_{\alpha})_{ijk}\text{,}\;\,\text{with}\,\alpha=1,...,3;\,\,i,j,k=1,...,N\}. \label{eq:Xin}
\end{equation}
Therefore, the number of batches considered during the training is equal to the number of elements in each snapshot sampled from the various simulations.
The outputs of the neural network, instead, are the weights $\theta_j^{\mathrm{ANN}}$ for the sharp filters in the modal space as defined in equation \eqref{eqn:LinearCombSharpModal}.
At each step of the training phase, the DNS is given as input to the network and the coefficients $\theta^{\mathrm{ANN}}_j$, provided as output of the network, are used to build a local filter that acts in each element. Then, the DNS is filtered according to equation \eqref{eqn:LinearCombSharpModal} and the filtered data for computing the loss function is defined component-wise as:
\begin{equation}
    \mathbf{Y_{\alpha}} = \{ \overline{(\rho u_{\alpha})}^{DD}_{ijk}\text{,}\;\,\text{with}\,\alpha=1,...,3;\,\,i,j,k=1,...,N\}. \label{eq:Yout}
\end{equation}
In this paper, we define a different loss function for each component of the momentum instead of defining a unique loss function for all the components. This allows us to speed up the training phase since the parameters of the network are updated every time we read each individual component.
The loss function for each component $\alpha$ is defined as the Mean Square Error (MSE) between each component of the filtered momentum vector and the ILES is computed as:
\begin{equation}
    \mathrm{MSE}_{\alpha}(\boldsymbol{\theta}^{\mathrm{ANN}})= \frac{1}{N_{el}} \sum_{i=1}^{N_{el}}\left( \mathbf{Y}_{\alpha}(\boldsymbol{\theta}^{\mathrm{ANN}}) - \mathbf{Y}_{\alpha}^{t} \right)^2,  \,\,\,\,\, \alpha=1,...,3,\label{eq:MSE}
\end{equation}
where $N_{el}$ is the number of elements of each DNS snapshot.
For clarity, in \eqref{eq:MSE} we have explicitly stated the dependence of the filtered quantity $\mathbf{Y}_{\alpha}$ from the outputs of the neural network $\theta^{\mathrm{ANN}}$.
\section{Numerical set-up} \label{sec:setup}
\subsection{The Taylor-Green vortex problem}
The Taylor-Green vortex~\cite{Brachet83,taylor1937mechanism} is a popular test case to evaluate the role played by the numerical scheme in under-resolved flows~\cite{Gassner2013OnTA,chapelier2016spectral,moura2017eddy,tonicello2023fully}. In the Taylor-Green vortex test case, the large vortices initialised in the flow field trigger a kinetic energy cascade towards the small scales. As time evolves, the developed small-scale structures approach the Kolmogorov length scale and the kinetic energy starts to decay, evolving towards fully turbulent motions. In later stages of its dynamics, the flow field resembles a turbulent energy decay regime similar to the one encountered in homogeneous isotropic turbulence.

The domain is a three-dimensional periodic box $[0,2\pi]^{3}$. The initial condition is given by:
\begin{align}
\rho (t_0) &= \rho_0,\\
u(t_0) &= u_0 \sin\left(\frac{x}{L}\right)\cos\left(\frac{y}{L}\right)\cos\left(\frac{z}{L}\right),\\
v(t_0) &= -u_0 \cos\left(\frac{x}{L}\right)\sin\left(\frac{y}{L}\right)\cos\left(\frac{z}{L}\right),\\
w(t_0) &= 0,\\
p(t_0) &= p_0 + \frac{\rho_0 u_0^2}{16} \left[ \cos\left(\frac{2x}{L}\right)+ \cos\left(\frac{2y}{L}\right)\right] \left[\cos\left(\frac{2z}{L}\right)+2\right],
\end{align}
with $p_0 = 100,\,\,\rho_0=1,\,\,u_0=1,\,\,L=1$. For these values, the Mach number is equal to $\Ma=0.08$. The Reynolds number $\mathrm{Re} = \rho_0 u_0 L/\mu_0$ is fixed at 1600. Time integration is carried out using the three-stages third-order Runge-Kutta scheme.
This test case is solved for a non-dimensional time $t^{*}=20$. 

Due to its tri-periodic formulation, the balance of the domain-averaged kinetic energy is classically considered to assess the impact of the numerical scheme and/or of explicit SGS models on turbulent motions.
Let us denote by $\Omega$ the measure of the volume of the domain; then the balance of domain averaged kinetic energy reads:
\begin{equation}
   -\frac{d E_k}{dt} = \epsilon_{v}+\epsilon_{d}+\epsilon_{num}, \label{eqn:Ekbalance}
\end{equation}
where $E_k$ is the domain-averaged kinetic energy,
\begin{equation}
    E_k = \frac{1}{\rho_0\Omega}\int_{\Omega}\frac{1}{2}\rho u_i u_i\, d\Omega, \label{eqn:ek}
\end{equation}
$\epsilon_{v}$ is the viscous dissipation term,
\begin{equation}
    \epsilon_{v}= \frac{1}{\rho_0\Omega}\int_{\Omega} \left[ 2\mu S_{ij}\,S_{ij}-\frac{2}{3} \mu \left( \frac{\partial u_i}{\partial x_i} \right)^2 \right] d\Omega, \label{eqn:diss_vis}
\end{equation}
with $S_{ij}=\left( \frac{\partial u_i}{\partial x_j} +\frac{\partial u_j}{\partial x_i} \right)$ the rate-of-strain tensor, $\epsilon_{d}$ is the pressure-dilatation term,
\begin{equation}
    \epsilon_{d}= -\frac{1}{\rho_0\Omega}\int_{\Omega} \left(p \frac{\partial u_i}{\partial x_i} \right) d\Omega, \label{eqn:diss_dil}
\end{equation}
$\epsilon_{num}$ is the numerical dissipation,
\begin{equation}
    \epsilon_{num}= -\frac{d E_k}{dt}-\epsilon_{v}-\epsilon_{d}. \label{eqn:diss_num}
\end{equation}
Finally, $\epsilon_k$ is the kinetic energy dissipation rate,
\begin{equation}
    \epsilon_{k} = -\frac{d E_k}{dt}. \label{eqn:diss_ek}
\end{equation}
For small values of the Mach number the pressure-dilatation term is expected to be negligible and it will not be considered in the following analyses.
\subsection{ILES of the Taylor-Green vortex problem}
Since we are interested in understanding the numerical footprint of the scheme, we start by considering different Implicit Large-Eddy Simulations of the Taylor-Green vortex case using different polynomial orders.
Additional details on both implicit Large-Eddy simulations and DNS are listed in Table~\ref{tab:space_disc}.
\begin{table}[!ht]
\begin{center}
\scalebox{0.8}{
\begin{tabular}{|c | c c c|} 
 \hline
 Discretization & Order & Elements & DoF \\ [0.5ex] 
 \hline
 $\mathrm{43p5}$ ($\mathrm{DNS}$) & 6 & 43 & $258^3$    \\ 
 \hline
 $\mathrm{21p5}$ & 6  & 21  & $126^3$ \\
 \hline
 $\mathrm{32p3}$ & 4 & 32  & $128^3$ \\
 \hline
 $\mathrm{43p2}$ & 3 & 43 & $129^3$ \\
 \hline
 $\mathrm{64p1}$ & 2 & 64 & $128^3$ \\ [1ex] 
 \hline
\end{tabular}
}
\end{center}
\vspace{-0.4cm}
\caption{details of the space discretization considered in this study.}
\label{tab:space_disc}
\end{table}

\begin{figure}[h!]
 \centering  
 \begin{subfigure}{0.32\textwidth}
     \includegraphics[width=\textwidth]{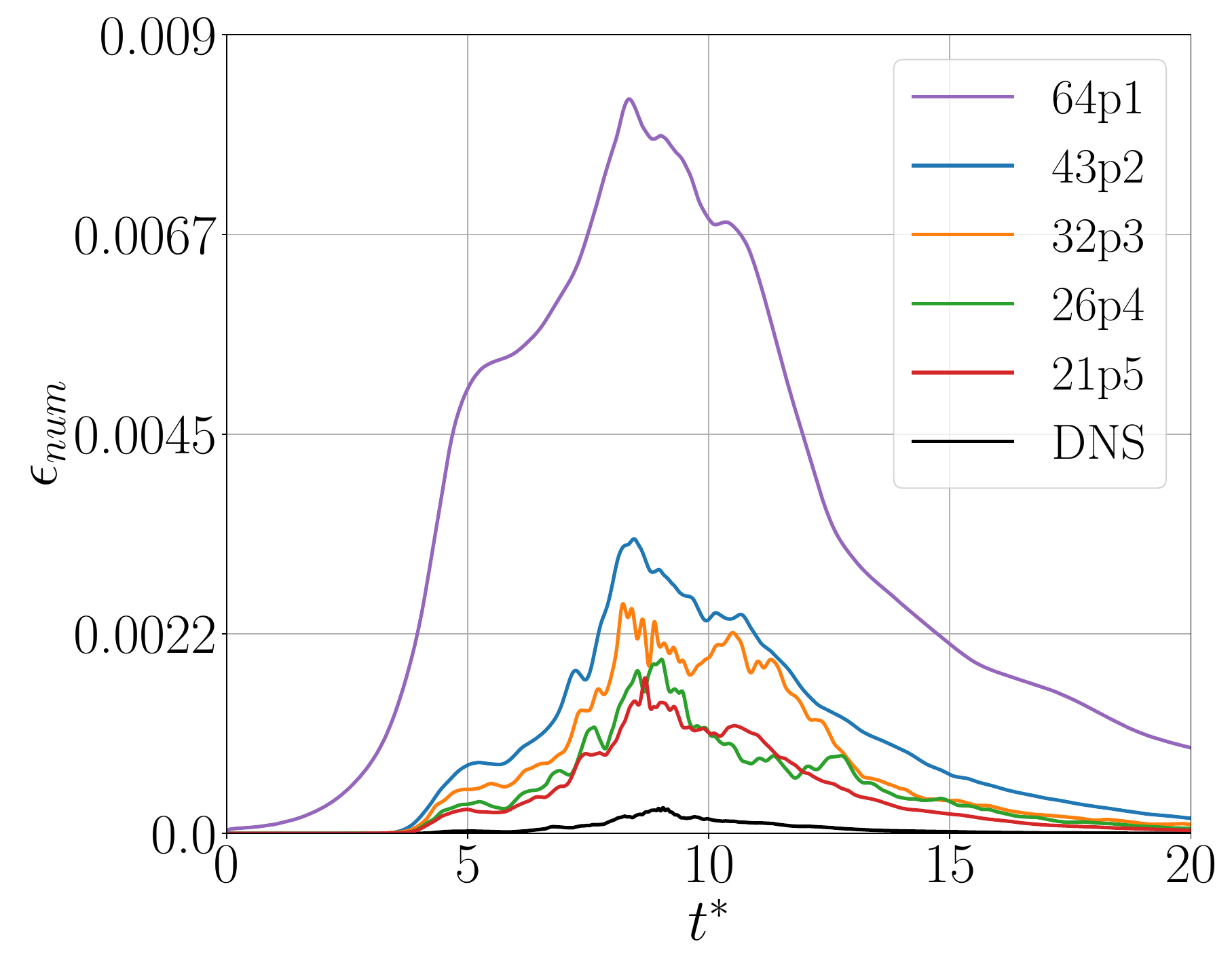}
     \caption{Numerical dissipation.}
     \label{fig:diss_iles:a}
 \end{subfigure}
 \begin{subfigure}{0.32\textwidth}
     \includegraphics[width=\textwidth]{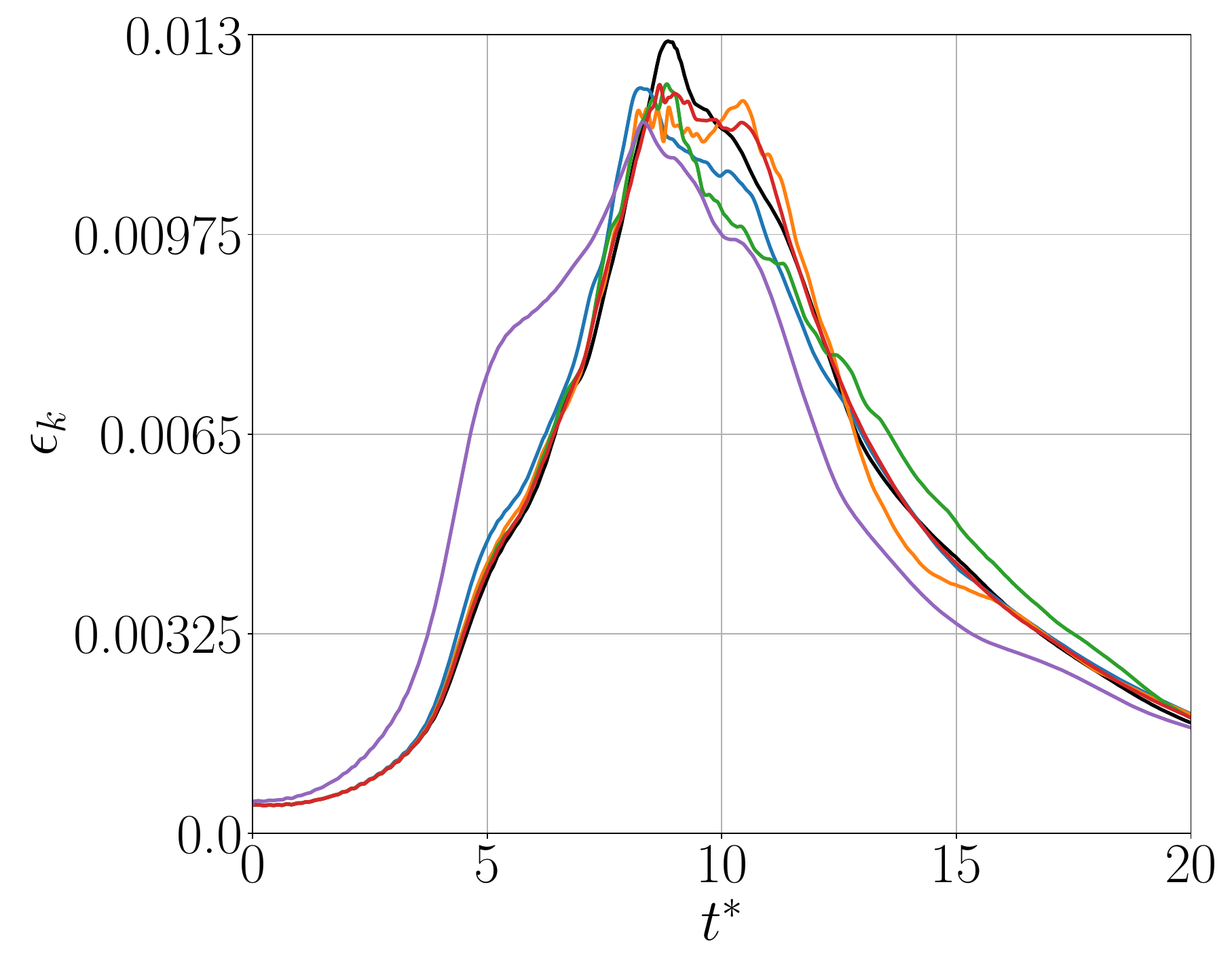}
     \caption{Kinetic energy dissipation rate.}
     \label{fig:diss_iles:b}
 \end{subfigure}
 \begin{subfigure}{0.32\textwidth}
     \includegraphics[width=\textwidth]{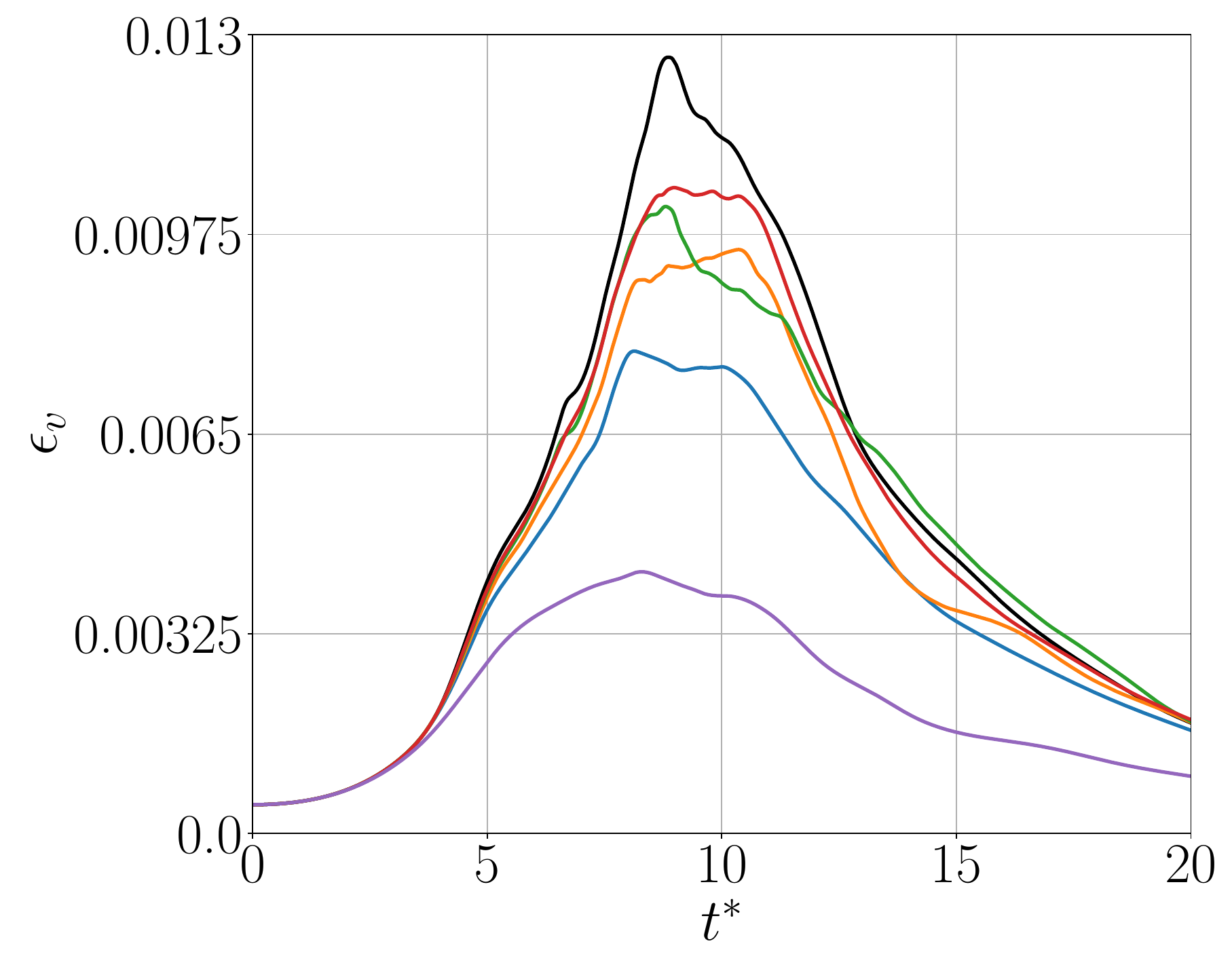}
     \caption{Viscous dissipation.}
     \label{fig:diss_iles:c}
 \end{subfigure}
\caption{Numerical dissipation (left), kinetic energy dissipation rate (middle) and viscous dissipation (right) for the Implicit Large-Eddy Simulations for different orders.}
   \label{fig:diss_iles}
\end{figure}

\begin{figure}[h!]
 \centering  
 \begin{subfigure}{0.43\textwidth}
     \includegraphics[width=\textwidth]{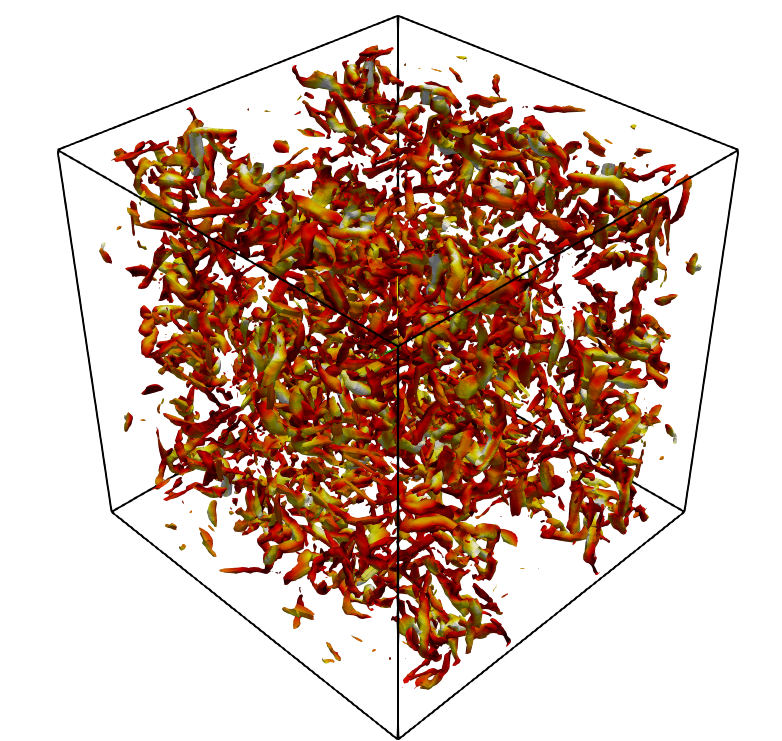}
     \caption{$\mathrm{p}=2$.}
     \label{fig:a}
 \end{subfigure}
 \begin{subfigure}{0.43\textwidth}
     \includegraphics[width=\textwidth]{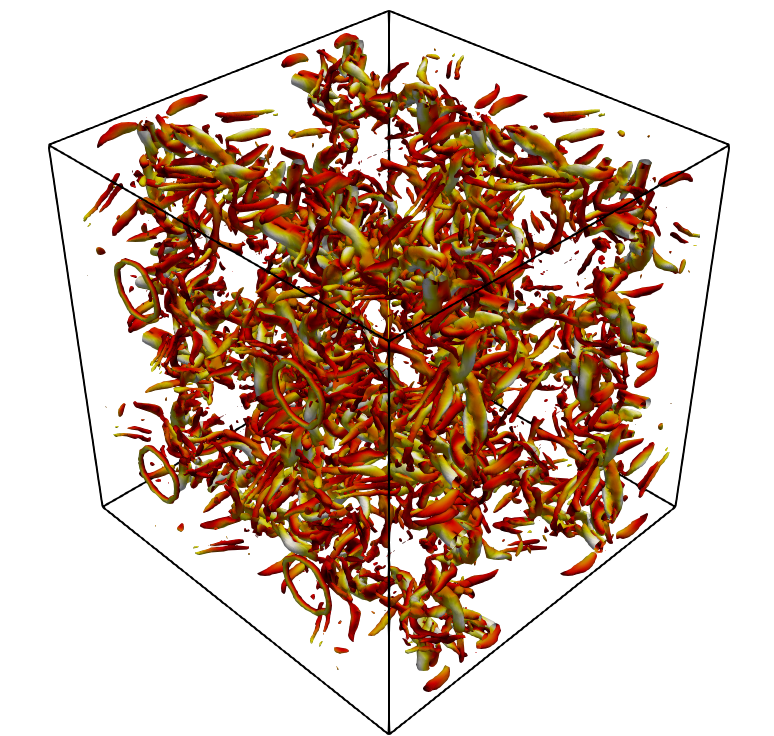}
     \caption{$\mathrm{p}=3$.}
     \label{fig:b}
 \end{subfigure}

 \medskip
 
 \begin{subfigure}{0.43\textwidth}
     \includegraphics[width=\textwidth]{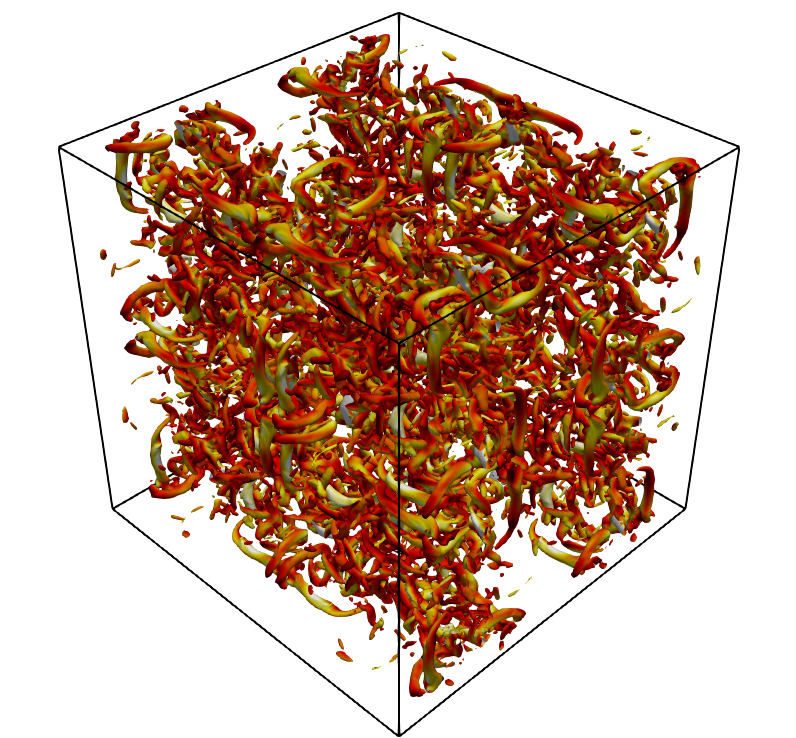}
     \caption{$\mathrm{p}=4$.}
     \label{fig:c}
 \end{subfigure}
 \begin{subfigure}{0.43\textwidth}
     \includegraphics[width=\textwidth]{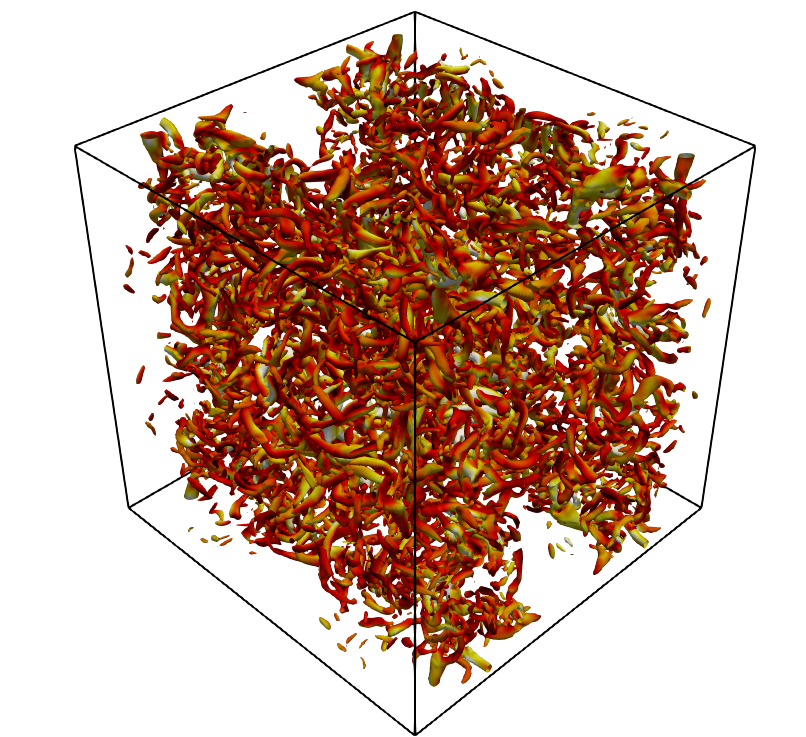}
     \caption{$\mathrm{p}=5$.}
     \label{fig:d}
 \end{subfigure}

 \caption{Iso-surface of Q-criterion at $t^*=15$ for different polynomial orders.}
 \label{Label}
 \label{fig:q-crt}
\end{figure}

Figure~\ref{fig:diss_iles} shows the numerical dissipation \eqref{eqn:diss_num}, the kinetic energy dissipation rate \eqref{eqn:diss_ek} and the viscous dissipation \eqref{eqn:diss_vis} for different polynomial orders.
Focusing on the time evolution of the numerical dissipation (Figure~\ref{fig:diss_iles:a}), the curve for $\mathrm{p}=1$ reveals that this case is the most dissipative one compared with the others as the numerical dissipation is significantly larger for this order. Instead, by comparing the curves for other polynomial orders, the behavior of the numerical dissipation is much less clear as the curves intersect each other at different times during the evolution of the system. By focusing carefully on the curves for $t^{*}>12$, the dissipation content for $\mathrm{p}=4$ is greater than $\mathrm{p}=5$.
The kinetic energy dissipation rate, shown in Figure~\ref{fig:diss_iles:b}, clearly shows that lower polynomial orders do not consistently result in a higher decay of kinetic energy throughout the entire simulation. After the peak of enstrophy, turbulence develops and nonlinear interactions can become predominant, leading to a scenario where the differences in the polynomial order become less distinguishable in the kinetic energy dissipation rate curves. This is further supported by the evolution of viscous dissipation (Figure~\ref{fig:diss_iles:c}), which shows that the curves for $\mathrm{p}=3$ and $\mathrm{p}=4$ intersect close to the enstrophy peak. Additionally, beyond the peak, there is no distinct pattern observed among the polynomial curves.
From these plots, we can infer that interpreting the numerical scheme is quite challenging since the numerical footprint appears to be more complex than merely dissipative ~\cite{Gassner2013OnTA,Beck2013,Beck2014}. Establishing the relationship given by \eqref{eqn:perfect_filter} between a fully resolved simulation (DNS) and the flow characteristics obtained from simulations with varying polynomial orders (ILES) is a difficult task, as the nonlinear dynamics and interactions across different temporal scales complicate the problem significantly.
In order to further highlight the impact of the different orders of approximation on flow structures, in Figure~\ref{fig:q-crt} we show Q-criterion iso-surfaces coloured by local velocity magnitude at $t^{*}=15$. Remember that the total number of degrees of freedom is approximately the same for all the computations and consequently the expected resolution capability should be the same across the different simulations. Although, it can be noticed that the structures captured by low orders are larger and many small scale details are lost due to the impact of the numerical scheme. As the order increases, instead, smaller and smaller turbulent structures are preserved, reflecting the scale-resolving capabilities of high-order spatial discretizations. Finally, notice that analyses such as averaged balance of kinetic energy, kinetic energy spectra and Q-criterion can provide some insight on the impact of the scheme but are clearly limited as there is no practical information regarding the implicit filter of the scheme. We aim at learning such filter directly from ILES data.

Ideally, the optimal filter operator \eqref{eqn:perfect_filter} should take into account the complex dynamics of the system, including the temporal non-locality that is a key characteristic of turbulence. At the same time, numerical schemes employed in compressible flows are usually local in time. A trade-off between these two aspects should be considered. 
In order to establish a relationship that accounts only for the space effects and thus decouples the temporal evolution and the accumulation error from the numerical footprint, we propose to start from DNS data, integrate in time for a certain time window $\Delta T$ on an LES resolution and try to link this simulation at the end of the time window with the DNS at that given time.
The extent of the time window should be chosen in such a way that it is small enough to reduce accumulation error but also large enough to trigger non-linear effects in the dynamics. A very short time window, for example, would simply coincide with a projection on the LES grid, whereas large time windows would lead back to a full ILES. In this work, we considered three different time windows, namely, $\Delta T=0.5,2,4$. For these analyses, each snapshot of the DNS and ILES (restarted or not) is stored with a sampling interval of $\delta t=0.5$.

\begin{figure}[h!]
 \centering  
 \begin{subfigure}{0.3\textwidth}
     \includegraphics[width=\textwidth]{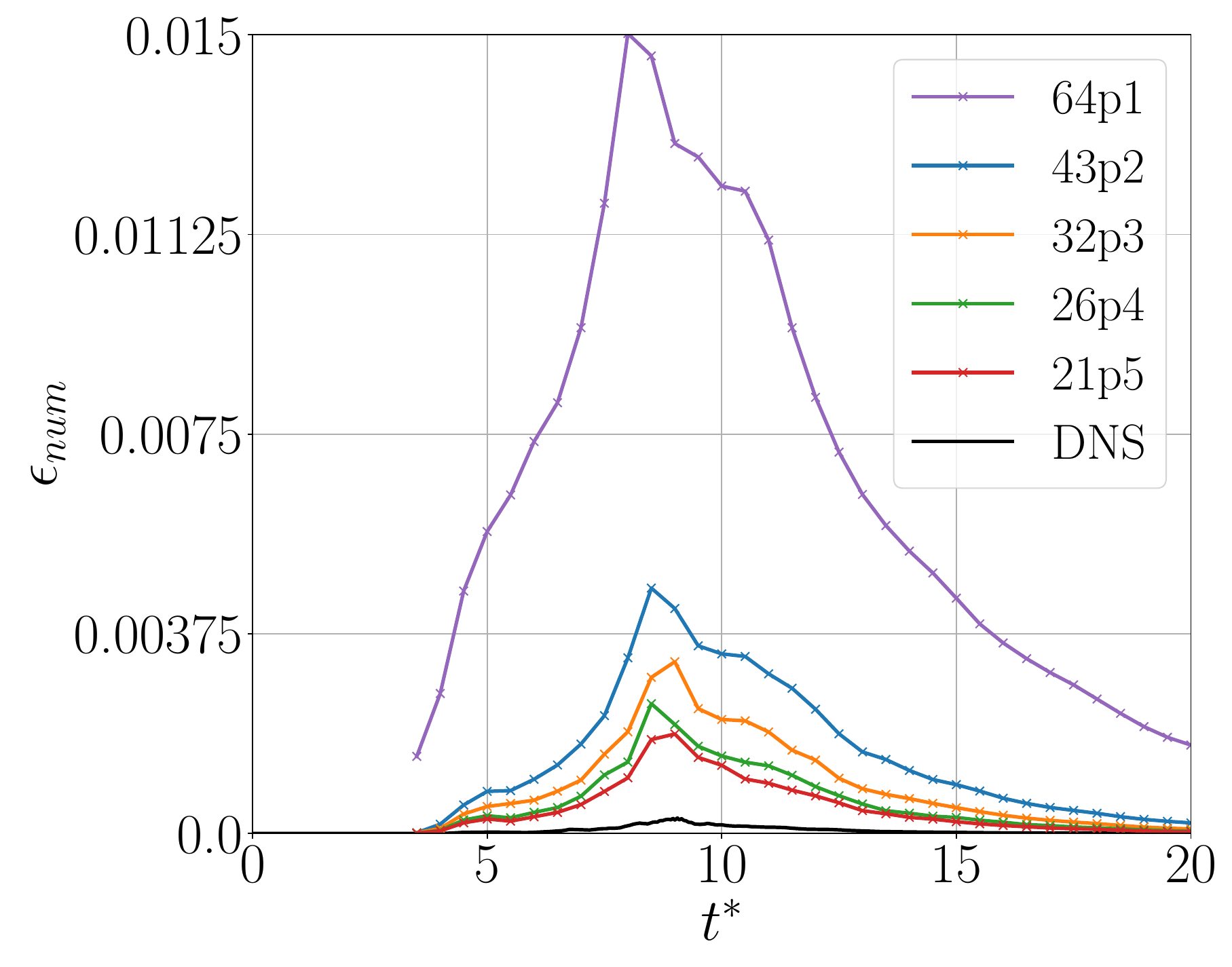}
     \caption{Numerical dissipation.}
     \label{fig:diss_res_0.5:a}
 \end{subfigure}
 \begin{subfigure}{0.3\textwidth}
     \includegraphics[width=\textwidth]{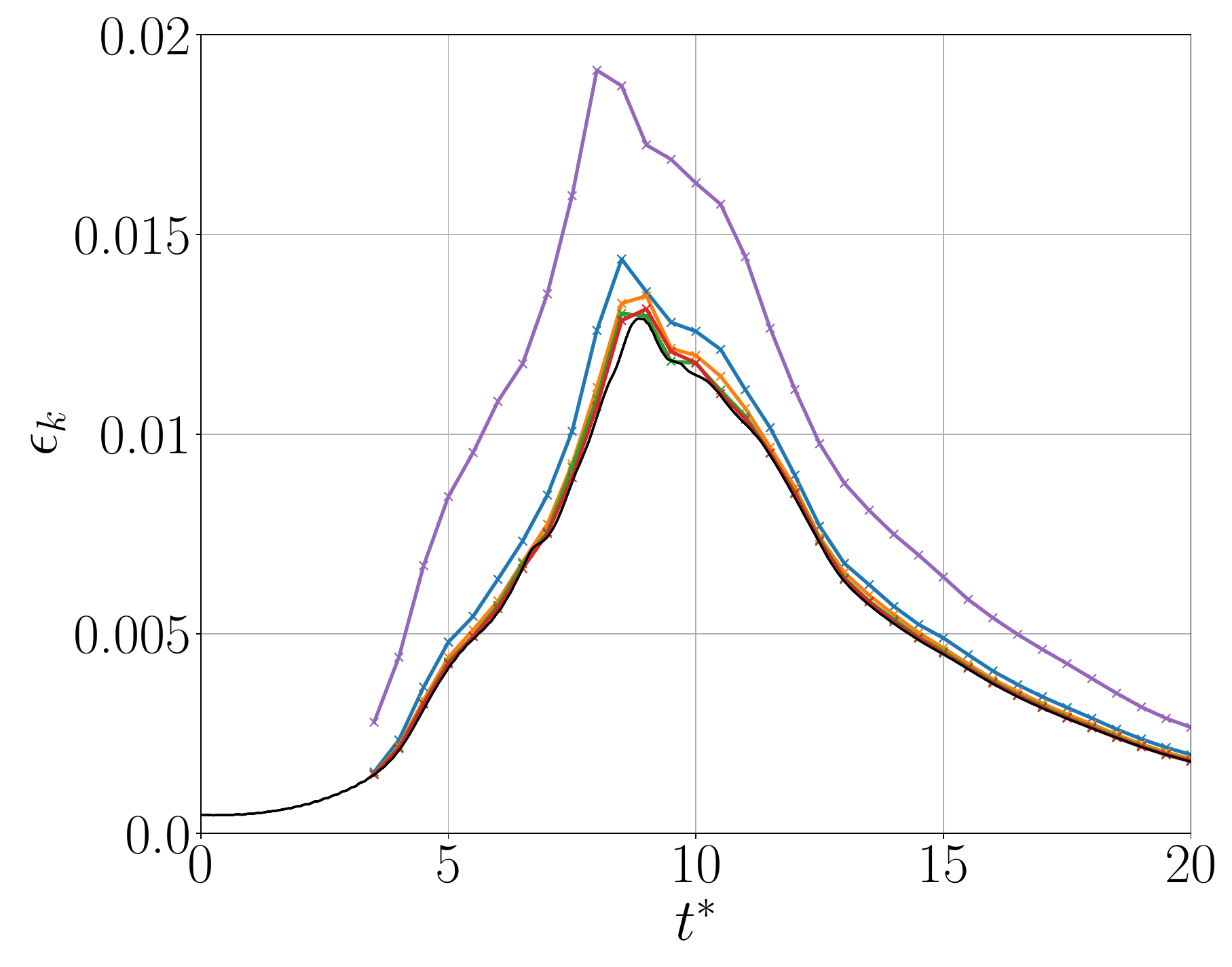}
     \caption{Kinetic energy dissipation rate.}
     \label{fig:diss_res_0.5:b}
 \end{subfigure}
 \begin{subfigure}{0.3\textwidth}
     \includegraphics[width=\textwidth]{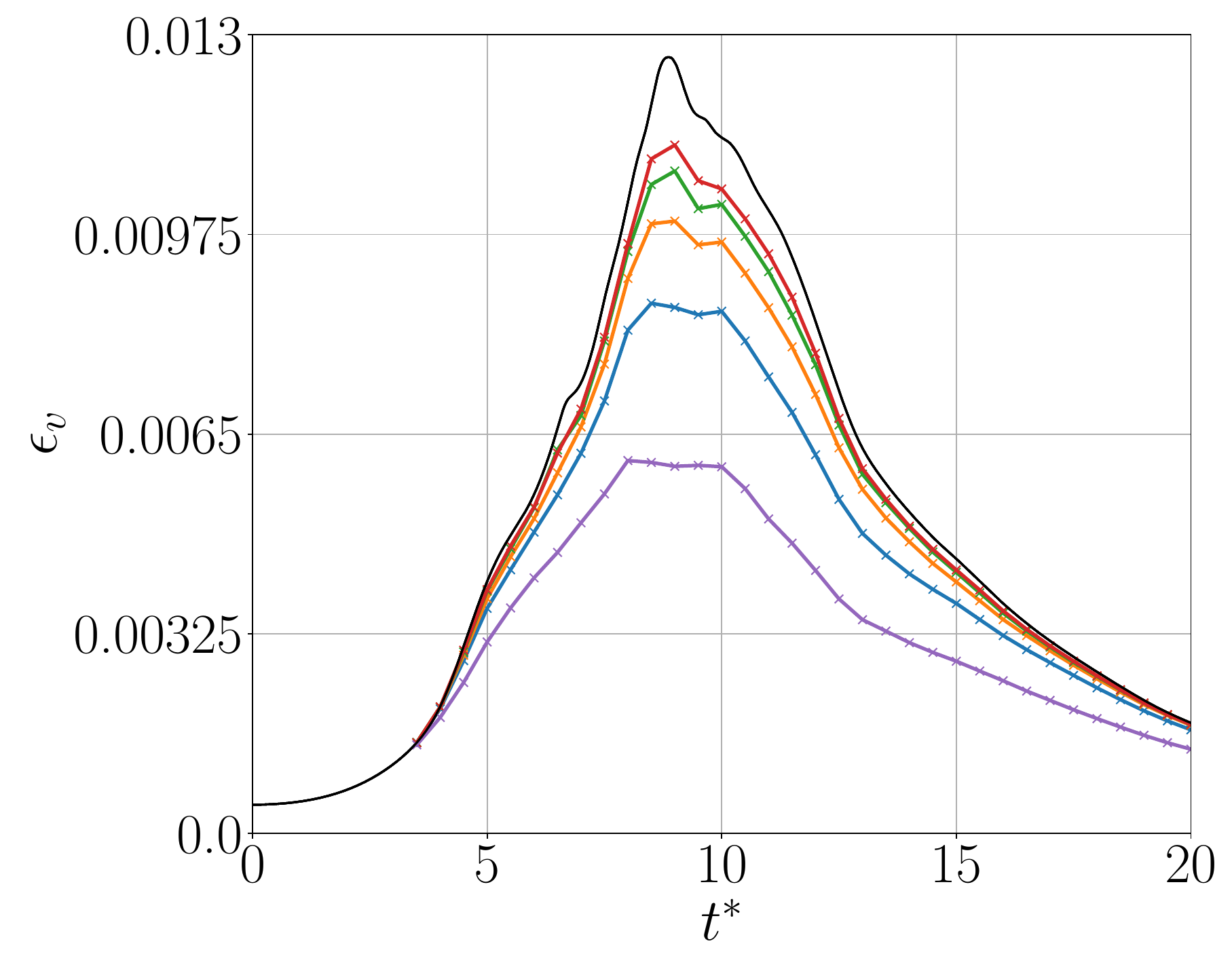}
     \caption{Viscous dissipation.}
     \label{fig:diss_res_0.5:c}
 \end{subfigure}
\caption{Numerical dissipation (left), kinetic energy dissipation rate (center) and viscous dissipation term (right) for thw ILES with a restart time window $\Delta T=0.5$.}
   \label{fig:diss_res_0.5}
\end{figure}

Figure~\ref{fig:diss_res_0.5} shows the numerical dissipation, the kinetic energy dissipation rate and the viscous dissipation for time window $\Delta T= 0.5$. In this scenario, the curves of numerical dissipation for various orders (Figure~\ref{fig:diss_res_0.5:a}) are monotonic. Lower polynomial degrees provide an amount of numerical dissipation which is larger with respect to the higher ones at all times. This is also confirmed by the kinetic energy dissipation rate~\ref{fig:diss_res_0.5:b} and the viscous dissipation~\ref{fig:diss_res_0.5:c}. This is in agreement with the results of the standard linear analysis, in which high order polynomials lead to better conservation of energy~\cite{moura2015linear,mengaldo2018spatial,mengaldo2018spatial2}. It is interesting to note that, throughout the whole simulation, the kinetic energy dissipation rate for all the ILES is greater than in the DNS. In the full ILES without restart, instead, this pattern was found only in the early stage of the evolution and lost closely after the enstrophy peak (see Figure~\ref{fig:diss_iles}). The primary underlying cause may be attributed to the high spatial gradients near the peak of enstrophy. Indeed, without restarting the simulation, high spatial gradients amplify the accumulation error which continues to grow for all the subsequent time steps due to the convective dominance of the flow. This is confirmed by comparing all the curves of the kinetic energy dynamics after the transition in the plots of Figure~\ref{fig:diss_iles}. As the transition to turbulence occurs, the dynamics becomes completely polluted by the accumulation errors.

\begin{figure}[h!]
 \centering  
 \begin{subfigure}{0.3\textwidth}
     \includegraphics[width=\textwidth]{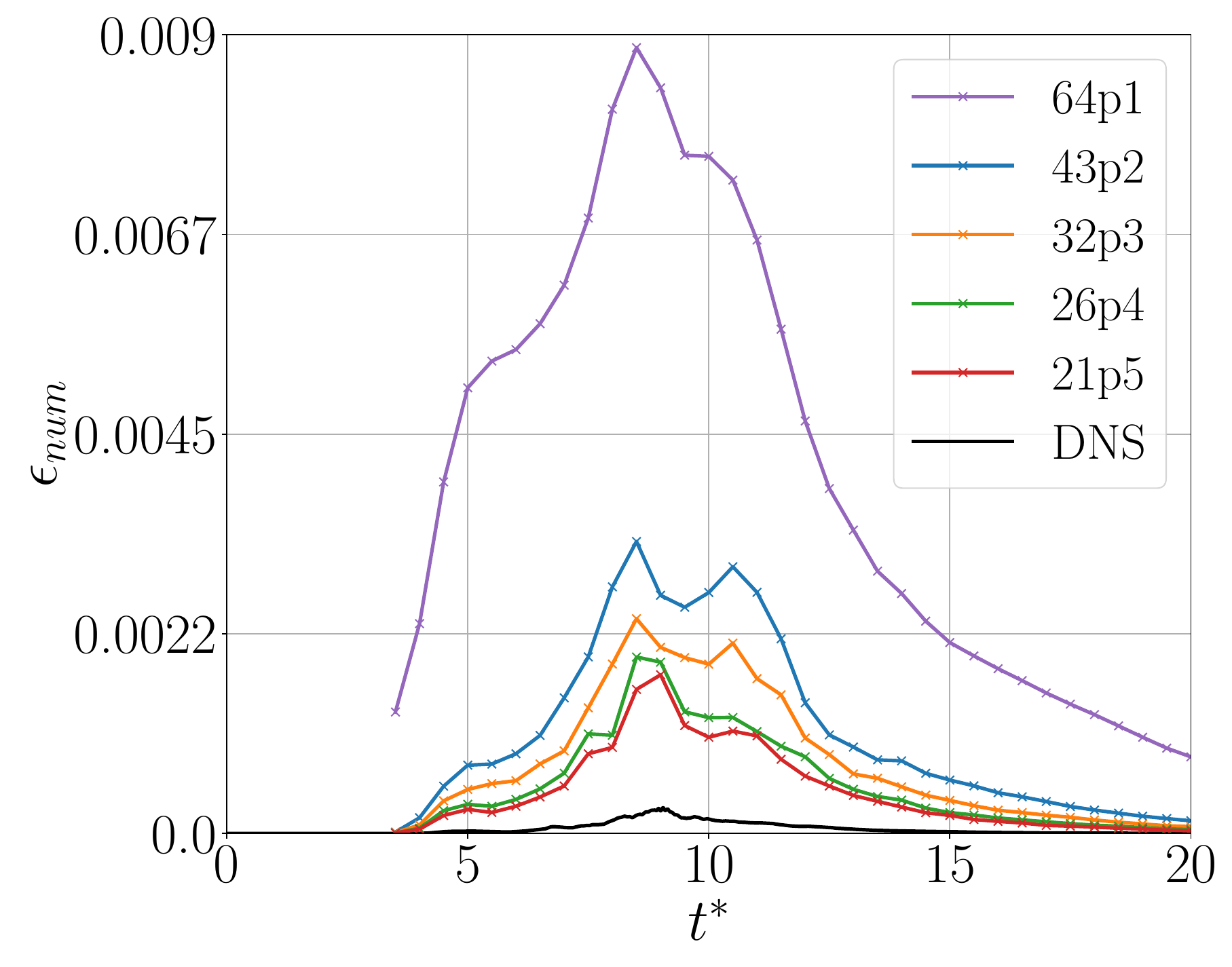}
     \caption{Numerical dissipation.}
     \label{fig:diss_res_4.0:a}
 \end{subfigure}
 \begin{subfigure}{0.30\textwidth}
     \includegraphics[width=\textwidth]{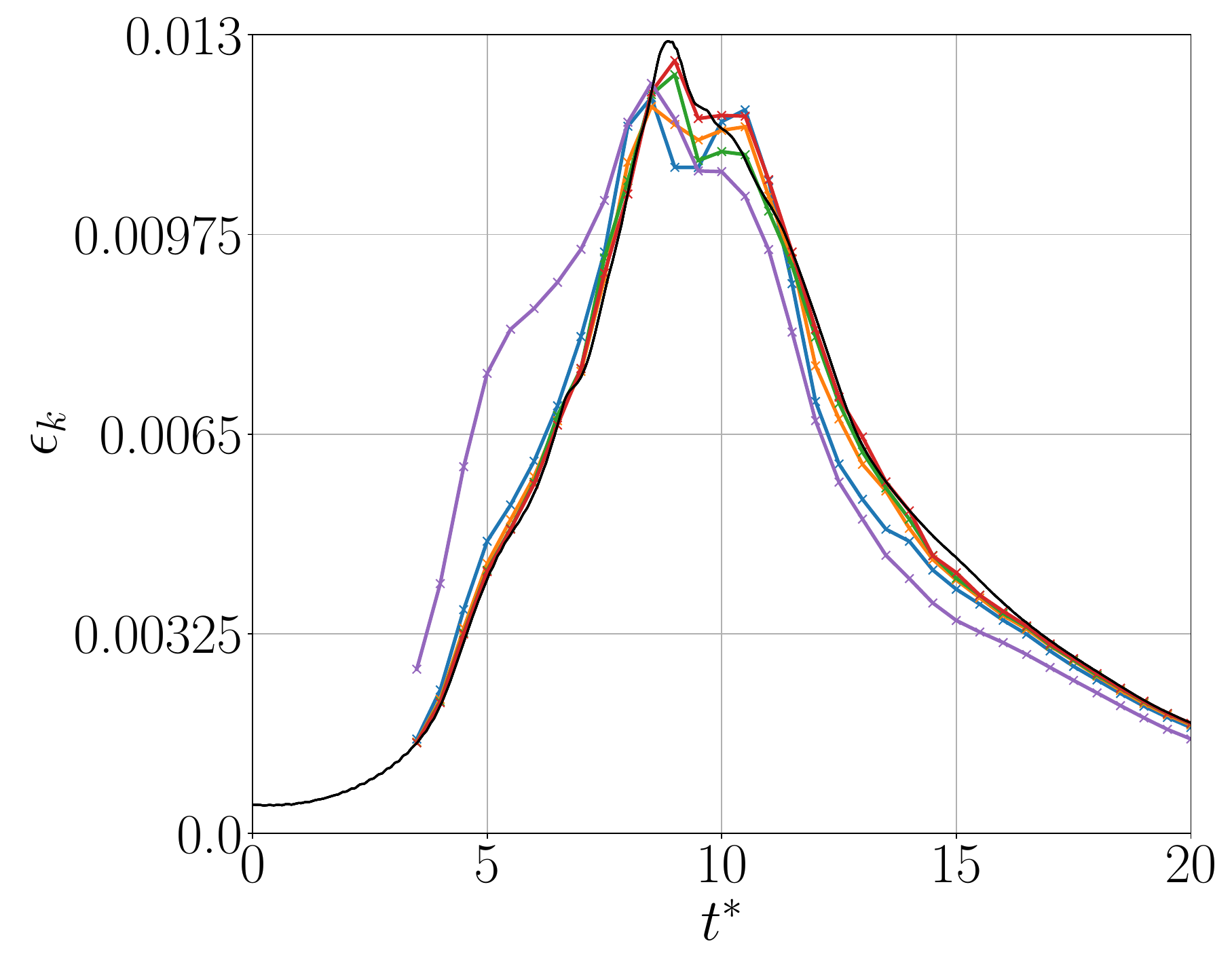}
     \caption{Kinetic energy dissipation rate.}
     \label{fig:diss_res_4.0:b}
 \end{subfigure}
 \begin{subfigure}{0.3\textwidth}
     \includegraphics[width=\textwidth]{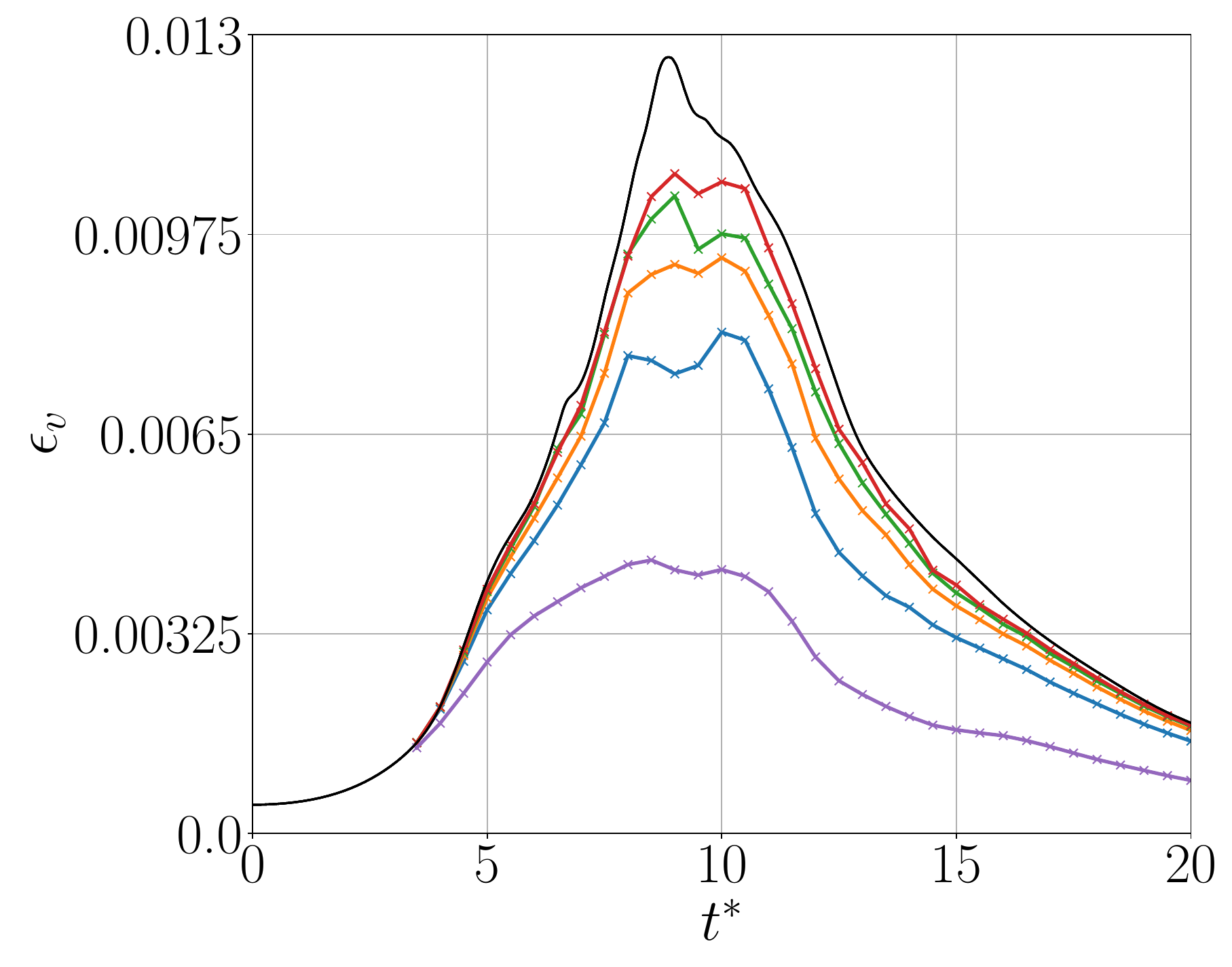}
     \caption{Viscous dissipation.}
     \label{fig:diss_res_4.0:c}
 \end{subfigure}
\caption{Numerical dissipation (left), kinetic energy dissipation rate (center) and viscous dissipation term (right) for thw ILES with a restart time window $\Delta T=4.0$.}
   \label{fig:diss_res_4.0}
\end{figure}

In Figure~\ref{fig:diss_res_4.0} the same quantities are plotted for time window $\Delta T= 4.0$. Note that a selected time window of $\Delta T = 4$ is already large enough to observe significant temporal effects. The accumulation error already starts to contaminate the dynamics, as revealed by the kinetic energy dissipation rate in Figure~\ref{fig:diss_res_4.0}. However, we can observe that, in this case, the ordering of the curves for various orders of accuracy for the viscous dissipation and the numerical dissipation is still maintained.

\begin{figure}[h!]
 \centering  
 \begin{subfigure}{0.3\textwidth}
     \includegraphics[width=\textwidth]{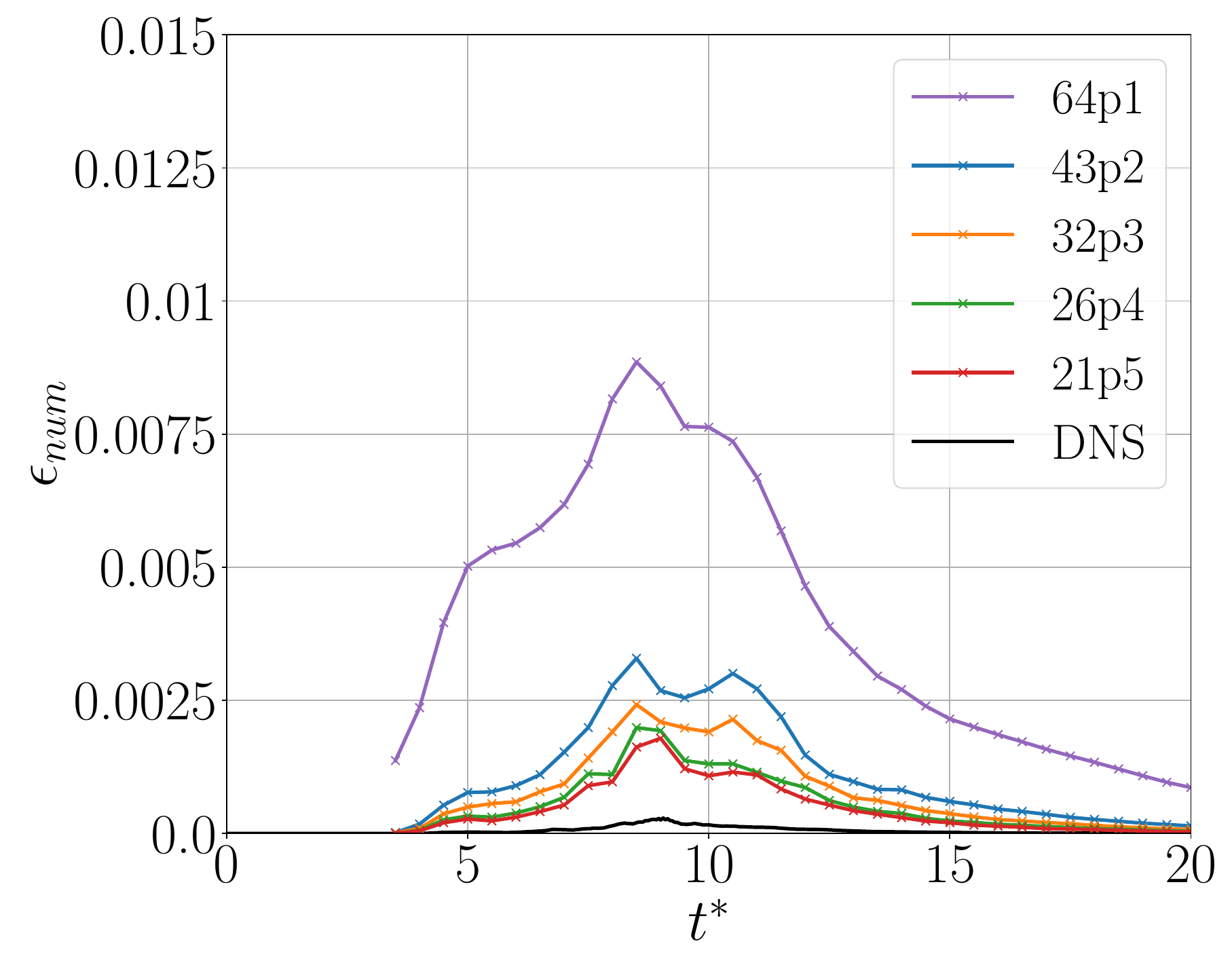}
     \caption{$\Delta T = 4.0$.}
     \label{fig:comp_num_diss:a}
 \end{subfigure}
 \begin{subfigure}{0.30\textwidth}
     \includegraphics[width=\textwidth]{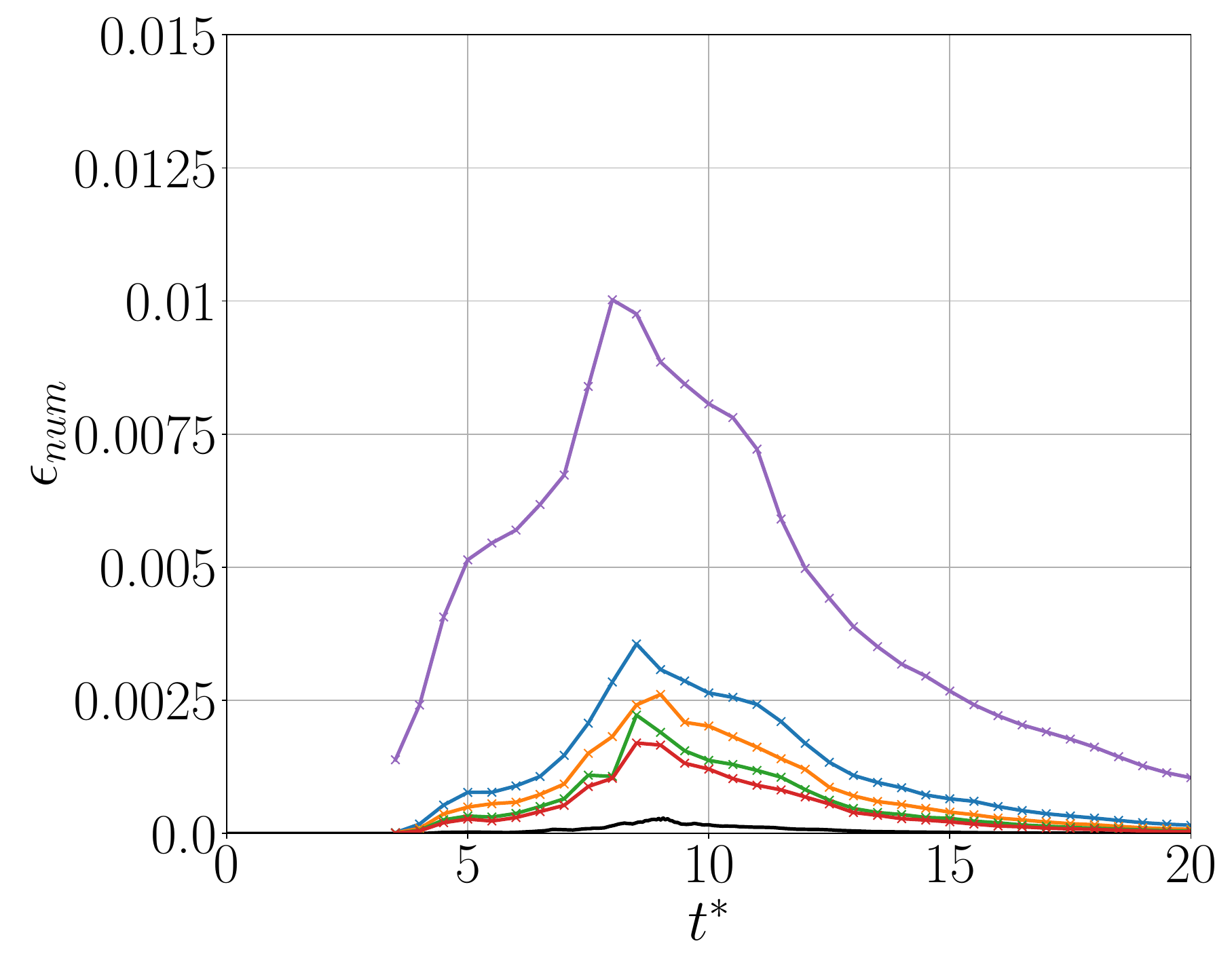}
     \caption{$\Delta T = 2.0$.}
     \label{fig:comp_num_diss:b}
 \end{subfigure}
 \begin{subfigure}{0.3\textwidth}
     \includegraphics[width=\textwidth]{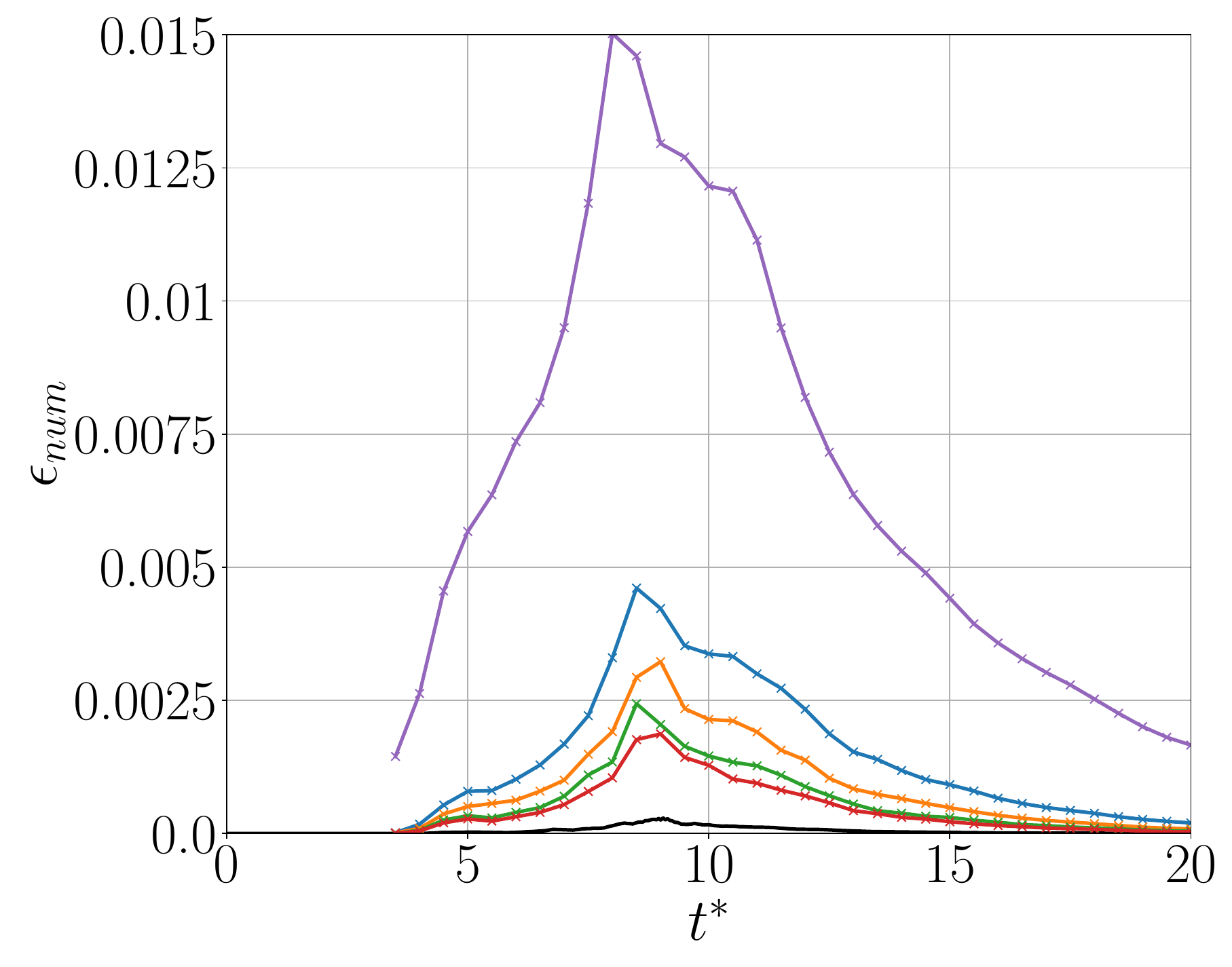}
     \caption{$\Delta T = 0.5$.}
     \label{fig:comp_num_diss:c}
 \end{subfigure}
\caption{Numerical dissipation evolution of various polynomial orders for different time windows.}
   \label{fig:comp_num_diss}
\end{figure}

\begin{figure}
    \centering
    \includegraphics[width=0.65\linewidth]{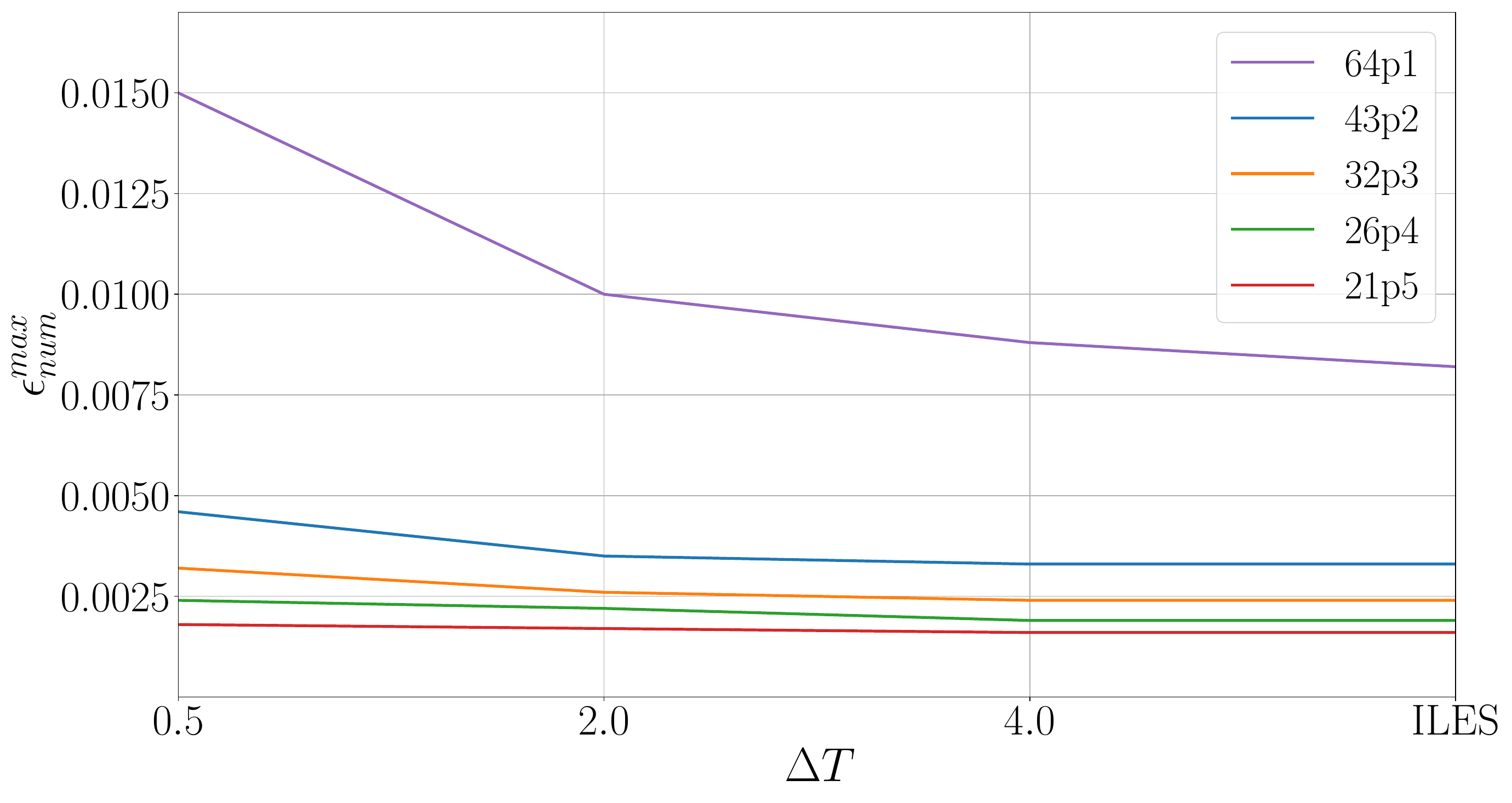}
    \caption{Maximum value of numerical dissipation \eqref{eqn:diss_num} measured throughout the whole time evolution for different time windows and for different polynomial orders.}
    \label{fig:MaxEPSnum}
\end{figure}

Finally, in Figure~\ref{fig:comp_num_diss} we compare the numerical dissipation for the various restart time windows.
First of all, it is clear that for the lowest order in consideration ($\mathrm{p}=1$) the numerical dissipation increases as the time window shortens. Although less evidently, the same trend is also found for the other orders.
In order to better inspect the behavior of the numerical dissipation, in Figure~\ref{fig:MaxEPSnum} we show its maximum value observed throughout the time evolution.
By focusing on the lowest polynomial order, $\mathrm{p}=1$, the numerical dissipation is significantly affected by the restart time window. Starting from a magnitude of $0.0150$ for $\Delta T=0.5$, it appears that it converges to a value of $0.009$ when the restart time window increases.
This value is close to the one obtained without restart. Instead, for high order polynomials, the influence of the restart time window on numerical dissipation becomes less significant, although each order of accuracy converges toward a certain value.
Given this trend, we can suppose that if the time window is large enough, the maximum numerical dissipation found during the whole simulation reaches an asymptotic value. Indeed, as the temporal effects become relevant, they obscure the purely dissipative effects of the space discretization. 
%
%
\subsection{Training data preparation}\label{subsec:trainigData}
The training data has been restricted in the time window $t^{*}$ in $[3.5,20]$ since the relevant dynamics starts to develop after $t^* =4$~\cite{taylor1937mechanism,Brachet83}. Furthermore, in this way we can also avoid the influence of initial conditions that may affect the training process and lead to biased results. As mentioned in the previous section, the snapshots are collected at intervals of $\delta t =0.5$. This interval ensures that the model captures key temporal features without introducing unnecessary data redundancy that may increase the risk of overfitting. The dataset was partitioned into two subsets: $80\%$ was used for training the model, while the remaining $20\%$ was reserved for validation of the model's performance and generalization capability. All the ILES snapshots (restarted or not) have been interpolated on the corresponding DNS grid in order to ensure the same number of elements for inputs and outputs in the training. 

It is important to highlight that although low polynomials orders are considered in this framework (as low-order ILES), the proposed modal filter is only applied to DNS data, which are characterised by a high-order discretization (\ie, $6^{\mathrm{th}}$-order). The application of modal filters to low-order simulations would be, in fact, ill-posed by definition.
\section{Results} \label{sec:results}
In the previous sections, we presented all the components required to construct the data-driven filter. We introduced the general structure of the filter as a linear combination of sharp modal filters and discussed the neural network architecture used for the approximation of the weights of the linear combination. Moreover, we assessed the ILES (restarted or not) for different polynomial orders by inspecting the various contributions of the domain-averaged kinetic energy equation.
\subsection{Evaluation of network's accuracy}\label{subsec:MSE-CC}
Now that the whole methodology has been presented, we conduct a quantitative analysis of the network's accuracy according to two metrics:  Mean Square Error and Cross-Correlation between the ILES snapshots and the filtered data obtained trough the data-driven filter.

\begin{figure}[h!]
        \centering
        \begin{subfigure}[t]{0.245\textwidth}
            \centering
            \includegraphics[width=\linewidth,height=50mm, keepaspectratio]{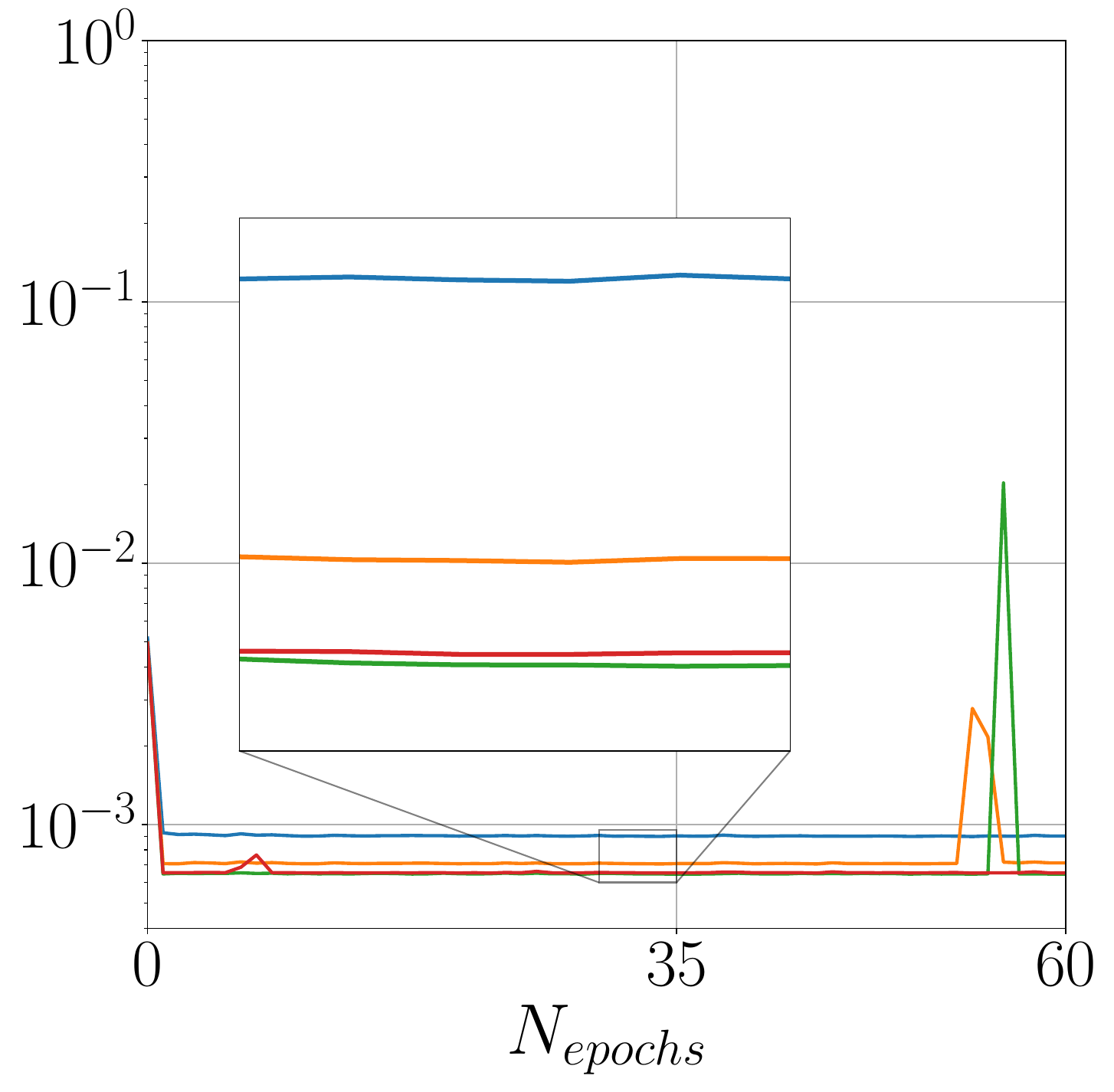}
            \caption{$\Delta T=0.5$.}
            \label{fig:Lt_05}
        \end{subfigure}
        \hfill
        \begin{subfigure}[t]{0.245\textwidth}
            \centering
            \includegraphics[width=\linewidth,height=40mm, keepaspectratio]{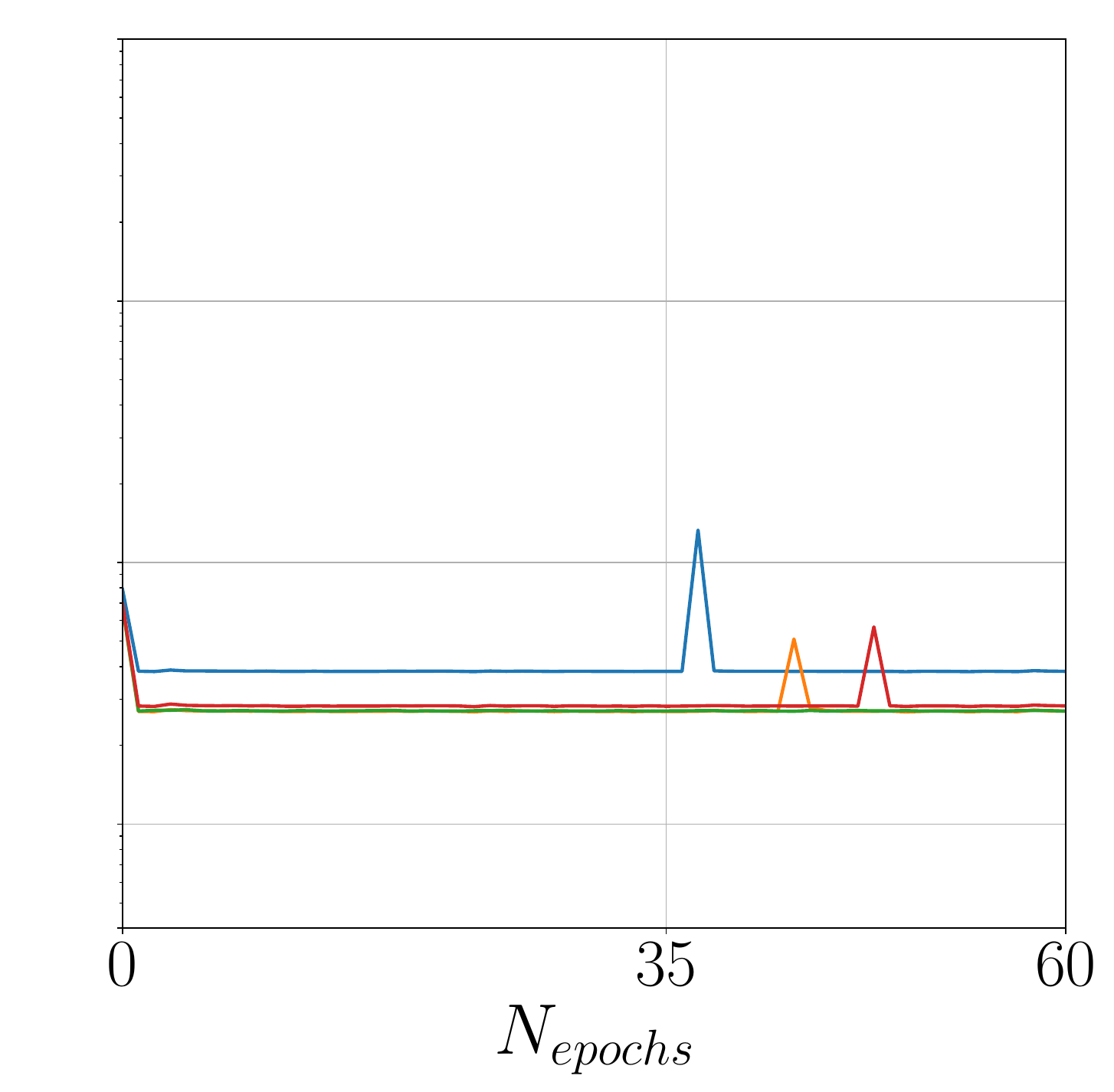}
            \caption{$\Delta T=2$.}
            \label{fig:Lt_2}
        \end{subfigure}
        \hfill
        \begin{subfigure}[t]{0.245\textwidth}
            \centering
            \includegraphics[width=\linewidth,height=50mm, keepaspectratio]{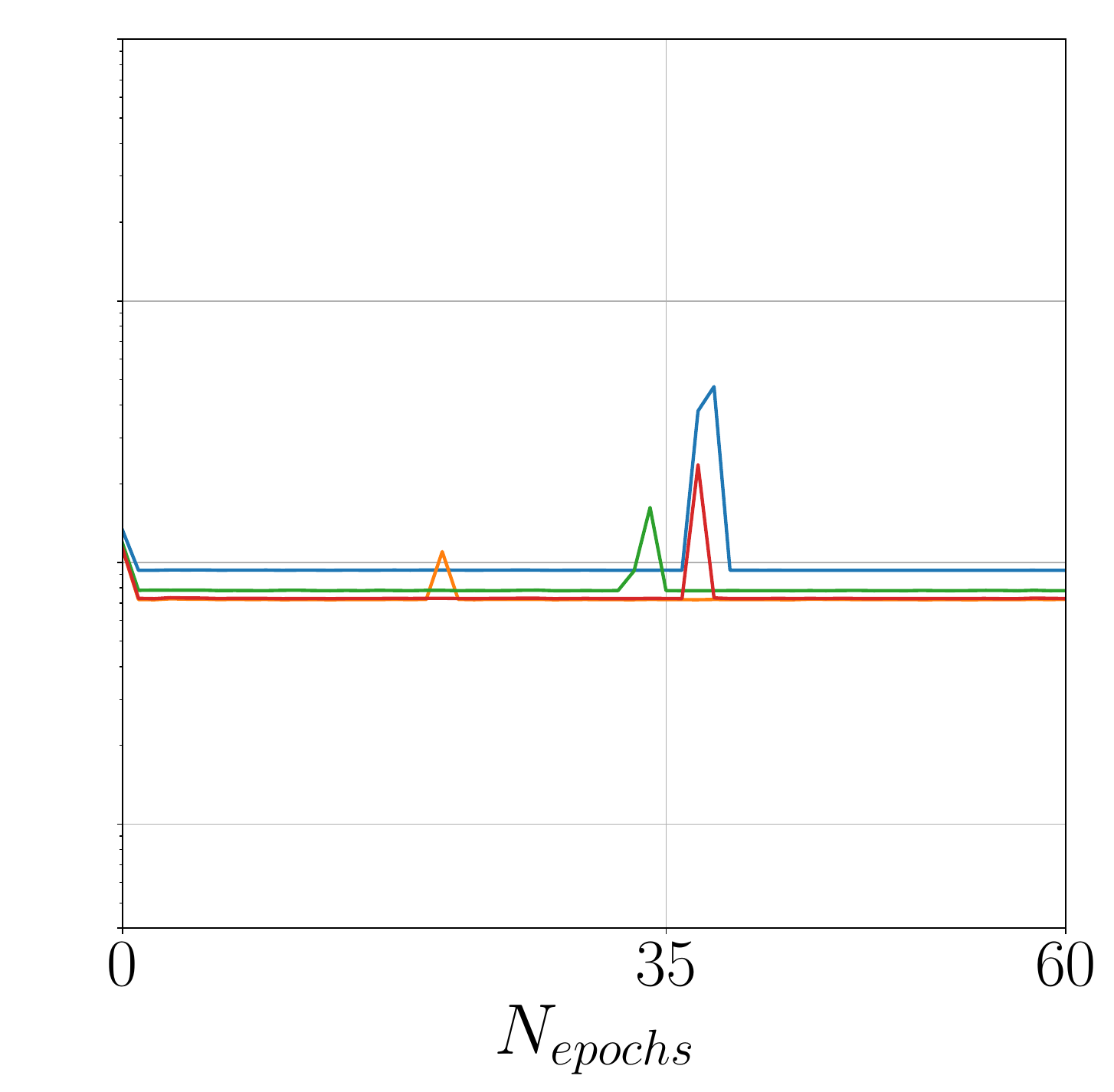}
            \caption{$\Delta T=4$.}
            \label{fig:Lt_4}
        \end{subfigure}
        \hfill
        \begin{subfigure}[t]{0.245\textwidth}
            \centering
            \includegraphics[width=\linewidth,height=40mm, keepaspectratio]{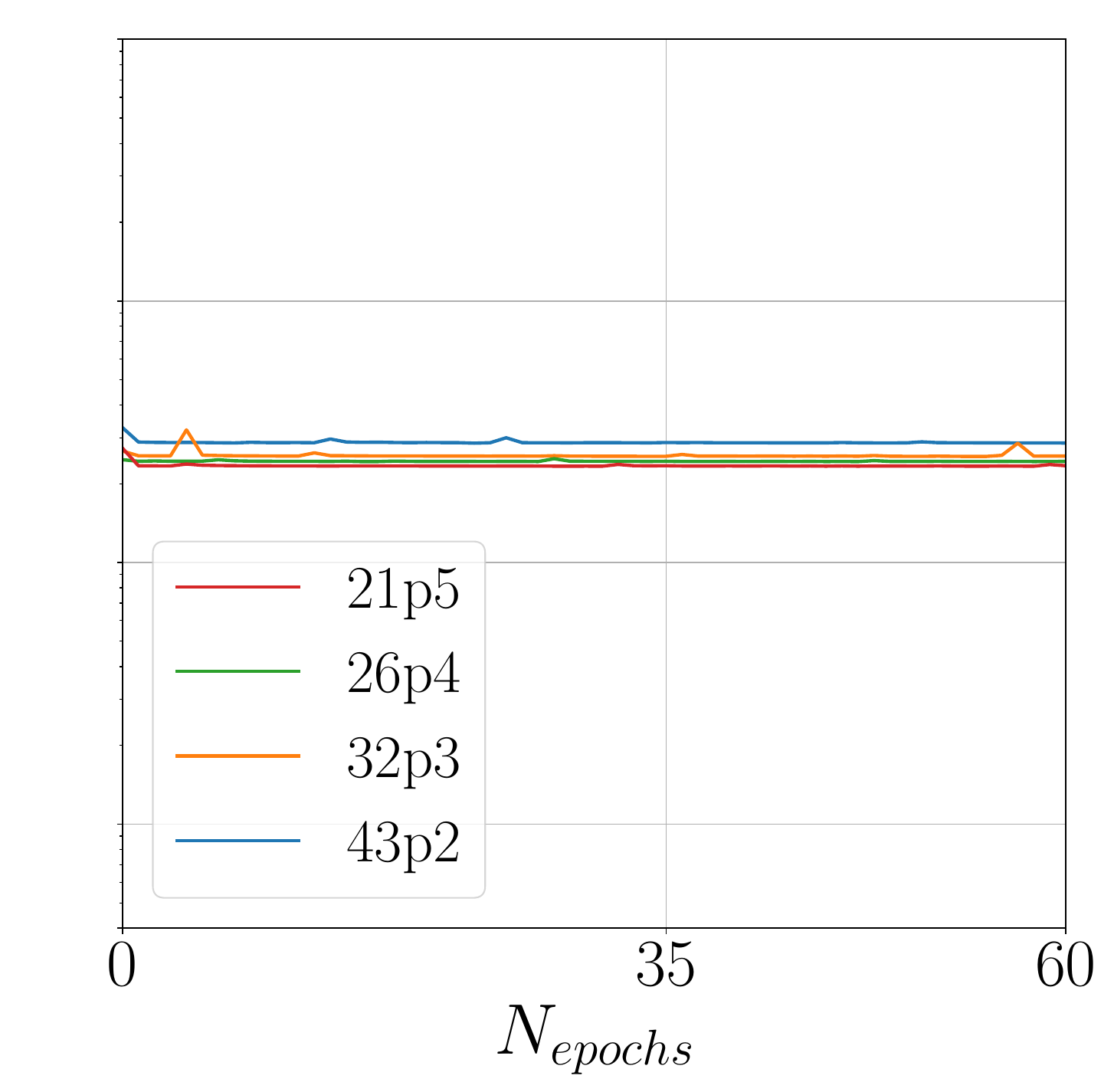}
            \caption{$\mathrm{ILES}$.}
            \label{fig:Lt_iles}
        \end{subfigure}
        \caption{Training MSE against epochs ($N_{epochs}$) for different polynomial orders. The integration time span increases from left to right.}
        \label{fig:TRAIN_LOSS_TIME_WINDOWS2}
\end{figure}

Figure~\ref{fig:TRAIN_LOSS_TIME_WINDOWS2} compares the training MSE obtained from different time windows for all the polynomial orders in consideration.

These plots show that as the time integration window lengthens, the error increases. For the network, linking the features of the filtered DNS to the ILES becomes increasingly difficult for long time-integration, as the accumulation error causes the trajectories of the DNS and the ILES to depart from each other.

Focusing on the smaller time window (figure \ref{fig:Lt_05}) the loss function increases as the polynomial degree decreases with the polynomial degrees $\mathrm{p}=4$-$5$ being almost overlapped. As the polynomial order decreases, the training becomes more difficult ($\mathrm{p}\leq 3$), resulting in slightly higher values for the loss functions.
As the order of approximation decreases, the flow features of the ILES deviate more significantly from those observed in the DNS, due to numerical dissipation. Thus, this discrepancy introduces challenges for the network in accurately mapping these flow features to the coefficients $\theta^{\mathrm{ANN}}$. Furthermore, the higher values of the loss function observed for low-order polynomials might be attributed to the simple modeling assumption made in equation \eqref{eqn:LinearCombSharpModal} which not only assumes that the resulting filter is a simple linear combination of sharp filters, but that it also acts in a purely ``local'' fashion, with only values inside each element interacting. Consequently, the jumps between elements are completely ignored by the model. It has been shown that the magnitude of the inter-element jumps increases as the polynomial degree decreases in the context of the Discontinuous Galerkin method~\cite{Fernndez2018OnTA,NguyenHybPDE,PeraireHyb,PeraireEmbedded}. A similar trend is also expected for the spectral difference scheme.
Consequently, the current methodology may experience some limitations when addressing lower polynomial orders.
Along similar lines, regarding the locality of the model, Benjamin \& Iaccarino~\cite{Iaccarino23LES1,benjamin2024neuralnetworkbasedclosuremodels} have shown that the input stencil size affects drastically the construction of a data-driven SGS stress tensor in the context of finite-differences for the plane channel flow benchmark. Therefore, we believe that an increase of the stencil for the filter would be beneficial.
However, at the same time, in order to preserve the compactness of the underlying discretization, local filters, acting element-wise, are more suitable within the framework of DSEMs.

It is interesting to note that the loss function becomes increasingly high when the time window increases. Indeed, for the case without restart (Figure~\ref{fig:Lt_iles}), the loss function is nearly twice as larger than the case with $\Delta T=0.5$ (Figure~\ref{fig:Lt_05}).
Moreover, note that the difference between $\mathrm{p=2}$ and the other orders gets smaller as $\Delta T$ increases. While for $\Delta T=0.5$ and $\Delta T=2$ there is a clear distinction between $\mathrm{p}=2$ and the other orders, the curve for $\mathrm{p=2}$ approaches the other curves as the time window increases. This occurs because, over extended time windows, accumulation errors due to time integration become dominant for all orders of approximation, resulting in similar values of the loss function. Indeed, we remark that the model itself is based on a spatial filtering operation which neglects temporal effects. These are although significant when full trajectories are considered (like in the ILES without restart).

The  peaks around $N_{epochs}=60$ for $\mathrm{p}=3$ and $\mathrm{p}=4$ could be attributed to the model difficulty in adapting to the full range of training data. This challenge likely arises from the inclusion of data from both the laminar and fully turbulent regimes in the training dataset that exhibit vastly different flow features. 
This results in some out-of-distribution peaks since the optimal parameters aim at characterizing both laminar and turbulent regimes, which can be a quite challenging task for the network.
To support this hypothesis, in \ref{:train_II} we report some of the results obtained by training only on data for $t^*$ in $[10,20]$.

\begin{figure}[h!]
        \centering
        \begin{subfigure}[t]{0.245\textwidth}
            \centering
            \includegraphics[width=\linewidth,height=50mm, keepaspectratio]{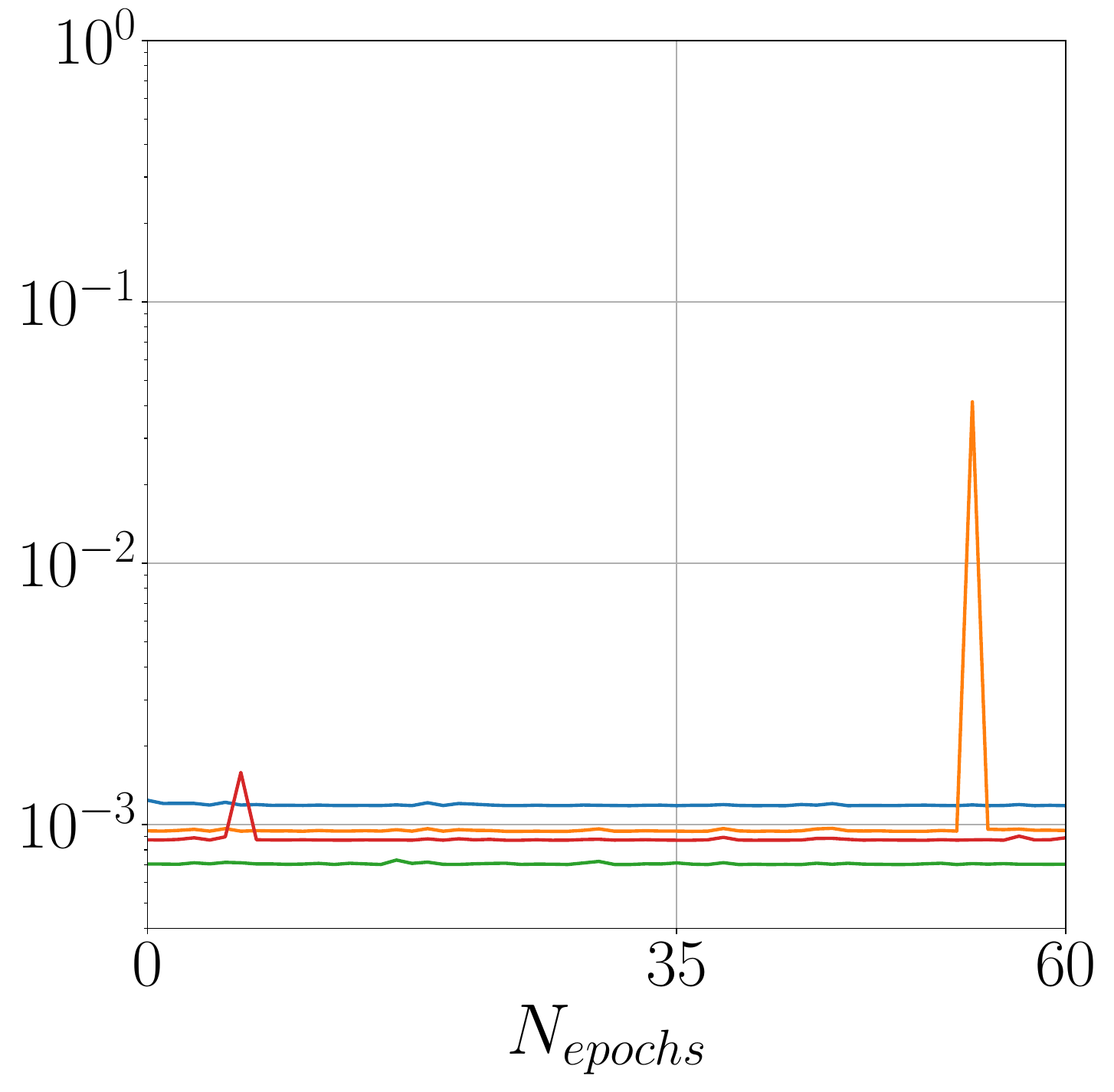}
            \caption{$\Delta T=0.5$.}
            \label{fig:Lv_05}
        \end{subfigure}
        \hfill
        \begin{subfigure}[t]{0.245\textwidth}
            \centering
            \includegraphics[width=\linewidth,height=40mm, keepaspectratio]{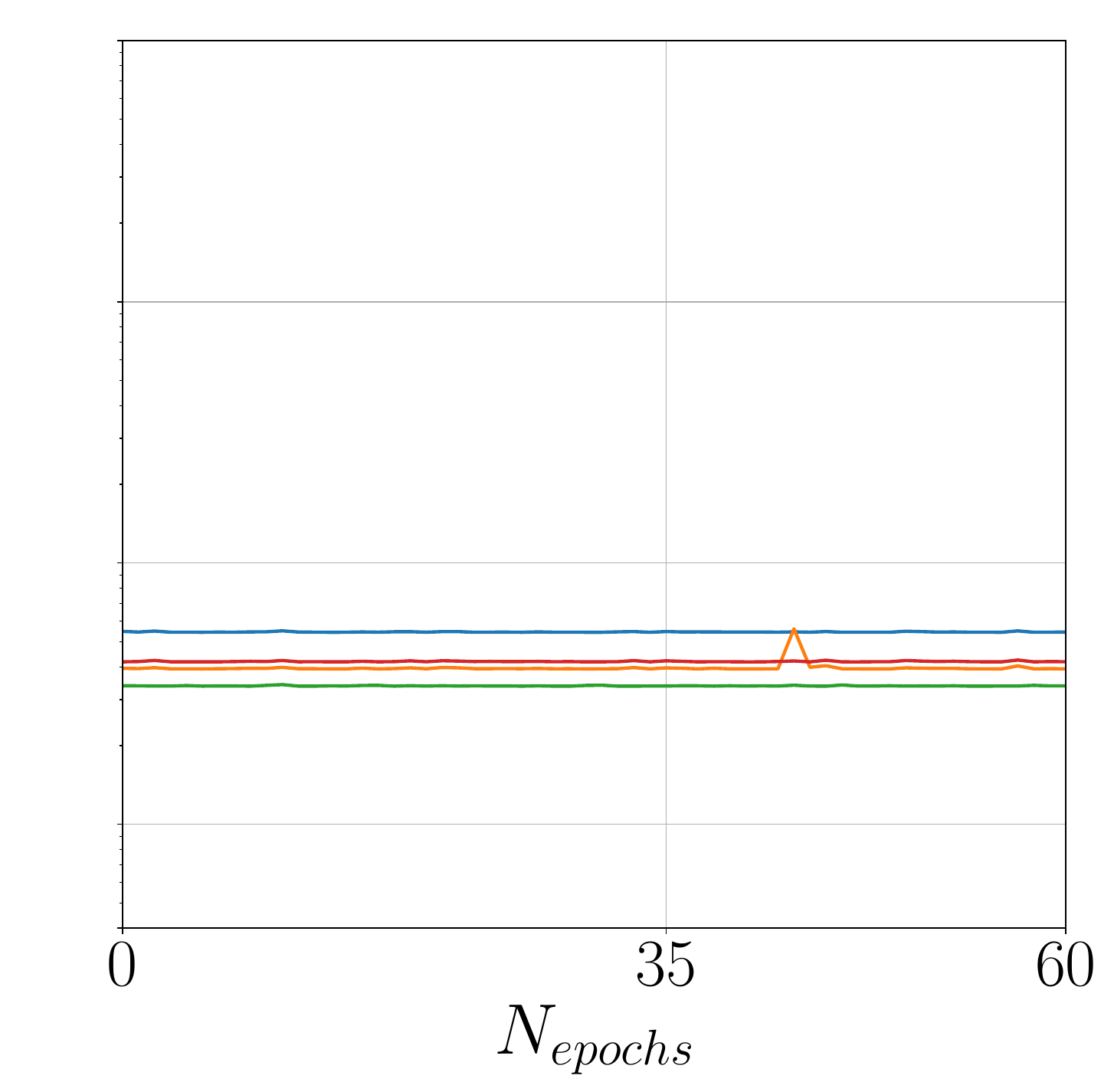}
            \caption{$\Delta T=2$.}
            \label{fig:Lv_2}
        \end{subfigure}
        \hfill
        \begin{subfigure}[t]{0.245\textwidth}
            \centering
            \includegraphics[width=\linewidth,height=50mm, keepaspectratio]{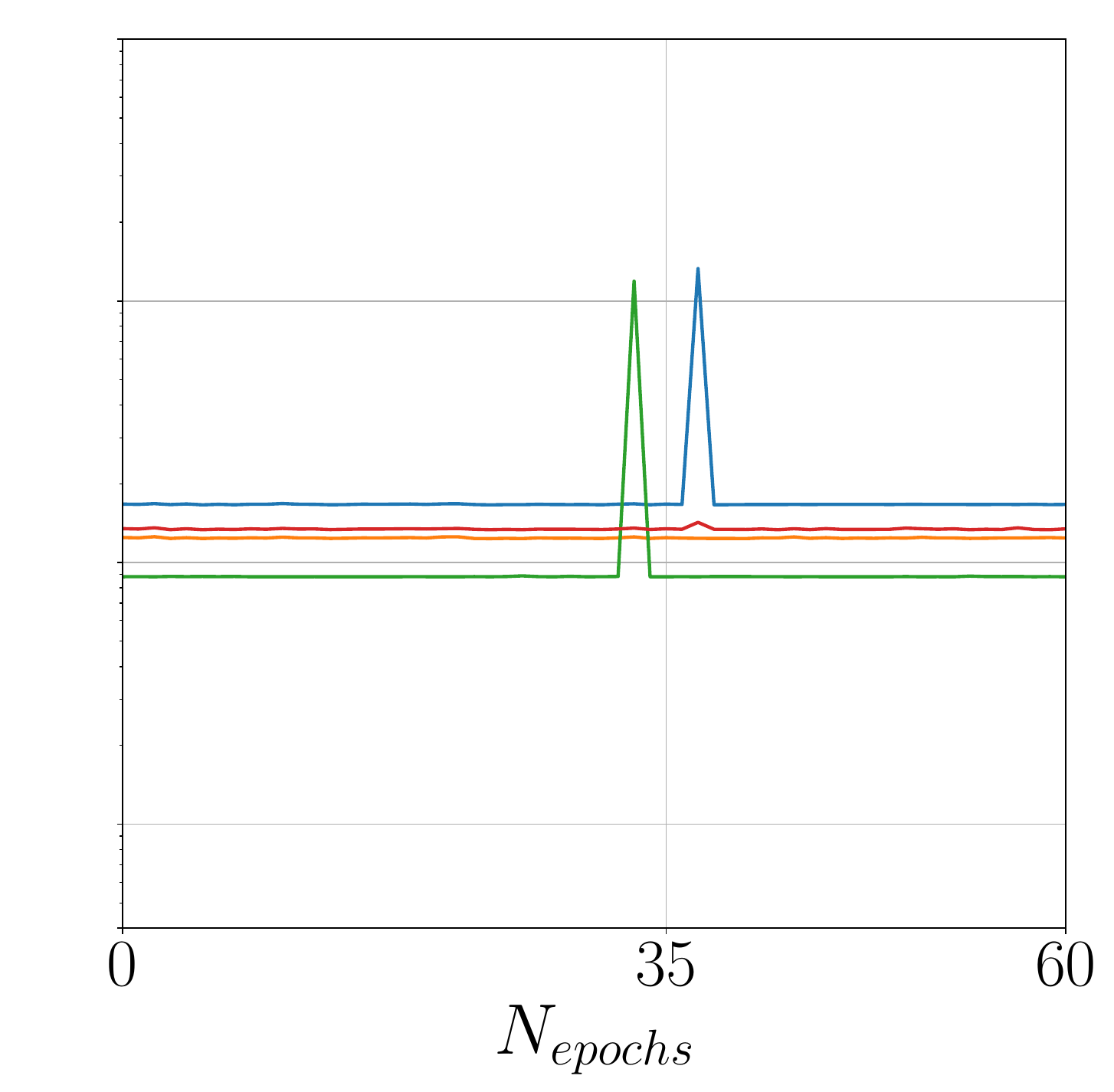}
            \caption{$\Delta T=4$.}
            \label{fig:Lv_4}
        \end{subfigure}
        \hfill
        \begin{subfigure}[t]{0.245\textwidth}
            \centering
            \includegraphics[width=\linewidth,height=40mm, keepaspectratio]{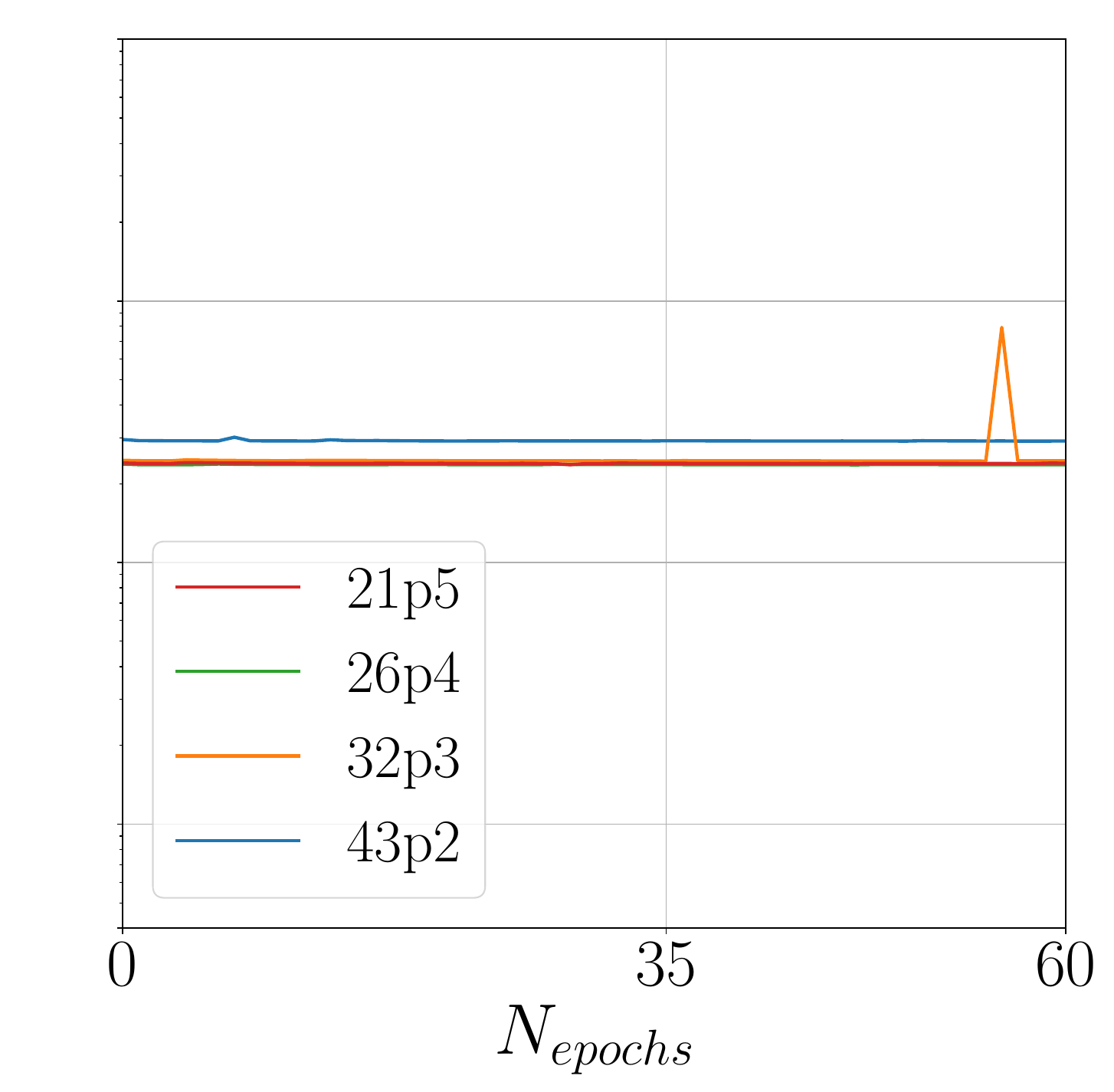}
            \caption{$\mathrm{ILES}$.}
            \label{fig:Lv_iles}
        \end{subfigure}
        \caption{Validation MSE against epochs ($N_{epochs}$) for different polynomial orders. The integration time span increases from left to right.}
        \label{fig:VAL_LOSS_TIME_WINDOWS2}
\end{figure}

For completeness, the validation curves for different time integration windows are plotted in Figure~\ref{fig:VAL_LOSS_TIME_WINDOWS2}. The general trend is similar to what found in the training set: the loss function becomes increasingly higher as the time window increases. However, for $\Delta T=0.5,2,4$, there is not a clear distinction between $\mathrm{p=2}$ and the other orders. Instead, for the case without restart, the network is able of differentiating $\mathrm{p=2}$ from the others.

\begin{figure}[htb]
 \centering  
 \begin{subfigure}{0.325\textwidth}
     \includegraphics[width=\textwidth]{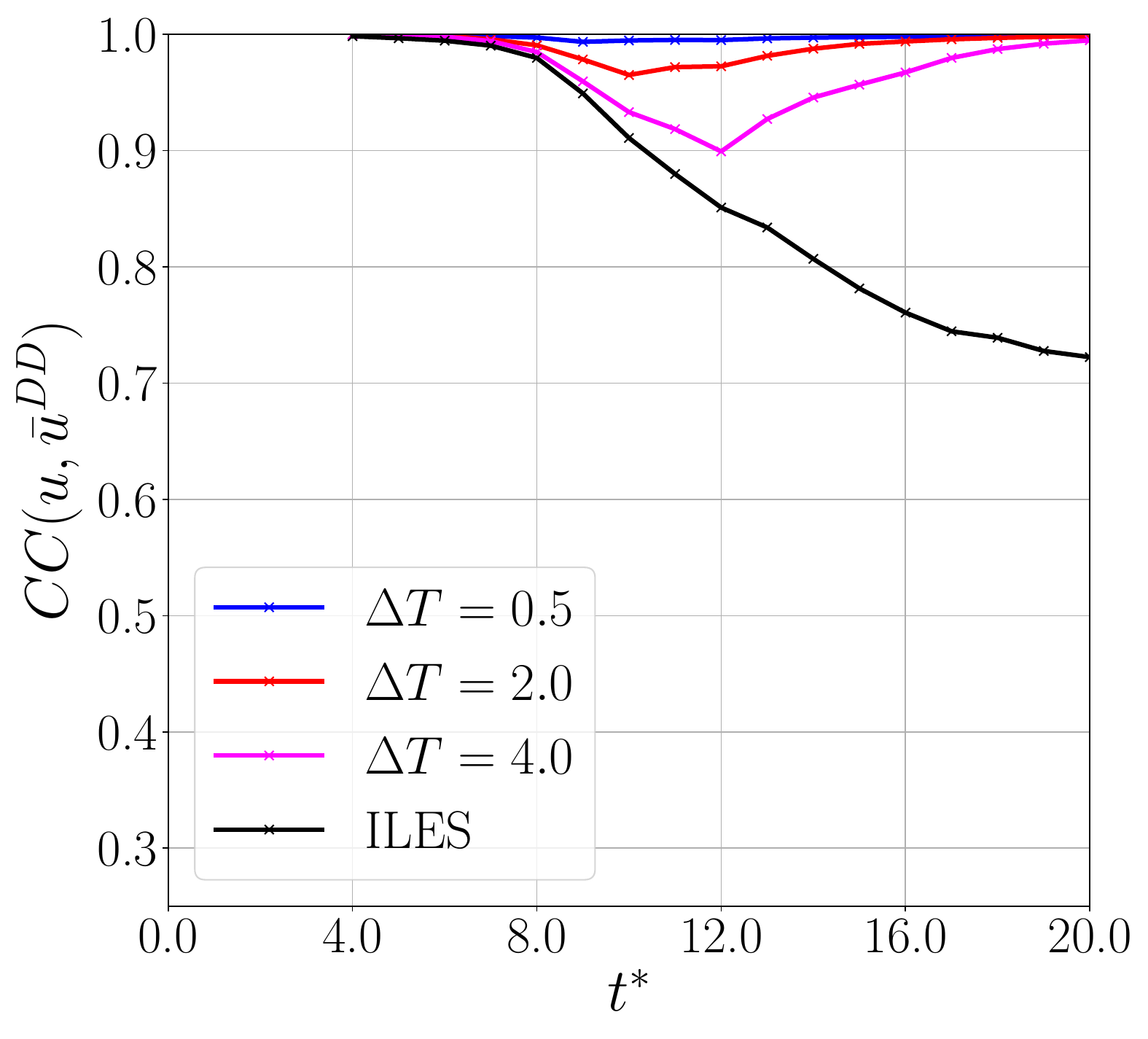}
     \caption{Component $u_1$.}
     \label{fig:corr_21p5_u}
 \end{subfigure}
 \begin{subfigure}{0.325\textwidth}
     \includegraphics[width=\textwidth]{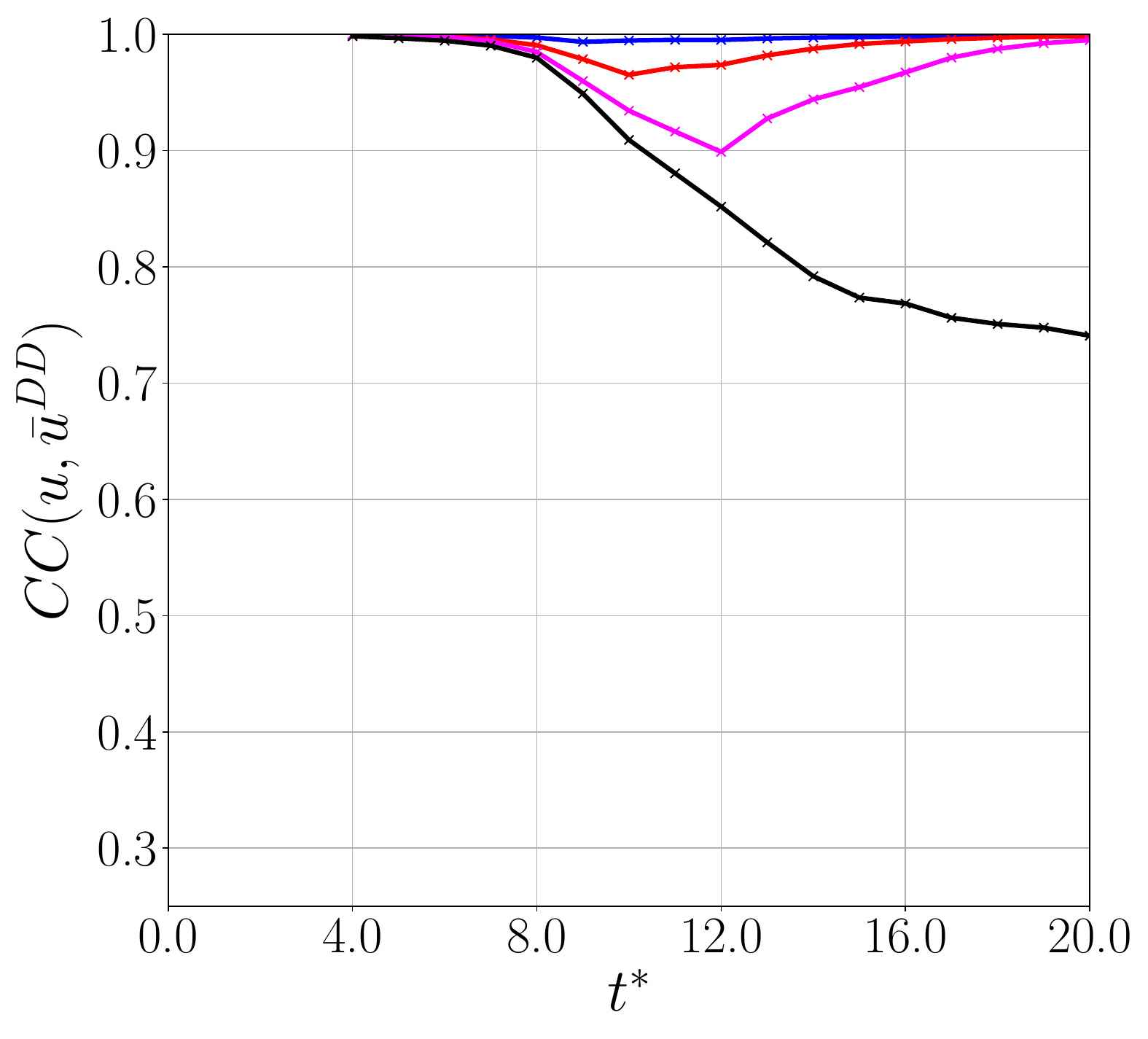}
     \caption{Component $u_2$.}
     \label{fig:corr_21p5_v}
 \end{subfigure}
 \begin{subfigure}{0.325\textwidth}
     \includegraphics[width=\textwidth]{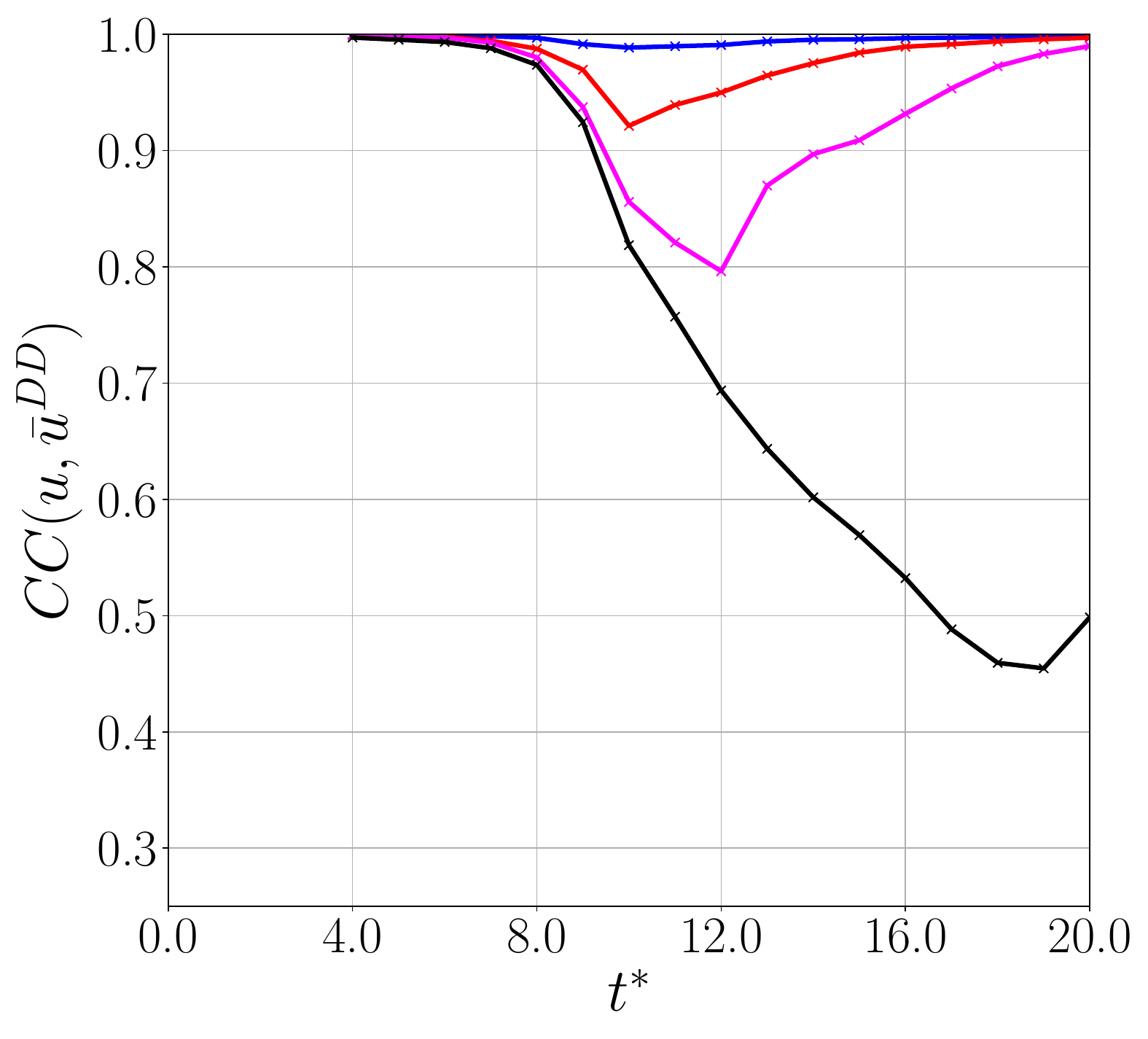}
     \caption{Component $u_3$.}
     \label{fig:corr_21p5_w}
 \end{subfigure}
 \vspace{-0.2cm}
\caption{Correlation coefficients for different velocity components for 21p5.}
\label{fig:Corr21p5}
\end{figure}

The loss function provides a ``global'' quantification of the accuracy over time, offering an overall assessment of the model's performance across the entire dataset. Instead, in order to assess the accuracy of the method at different times, and consequently different flow regimes, we compute the cross correlation $CC$ between the targets $\mathbf{u}$ and the filtered data-driven solution $\overline{\mathbf{u}}^{DD}$ as:
\begin{equation}
    CC(\mathbf{u},\overline{\mathbf{u}}^{DD}) = \frac{\sum_{ijk}\left(\mathbf{u}_{ijk} -\langle \mathbf{u}_{ijk} \rangle)(\overline{\mathbf{u}}_{ijk}^{DD} -\langle \overline{\mathbf{u}}_{ijk}^{DD} \rangle \right)}{\sqrt{\sum_{ijk}( \mathbf{u}_{ijk} -\langle \mathbf{u}_{ijk} \rangle)^2} \sqrt{\sum_{ijk}(\overline{\mathbf{u}}_{ijk}^{DD} -\langle \overline{\mathbf{u}}_{ijk}^{DD} \rangle)^2}}, \label{eq:cross_corr}
\end{equation}
where $i,j,k=1,...,N$. The relation \eqref{eq:cross_corr} is applied component-wise.
The plots in Figure~\ref{fig:Corr21p5} illustrate the correlation coefficients computed at different times for the highest polynomial order in consideration (\ie, $\mathrm{p}=5$) for all the training cases considered.

For $\Delta T=0.5$ a strong positive correlation is evident for all times (around $99,9\%$), with a slight decrease close to $t^{*}=10$. This also happens for all the other time windows, though the values decrease as the time window lengthens. The primary reason for the overall decrease during the initial time span is that the fields become increasingly less smooth as we approach the peak of enstrophy, where the spatial gradients reach their maximum and the network is the most challenged in accurately predicting the mapping between DNS and ILES. After $t^* \simeq 10$, turbulence starts to decay under the effect of viscosity, allowing the neural network to better extrapolate the main features of the fields. For the models trained on data without restarting the simulations, instead, the correlation continues to decrease over time as the accumulation errors keep growing. For restarted cases, instead, at each restart they are set to zero. 

Note that, for the third component $u_3$, the correlation is lower with respect to the other two. This is likely related to the training process itself. For the particular regime of the test case herein considered, compressibility effects are negligible and, as a results, the symmetry in the first two velocity components holds throughout the whole simulation \cite{taylor1937mechanism,Brachet83}, namely; $u_1(x_1,x_2,x_3,t^{*})=u_2(x_2,\pi-x_1,x_3,t^{*})$. This relation influences the interpretability of the network, causing a bias toward the first two components and resulting in smaller correlations for the third one.

\begin{figure}[htb]
 \centering  
 \begin{subfigure}{0.32\textwidth}
     \includegraphics[width=\textwidth]{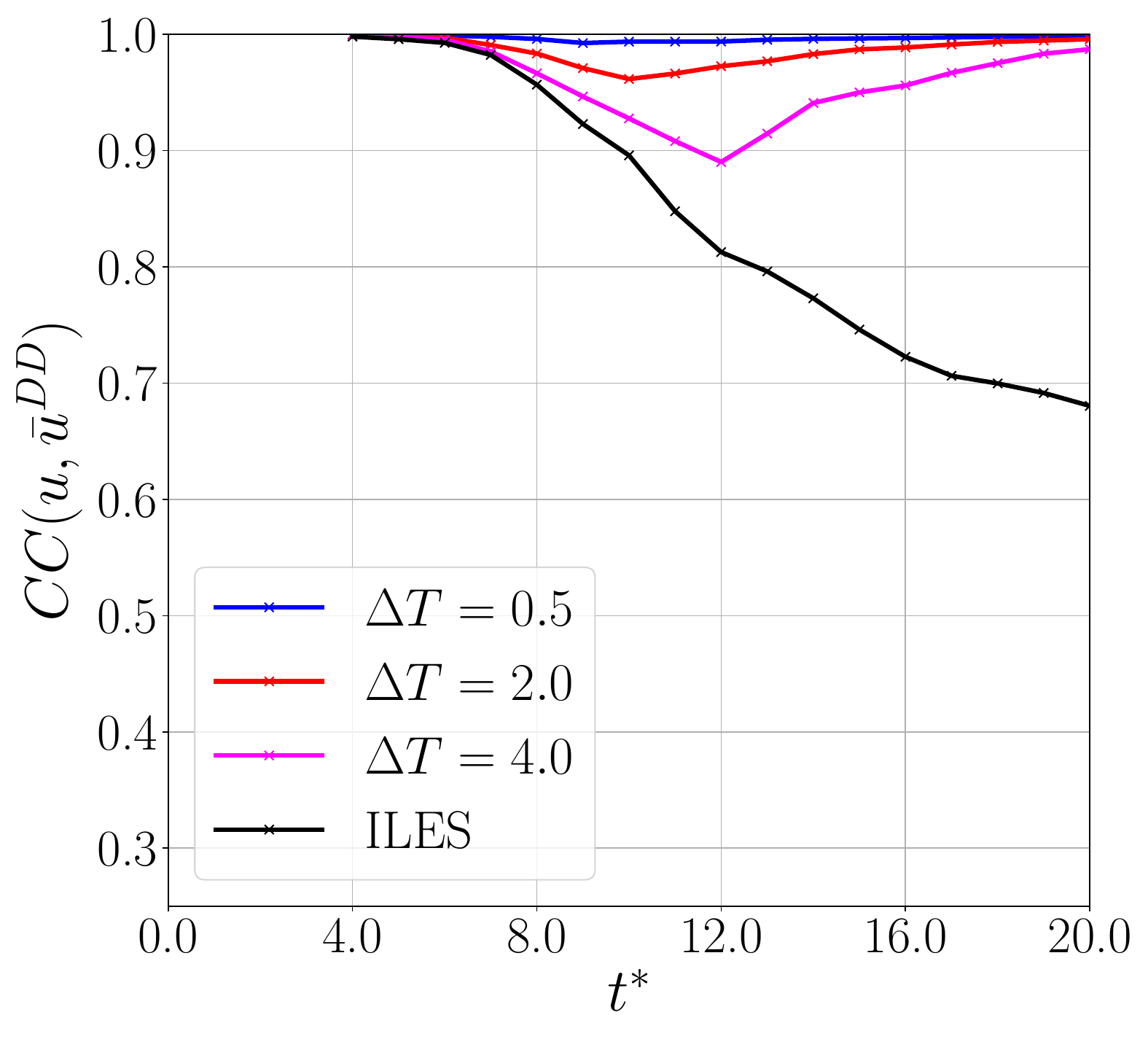}
     \caption{Component $u_1$.}
     \label{fig:corr_43p2_u}
 \end{subfigure}
 \begin{subfigure}{0.32\textwidth}
     \includegraphics[width=\textwidth]{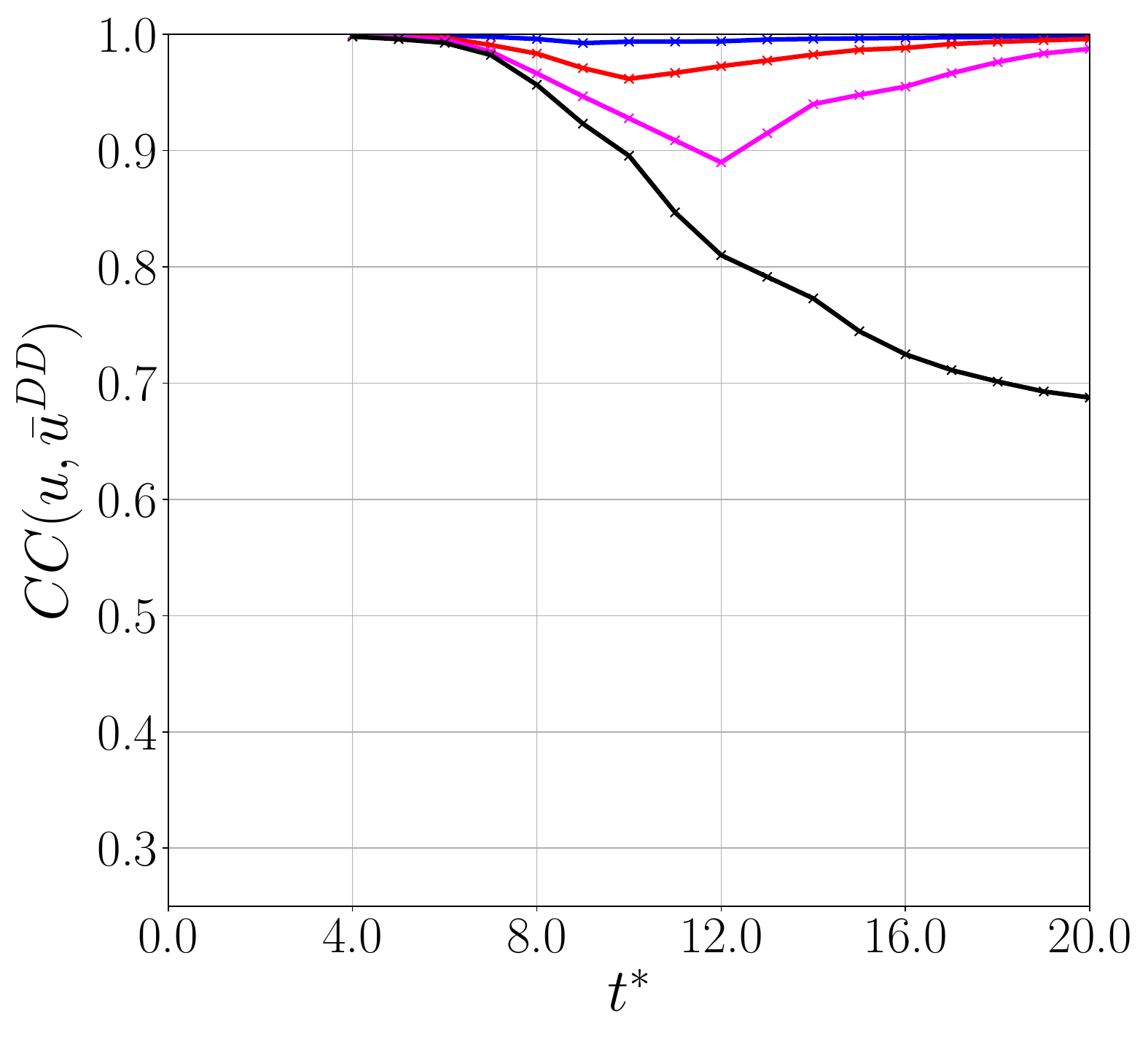}
     \caption{Component $u_2$.}
     \label{fig:corr_43p2_v}
 \end{subfigure}
 \begin{subfigure}{0.32\textwidth}
     \includegraphics[width=\textwidth]{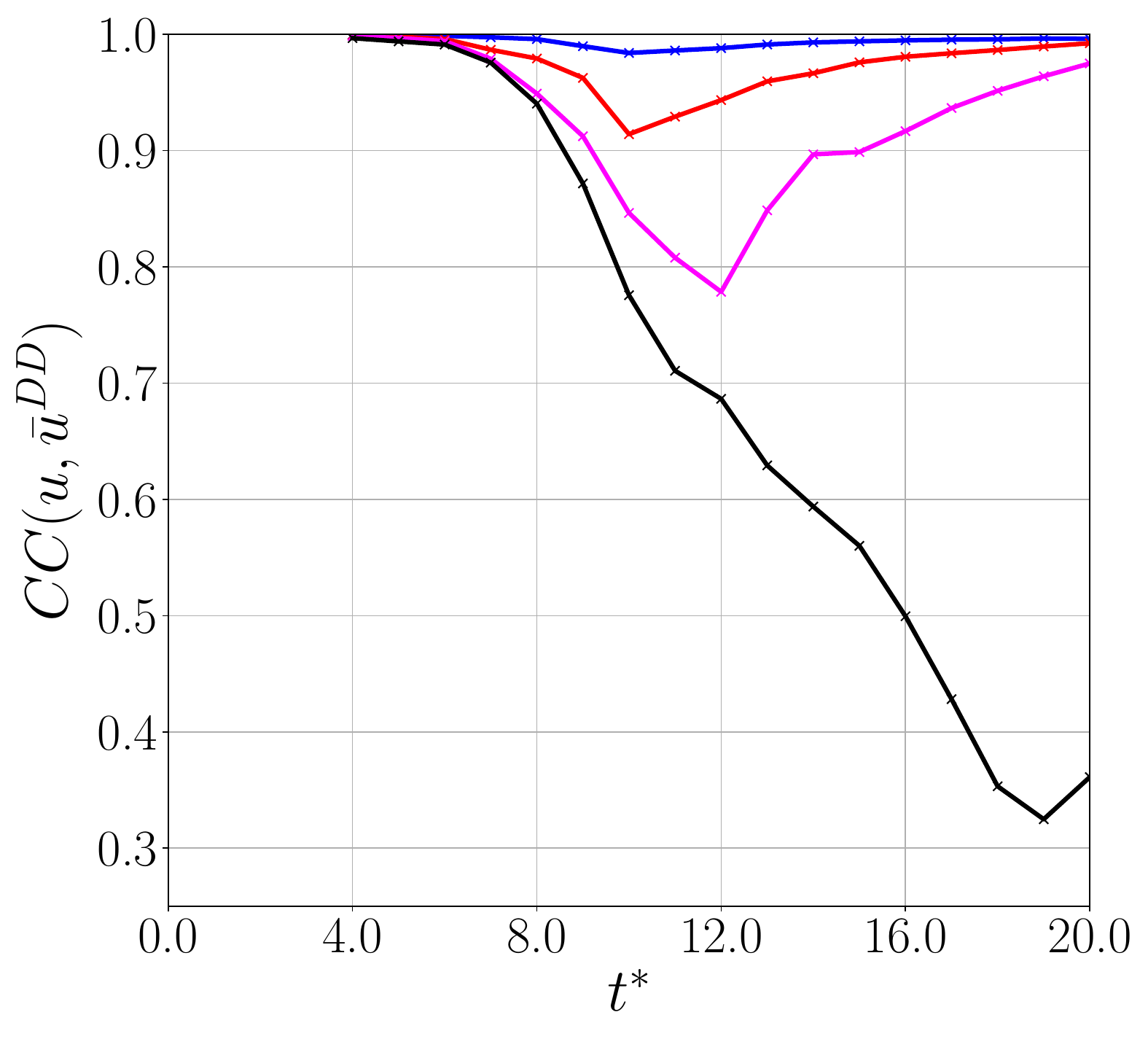}
     \caption{Component $u_3$.}
     \label{fig:corr_43p2_w}
 \end{subfigure}
 \vspace{-0.2cm}
\caption{Correlation coefficients for different velocity components for 43p2.}
   \label{fig:Corr43p2}
\end{figure}

Figure~\ref{fig:Corr43p2} shows instead the correlation in time for the lowest polynomial order in consideration. Qualitatively speaking, the trend is similar to $\mathrm{p}=5$, with a decrease in the first part and an increase after the enstrophy peak, in the second regime. However, upon closer inspection, it is evident that the correlation values are generally lower with respect to $\mathrm{p}=5$, indicating some challenges for the network in this scenario. This is consistent with the behavior of the loss function we have found above.

It is interesting to note that the point of minimum correlation shifts at later times with the increase of the restarting time window. We hypothesize that this is due to larger time windows allowing accumulation errors to grow for longer times.

\begin{figure}[htb]
 \centering  
 \begin{subfigure}{0.32\textwidth}
     \includegraphics[width=\textwidth]{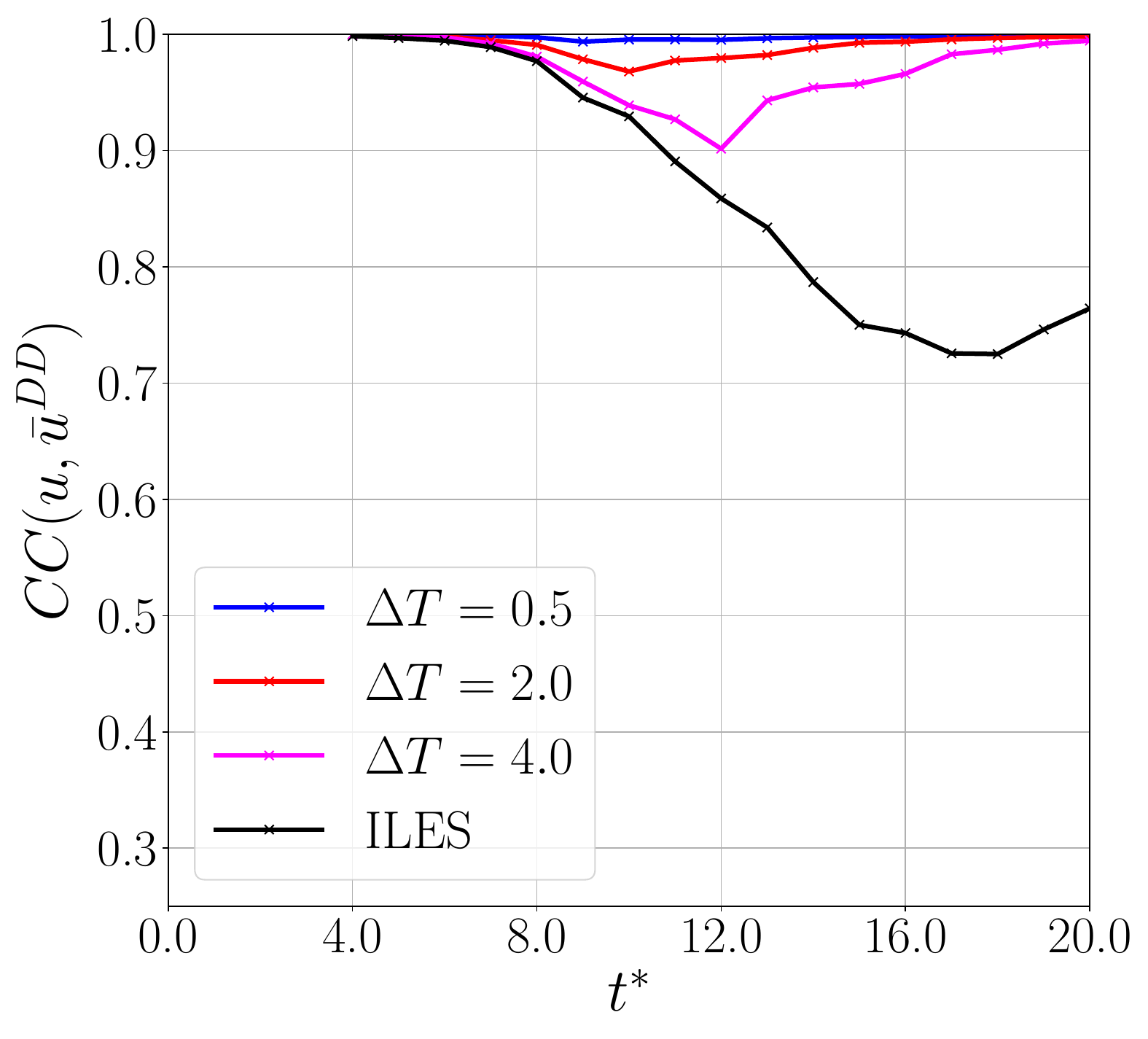}
     \caption{Component $u_1$.}
     \label{fig:corr_26p4_u}
 \end{subfigure}
 \begin{subfigure}{0.32\textwidth}
     \includegraphics[width=\textwidth]{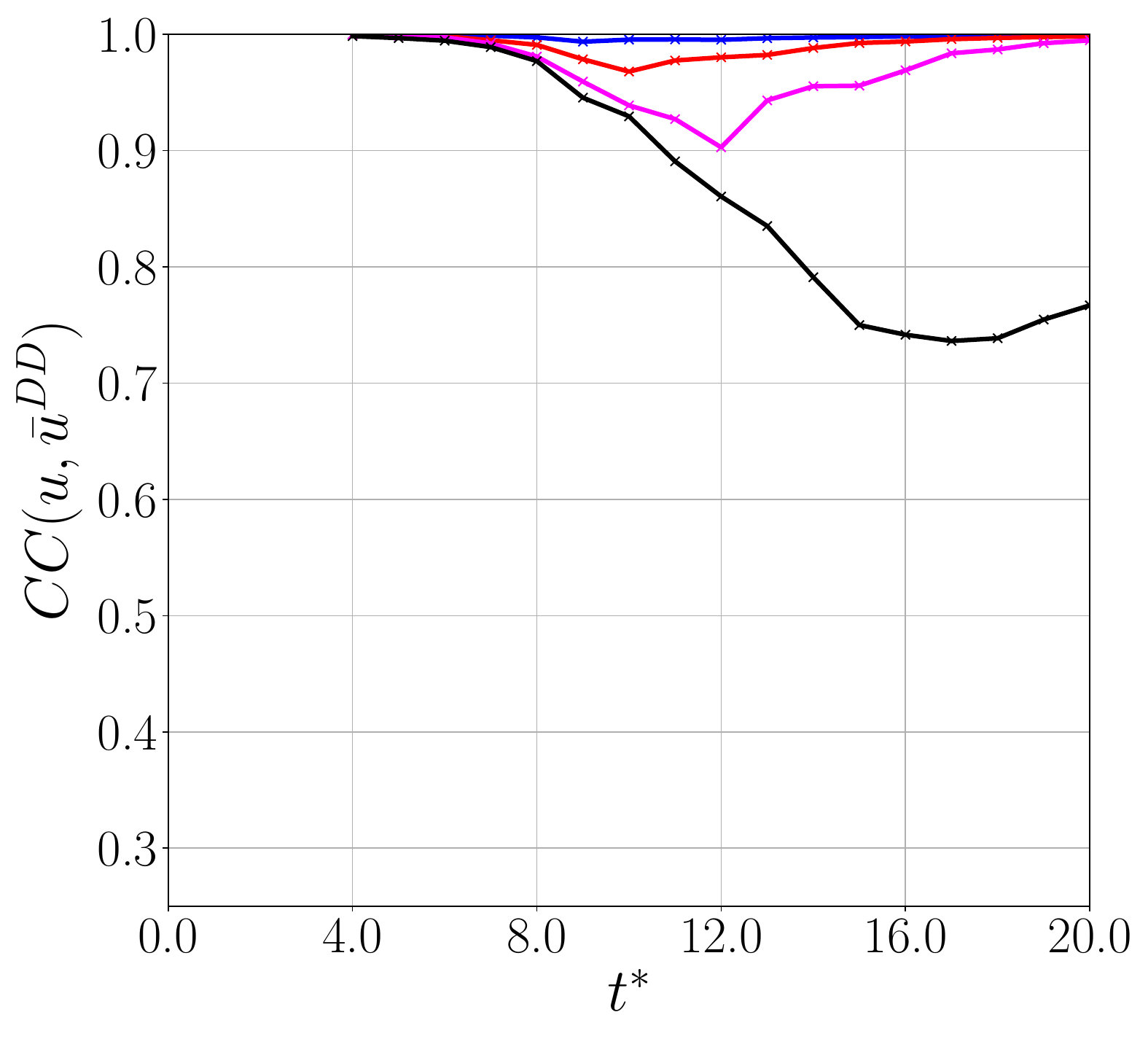}
     \caption{Component $u_2$.}
     \label{fig:corr_26p4_v}
 \end{subfigure}
 \begin{subfigure}{0.32\textwidth}
     \includegraphics[width=\textwidth]{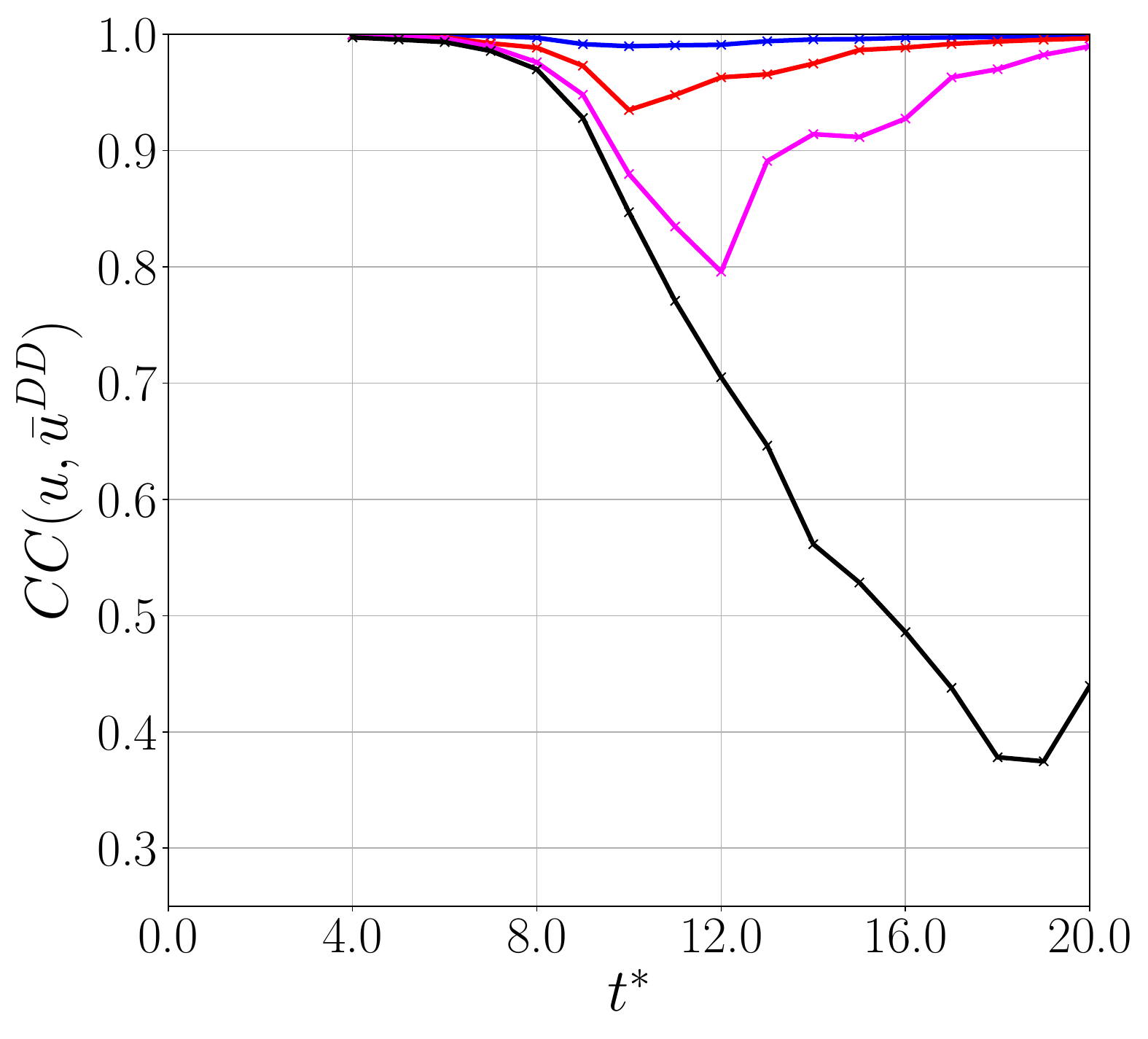}
     \caption{Component $u_3$.}
     \label{fig:corr_26p4_w}
 \end{subfigure}
 \vspace{-0.2cm}
\caption{Correlation coefficients for different velocity components for 26p4.}
   \label{fig:Corr26p4}
\end{figure}

\begin{figure}[htb]
 \centering  
 \begin{subfigure}{0.32\textwidth}
     \includegraphics[width=\textwidth]{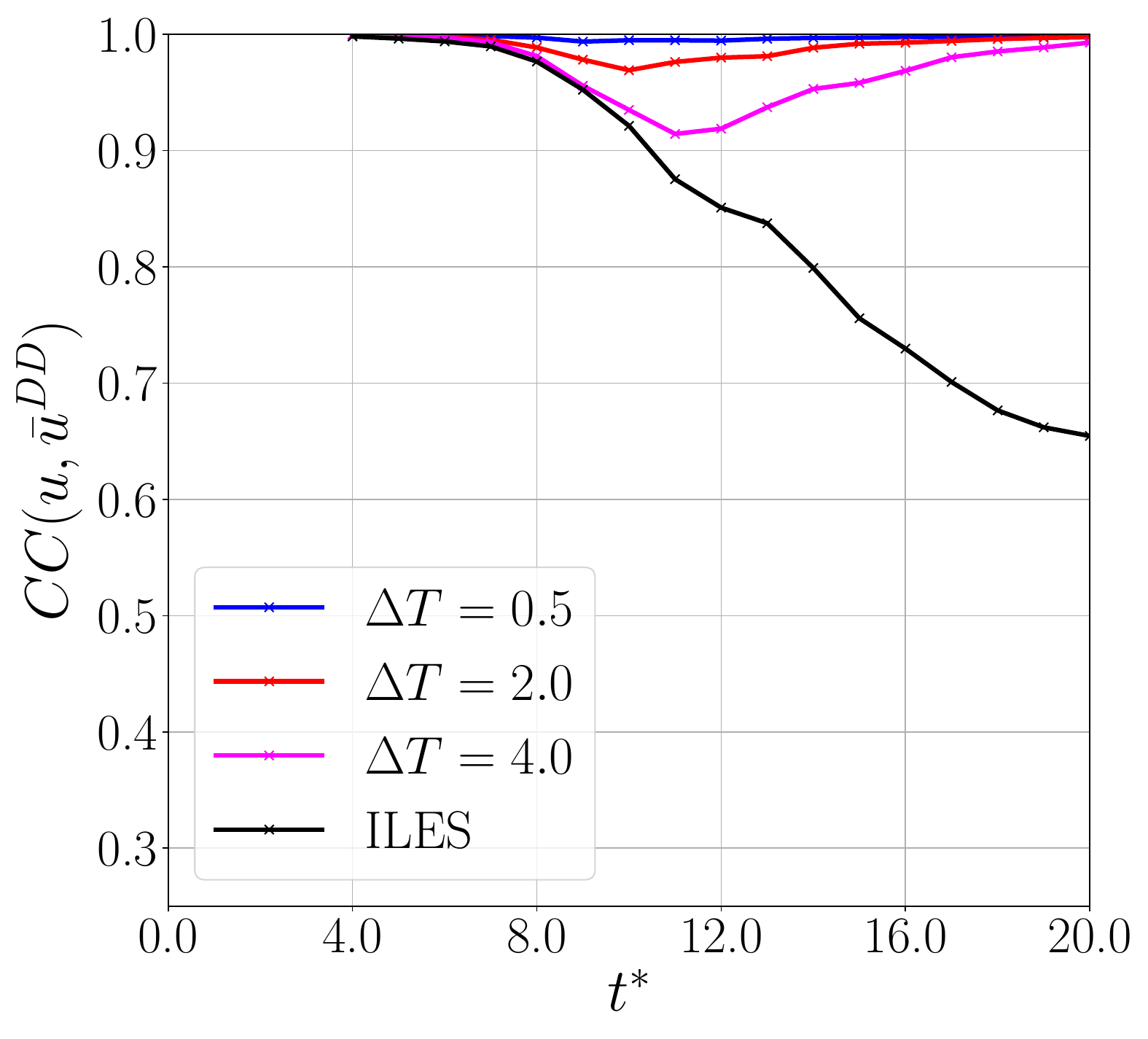}
     \caption{Component $u_1$.}
     \label{fig:corr_32p3_u}
 \end{subfigure}
 \begin{subfigure}{0.32\textwidth}
     \includegraphics[width=\textwidth]{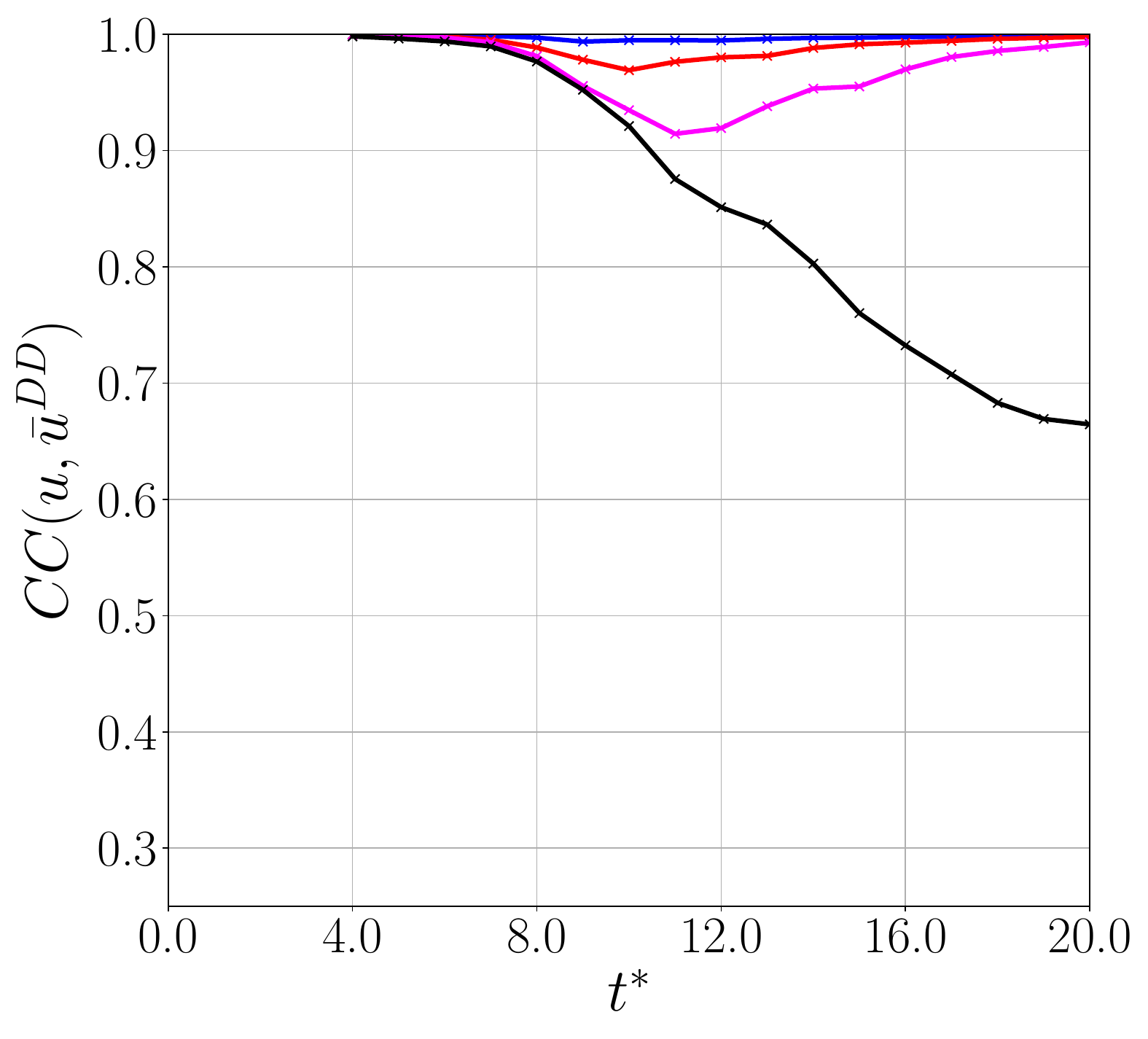}
     \caption{Component $u_2$.}
     \label{fig:corr_32p3_v}
 \end{subfigure}
 \begin{subfigure}{0.32\textwidth}
     \includegraphics[width=\textwidth]{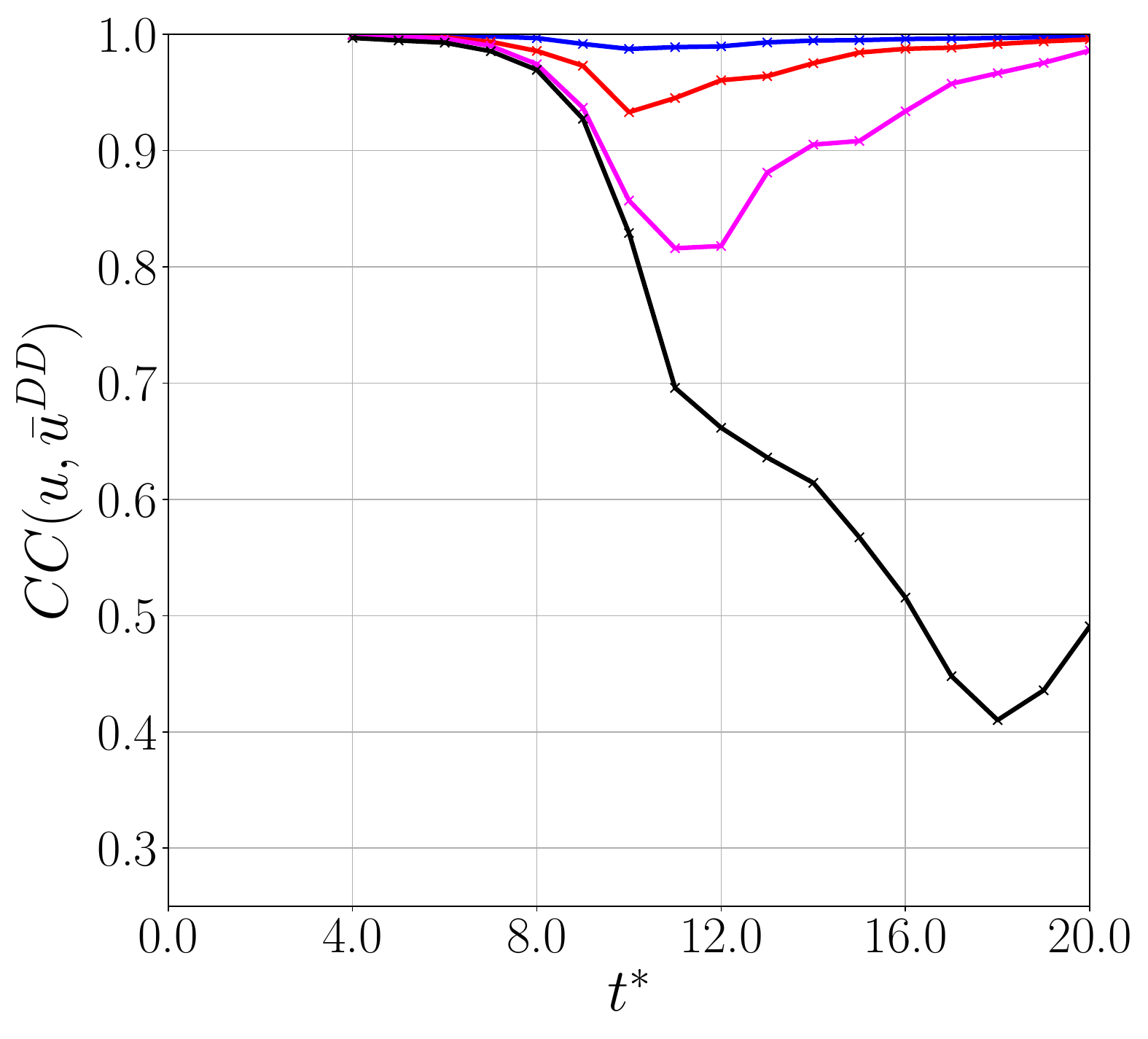}
     \caption{Component $u_3$.}
     \label{fig:corr_32p3_w}
 \end{subfigure}
 \vspace{-0.2cm}
\caption{Correlation coefficients for different velocity components for 32p3.}
   \label{fig:Corr32p3}
\end{figure}

The correlations between polynomials $\mathrm{p}\geq 3$ are quite similar both qualitatively and quantitatively with respect to $\mathrm{p}=5$, confirming the consistency of the models with the analysis of the previous section. Such correlations are shown in figures~\ref{fig:Corr26p4} and \ref{fig:Corr32p3} for completeness.

%

\subsection{Modal coefficients and kinetic energy spectra}\label{subsec:MD-KES}
After evaluating the global and local accuracy of the method across various restart time windows, we now present the models' capabilities in providing energetically significant information. We will use the modal coefficients behavior \eqref{eqn:ufilteredmodal} and the kinetic energy spectrum to have an effective interpretation of various ILES orders. These quantities are shown for $\Delta T=0.5$ and $\Delta T=4$ to compare the effects of the restart time window. Regarding the modal decay, we anticipate that the differences are of the order of $10^{-2}$ between the different components of the momentum, and as such, they have not been considered separately. Instead, the modal decay is averaged between all the three components and between all the elements in order to have a more insightful evaluation of the model.

\begin{figure}[h!]
 \centering  
 \begin{subfigure}{0.45\textwidth}
     \includegraphics[width=\textwidth]{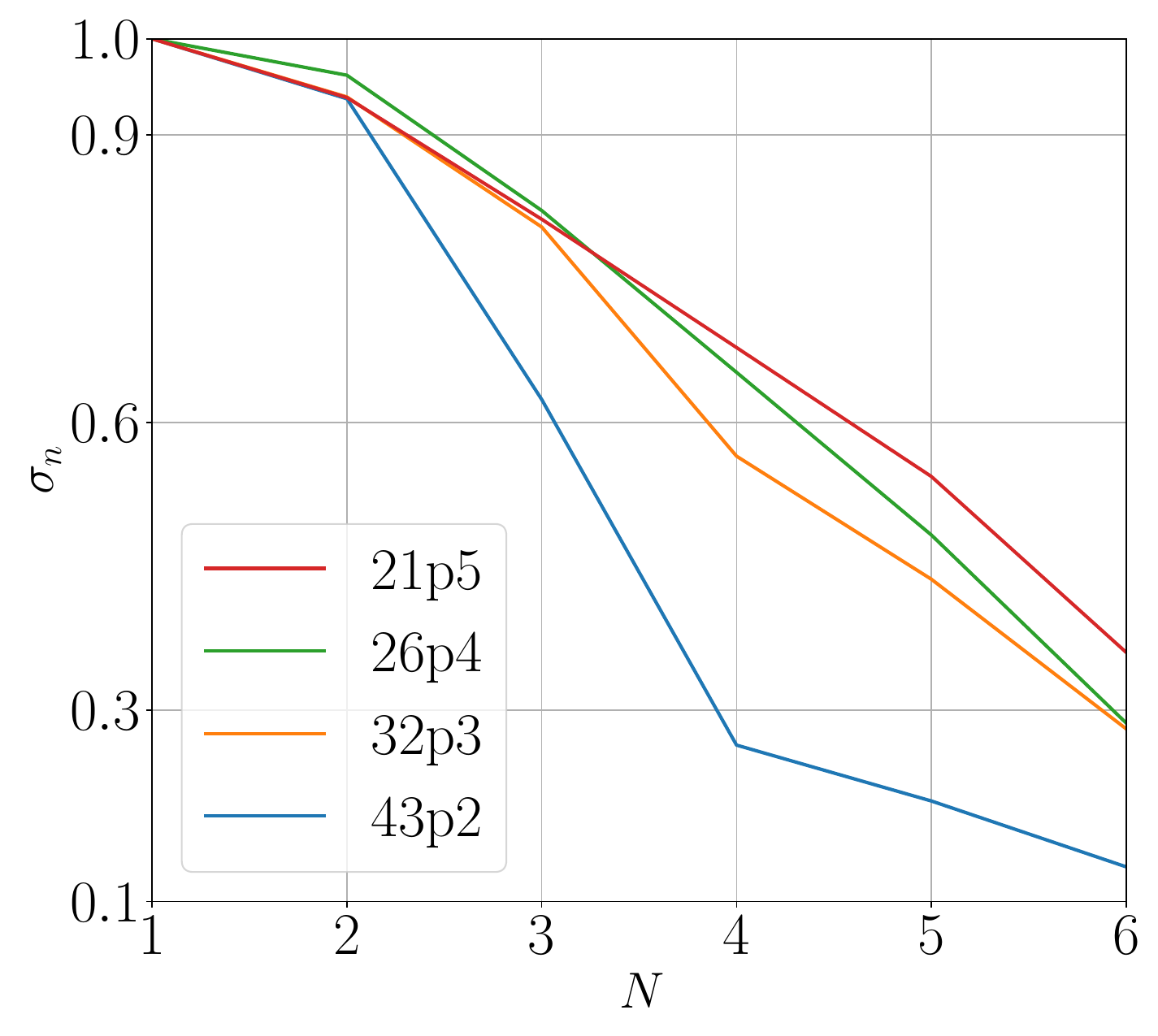}
     \caption{Averaged modal coefficients.}
     \label{fig:CNN_DT_05:a}
 \end{subfigure}
 \begin{subfigure}{0.45\textwidth}
     \includegraphics[width=\textwidth]{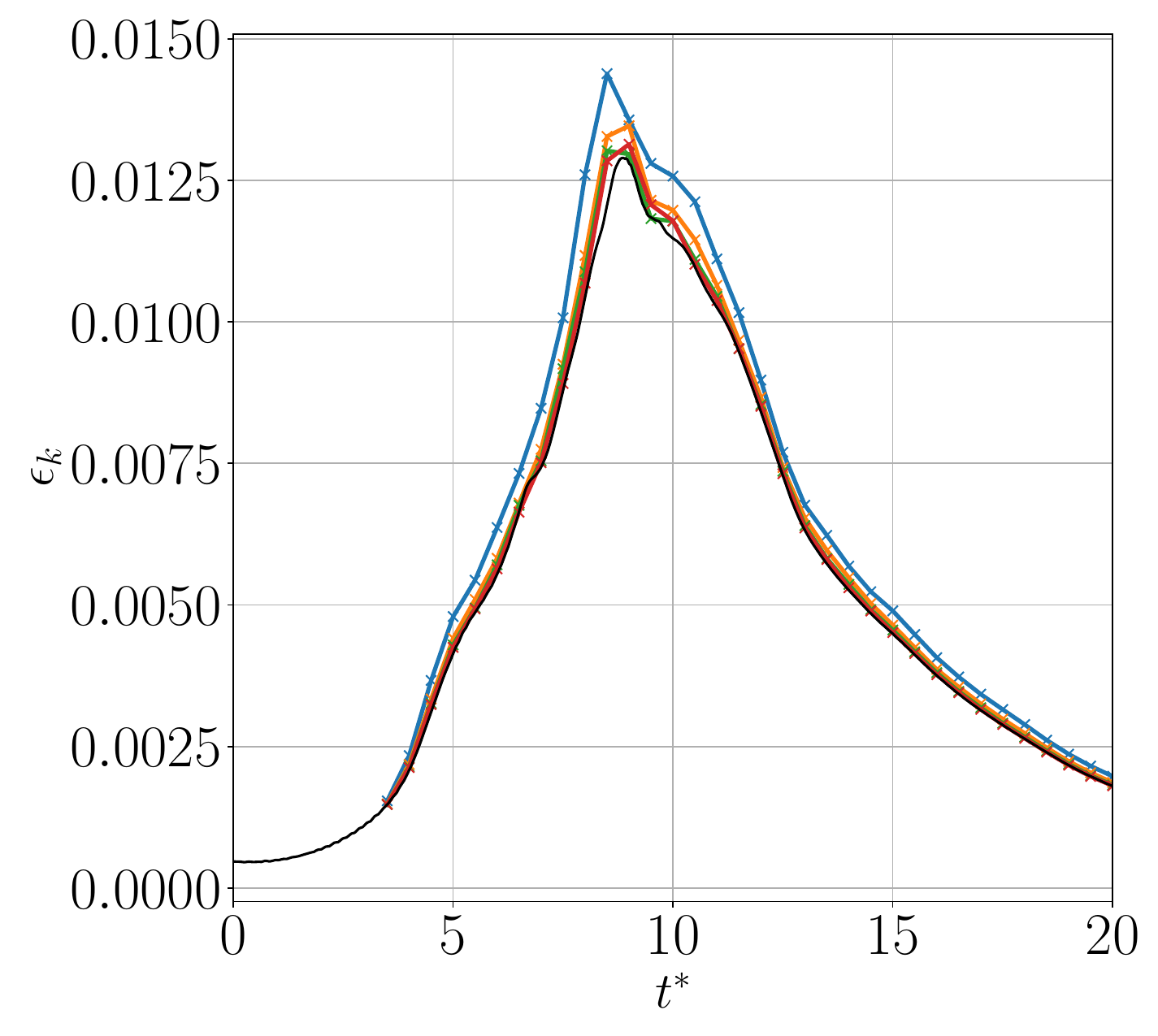}
     \caption{Kinetic energy dissipation rate.}
     \label{fig:CNN_DT_05:b}
 \end{subfigure}
\caption{Averaged modal coefficients (left) and time evolution of the kinetic energy dissipation rate (right) for data-driven filtered data with $\Delta T=0.5$.}
   \label{fig:CNN_DT_05}
\end{figure}

In Figure~\ref{fig:CNN_DT_05:a} it is shown the behavior of the modal coefficients obtained by training the model with $\Delta T = 0.5$. Qualitatively speaking, the results reveal that the curves corresponding to higher polynomial orders (\ie, $\mathrm{p}\geq 3$) exhibit a similar profile, indicating a consistent attenuation of the coefficients. 
The curves for $\mathrm{p}=4$ and $\mathrm{p}=5$ are almost overlapped, with only slight differences in the first three modes and more deviations in the highest modes. These findings explain the overlapping of the curves for polynomials $\mathrm{p}>2$ in Figure~\ref{fig:CNN_DT_05:b}, where the kinetic energy dissipation is shown. Since the modal decay is similar for $\mathrm{p} \geq 3$, the kinetic energy dissipation rates are closely aligned.
In contrast, the curve for $\mathrm{p}=2$ shows a rapid decay after the second mode with an attenuation after the fourth mode. 
Note that, for all the polynomials, the first coefficient has a unitary value, showing that the most relevant differences between DNS and ILES lie in the small scale features which are represented by high-order modes \cite{Gassner2013OnTA,Beck2013}.

\begin{figure}[h!]
 \centering  
 \begin{subfigure}{0.45\textwidth}
     \includegraphics[width=\textwidth]{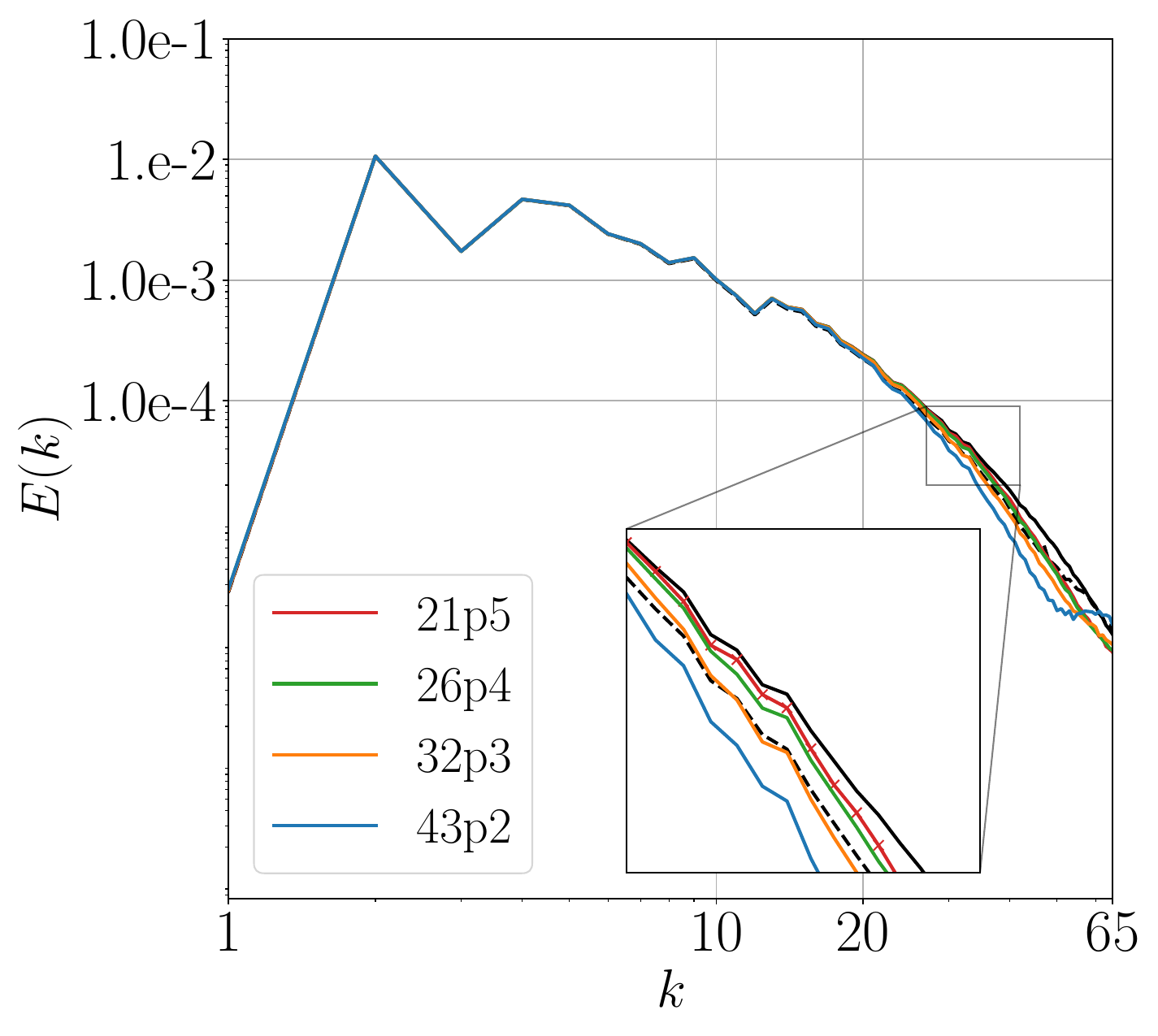}
     \caption{Baseline ILES.}
     \label{fig:spectrum_dt_05:a}
 \end{subfigure}
 \begin{subfigure}{0.45\textwidth}
     \includegraphics[width=\textwidth]{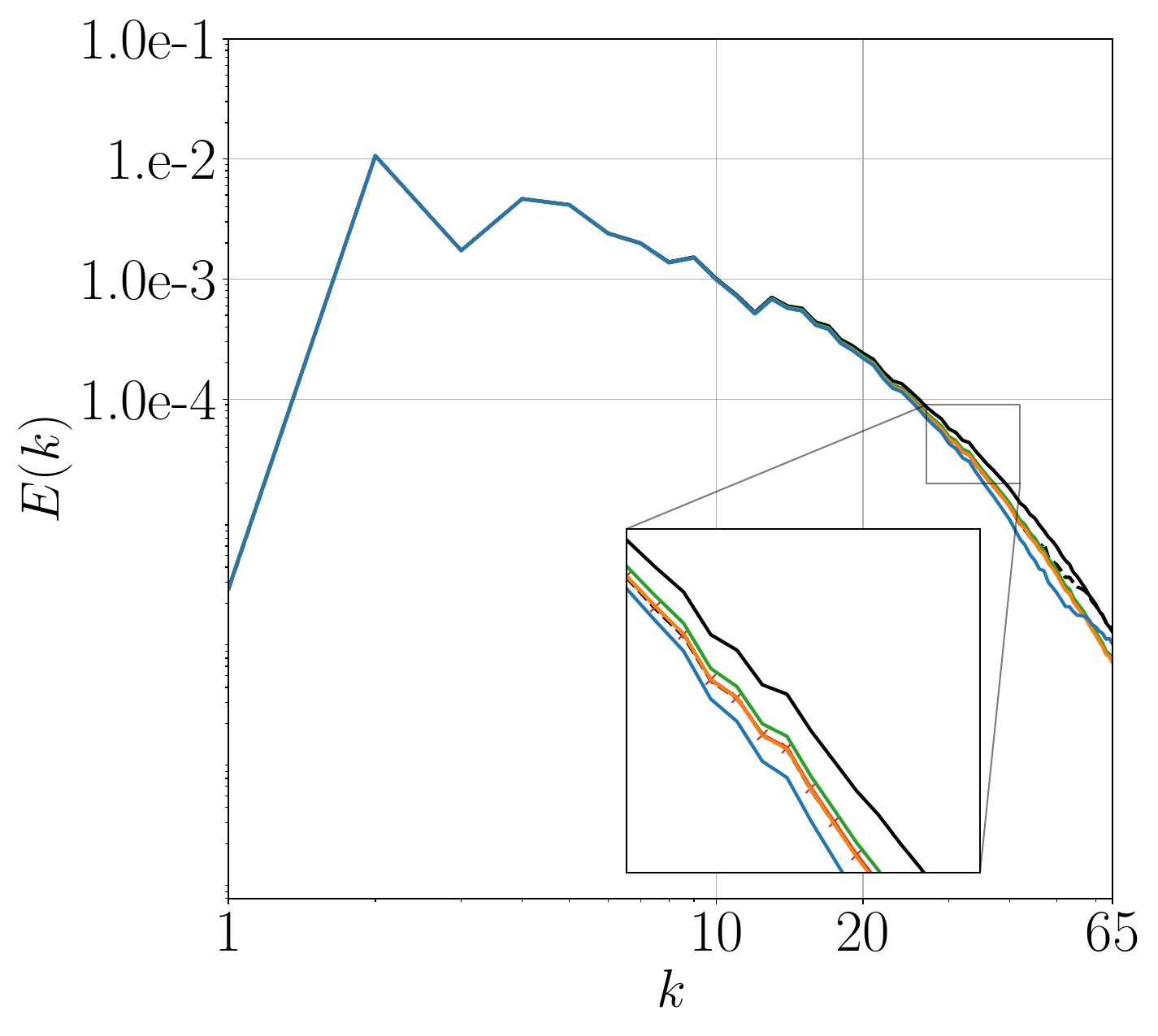}
     \caption{Data-driven filtered DNS data.}
     \label{fig:spectrum_dt_05:b}
 \end{subfigure}
\caption{Kinetic energy spectra for different polynomial orders obtained at $t^*=15$ for the baseline ILES (left) and from data-driven filtered DNS data with $\Delta T= 0.5$ (right). Solid black line, DNS; dashed line, filtered with $\Delta = 2h$.}
   \label{fig:spectrum_dt_05}
\end{figure}

As an additional analysis, in Figure~\ref{fig:spectrum_dt_05} we compare the kinetic energy spectra of the ILES (restarted with $\Delta T=0.5$) and the filtered DNS at $t^{*}=15$ with the various data-driven filters. Notice that since the data-driven filters have been trained to reproduce ILES data, a perfect filtering operation should produce exactly the same curves in figures ~\ref{fig:spectrum_dt_05:a} and ~\ref{fig:spectrum_dt_05:b}.
For reference, the DNS is also filtered using an exponential modal filter with a cutoff width of $\Delta=2h$, where $2\pi/N_{\mathrm{DoF}}$ and $N_{\mathrm{DoF}}=N\times N^{1D}_{el}$ are the total degrees of freedom in one direction.
We can observe that the approach herein considered is able to reproduce quite accurately the target ILES solution.
The curves for $\mathrm{p}=2$ have a similar trend, confirming that the model is able to differentiate the scales when the variation between polynomials is large enough. However, the predictions for $\mathrm{p}\geq 3$ are not perfectly reproduced. As depicted in the zoomed view of Figure~\ref{fig:spectrum_dt_05:b}, the curve for $\mathrm{p}=4$ is located slightly above the curve for $\mathrm{p}=5$, that is overlapped to $\mathrm{p}=3$. This behavior is easily explained by looking at the modal decay in Figure~\ref{fig:CNN_DT_05:a}: the coefficients for the most energetic modes ($N\leq 3$) for $\mathrm{p}=4$ are slightly larger than the ones of $\mathrm{p}=5$ and $\mathrm{p}=3$, leading to a better preservation of the kinetic energy. This suggests that the network has difficulties in classifying the small differences between high order polynomials.

Next, we assess the model trained on the data based on the largest time window considered, namely $\Delta T=4$. 
\begin{figure}[h!]
 \centering  
 \begin{subfigure}{0.45\textwidth}
     \includegraphics[width=\textwidth]{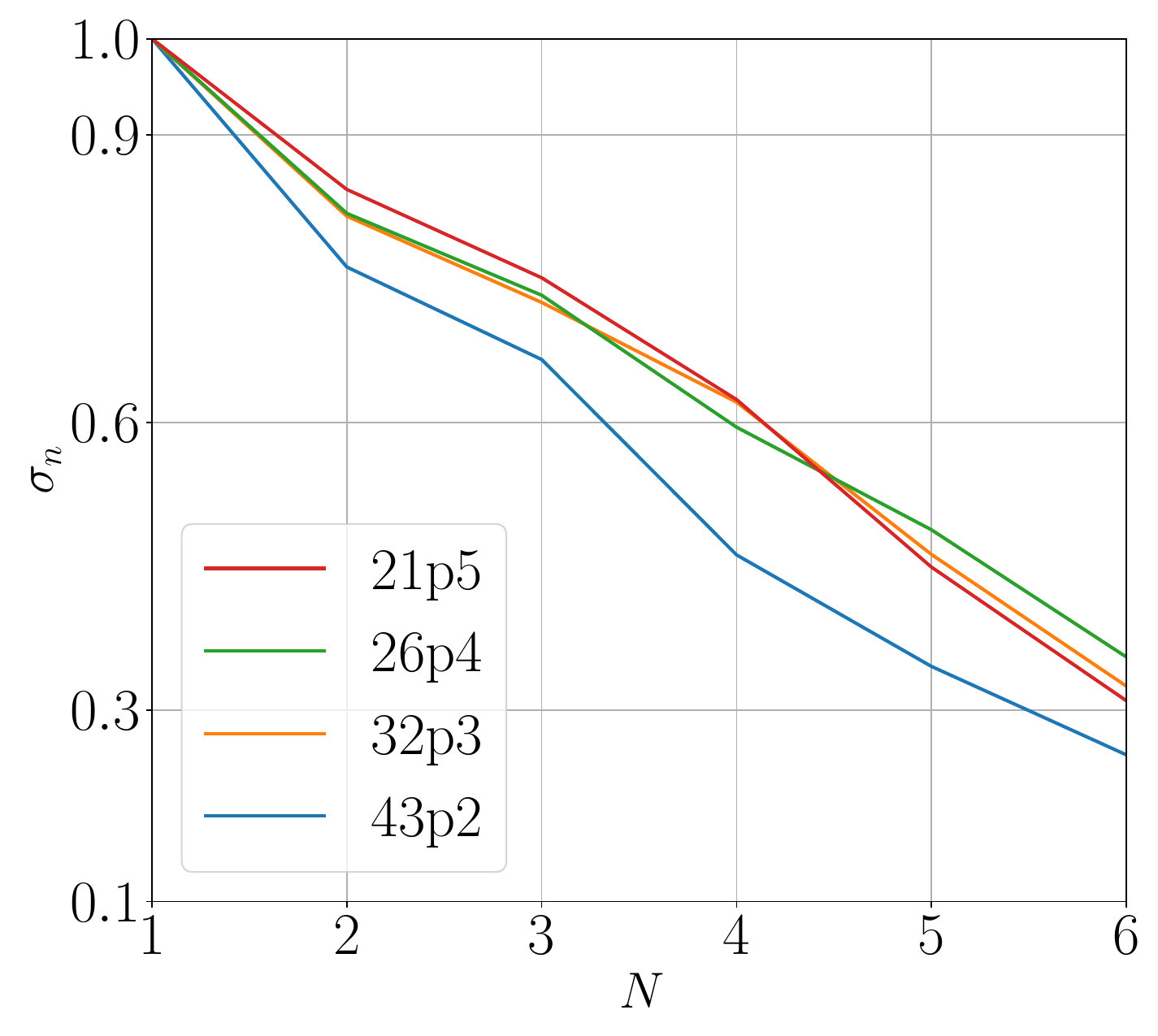}
     \caption{Averaged modal coefficients.}
     \label{fig:CNN_DT_4:a}
 \end{subfigure}
 \begin{subfigure}{0.44\textwidth}
     \includegraphics[width=\textwidth]{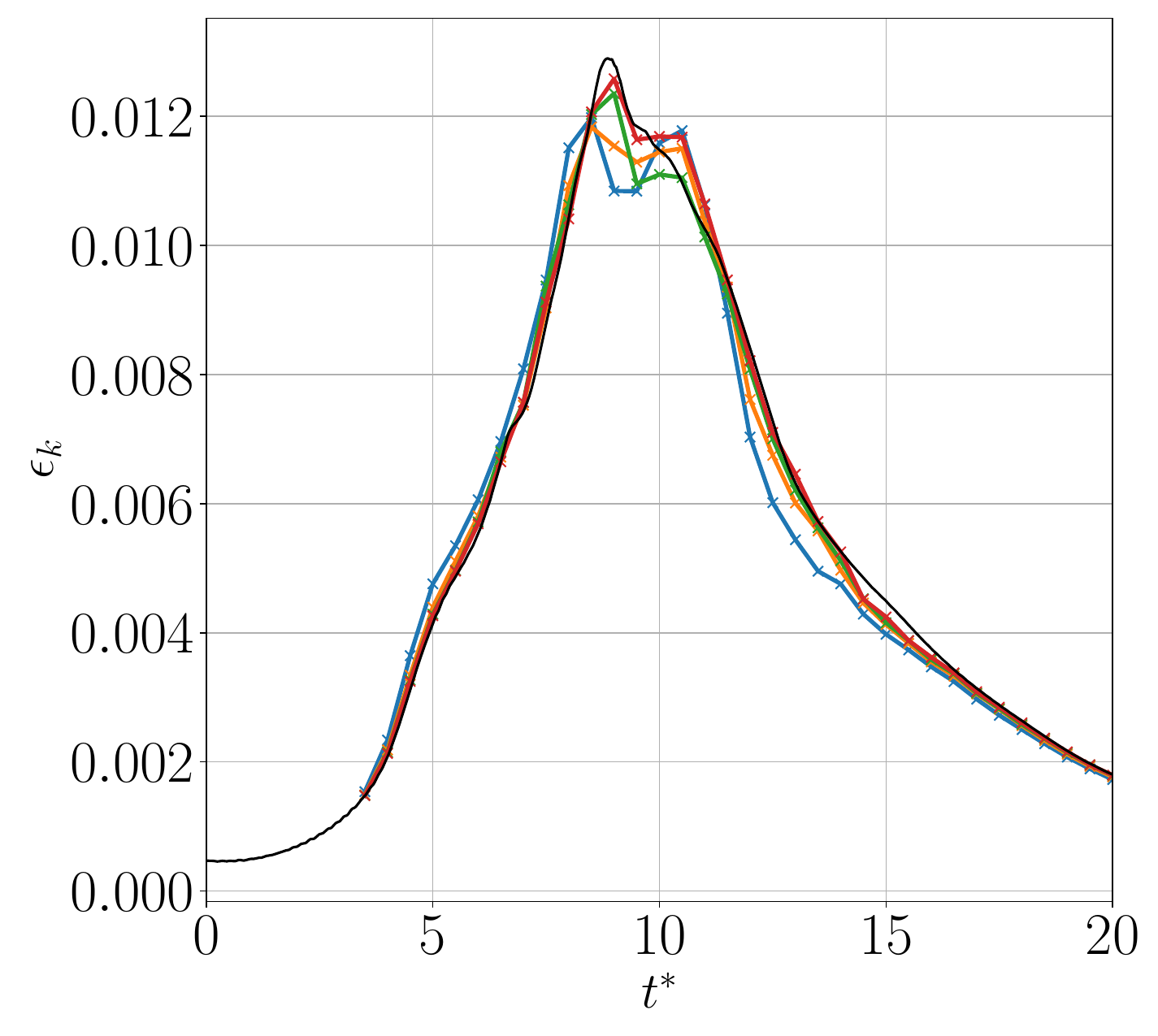}
     \caption{Kinetic energy dissipation rate.}
     \label{fig:CNN_DT_4:b}
 \end{subfigure}
\caption{Averaged modal coefficients (left) and time evolution of the kinetic energy dissipation rate (right) for data-driven filtered data with $\Delta T=4.0$.}
   \label{fig:CNN_DT_4}
\end{figure}
In Figure~\ref{fig:CNN_DT_4} the modal coefficients and kinetic energy spectra are shown for $\Delta T=4$. 
Similarly with respect to Figure~\ref{fig:CNN_DT_05}, the curves for $\mathrm{p}\geq 3$ are close together and they differ considerably from $\mathrm{p}=2$. This confirms that the network is able to differentiate between $\mathrm{p}=2$ and $\mathrm{p}\geq 3$ polynomials. Also, the qualitative modal decay is similar among all the orders. This behavior, on the contrary, is different from what we observed for $\Delta T =0.5$, where  $\mathrm{p}=2$ showed a clear decay after the second mode. It is possible that the significant accumulation error outweighs the influence of space discretization, thereby forcing the modal decay to follow a qualitative pattern which is similar across the different orders.

\begin{figure}[h!]
 \centering  
 \begin{subfigure}{0.45\textwidth}
     \includegraphics[width=\textwidth]{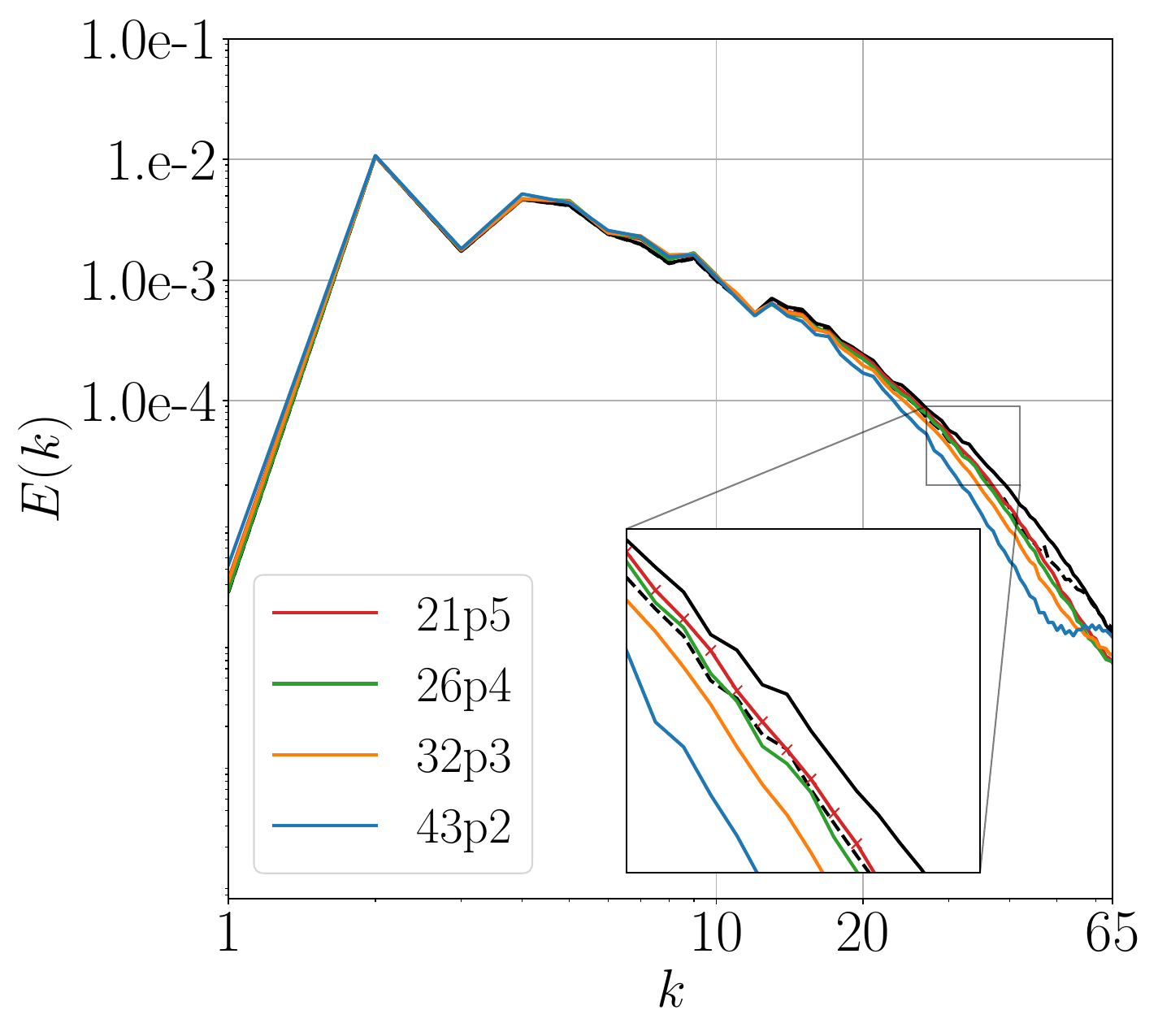}
     \caption{Baseline ILES.}
     \label{fig:spectrum_dt_4:a}
 \end{subfigure}
 \begin{subfigure}{0.45\textwidth}
     \includegraphics[width=\textwidth]{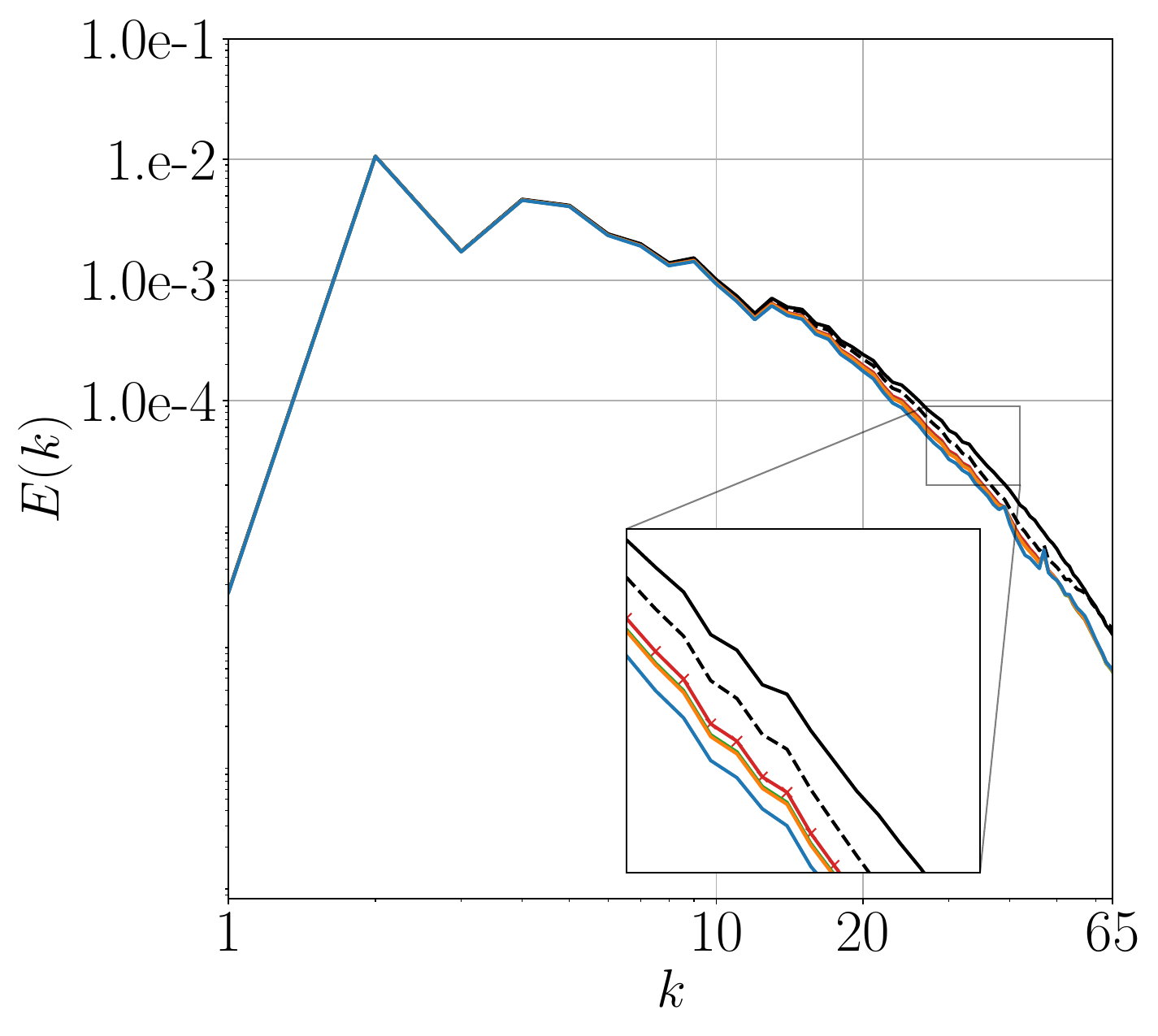}
     \caption{Data-driven filtered DNS data.}
     \label{fig:spectrum_dt_4:b}
 \end{subfigure}
\caption{Kinetic energy spectrum for different polynomial orders obtained at $t^*=15$ for the baseline ILES (left) and from data-driven filtered DNS data with $\Delta T= 4.0$ (right). Solid black line, DNS; dashed line, filtered with $\Delta = 2h$.}
   \label{fig:spectrum_dt_4}
\end{figure}

By comparing the kinetic energy spectrum for the $\mathrm{ILES}$ (restarted with $\Delta T=4.0$) and the filtered DNS in figures~\ref{fig:spectrum_dt_4}, it becomes evident that they do not align well. This confirms that already for this time window, the accumulation error drastically affects the ability of the network in compressing the features of a fully resolved simulation in modal coefficients. 

To better understand how the restart time window influences the different orders of accuracy, we inspect the modal decay and the kinetic energy spectrum for $\mathrm{p}=5$ and $\mathrm{p=2}$ for all the time windows considered. We will not show other orders as we have seen that orders of $\mathrm{p}\geq 3$  give very close results. This analysis helps to clarify how the time window affects the highest order (least dissipative) with respect to the lowest one (most dissipative).
\begin{figure}[h!]
 \centering  
 \begin{subfigure}{0.443\textwidth}
     \includegraphics[width=\textwidth]{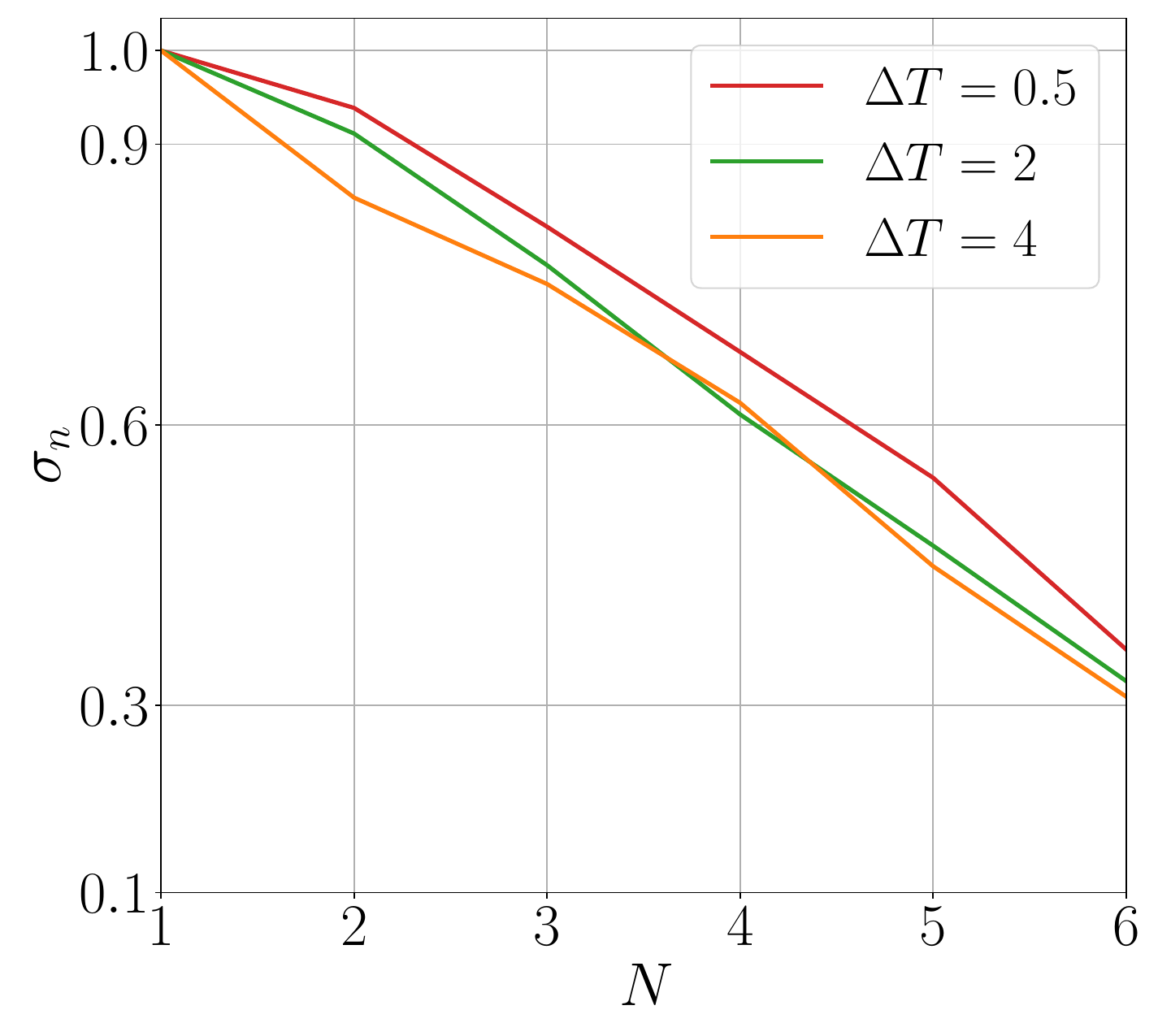}
     \caption{Averaged modal coefficients.}
     \label{fig:21p5_comp_dt:a}
 \end{subfigure}
 \begin{subfigure}{0.45\textwidth}
     \includegraphics[width=\textwidth]{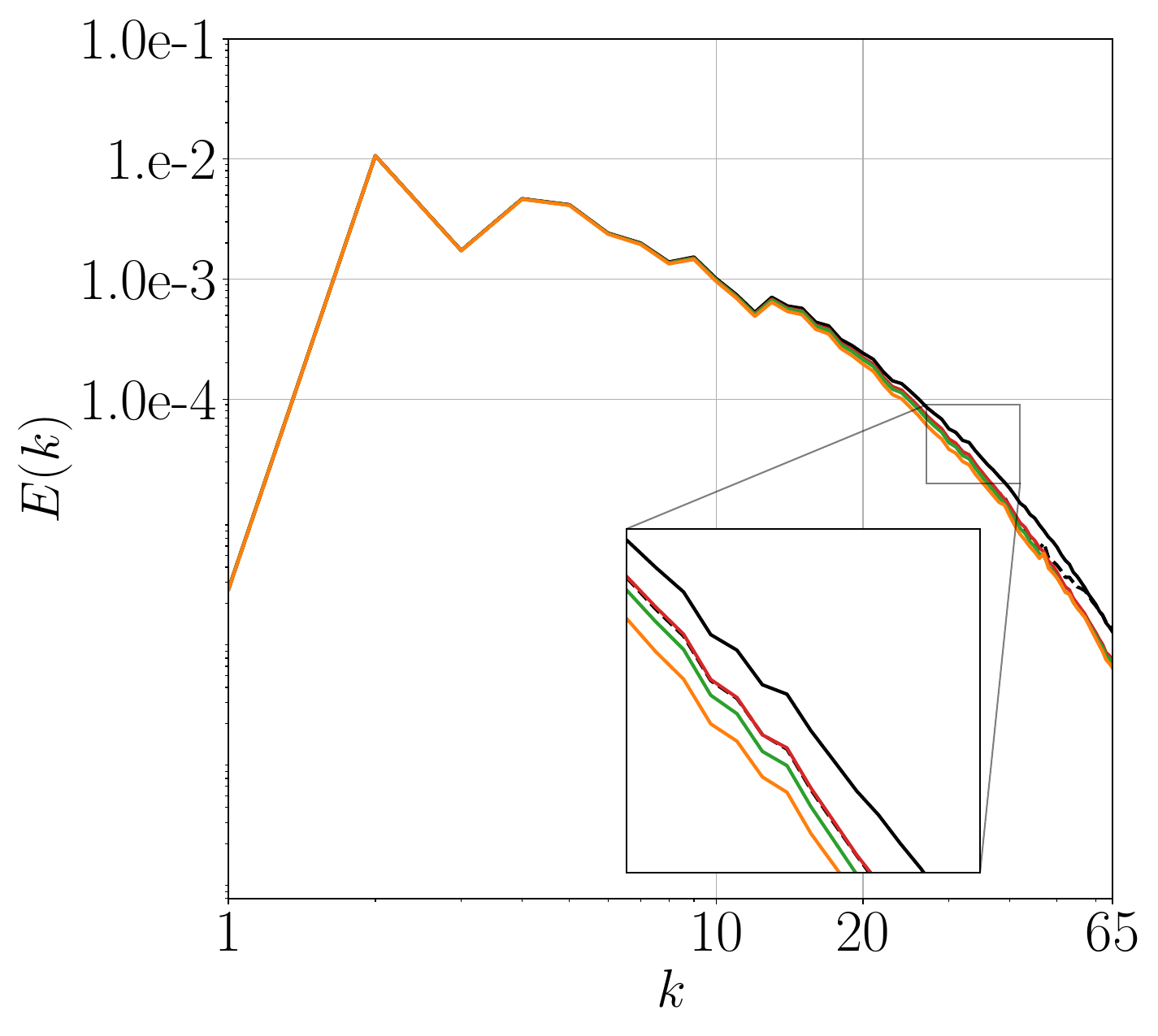}
     \caption{Kinetic energy spectrum.}
     \label{fig:21p5_comp_dt:b}
 \end{subfigure}
\caption{Averaged modal coefficients (left) and kinetic energy spectrum (right) for different time windows for $\mathrm{p}=5$. Solid black line, DNS; dashed line, filtered with $\Delta = 2h$.}
   \label{fig:21p5_comp_dt}
\end{figure}

In Figure~\ref{fig:21p5_comp_dt:a} we compare the modal coefficients for the highest polynomial degree (\ie, $\mathrm{p}=5$) for the various restart time windows in consideration. As the time window increases, there is a greater damping effect, resulting in a more pronounced decay. 
However, the differences between the different time windows are not particularly pronounced, and this results in a similar kinetic energy spectrum in Figure \ref{fig:21p5_comp_dt:b}. 
\begin{figure}[h!]
 \centering  
 \begin{subfigure}{0.442\textwidth}
     \includegraphics[width=\textwidth]{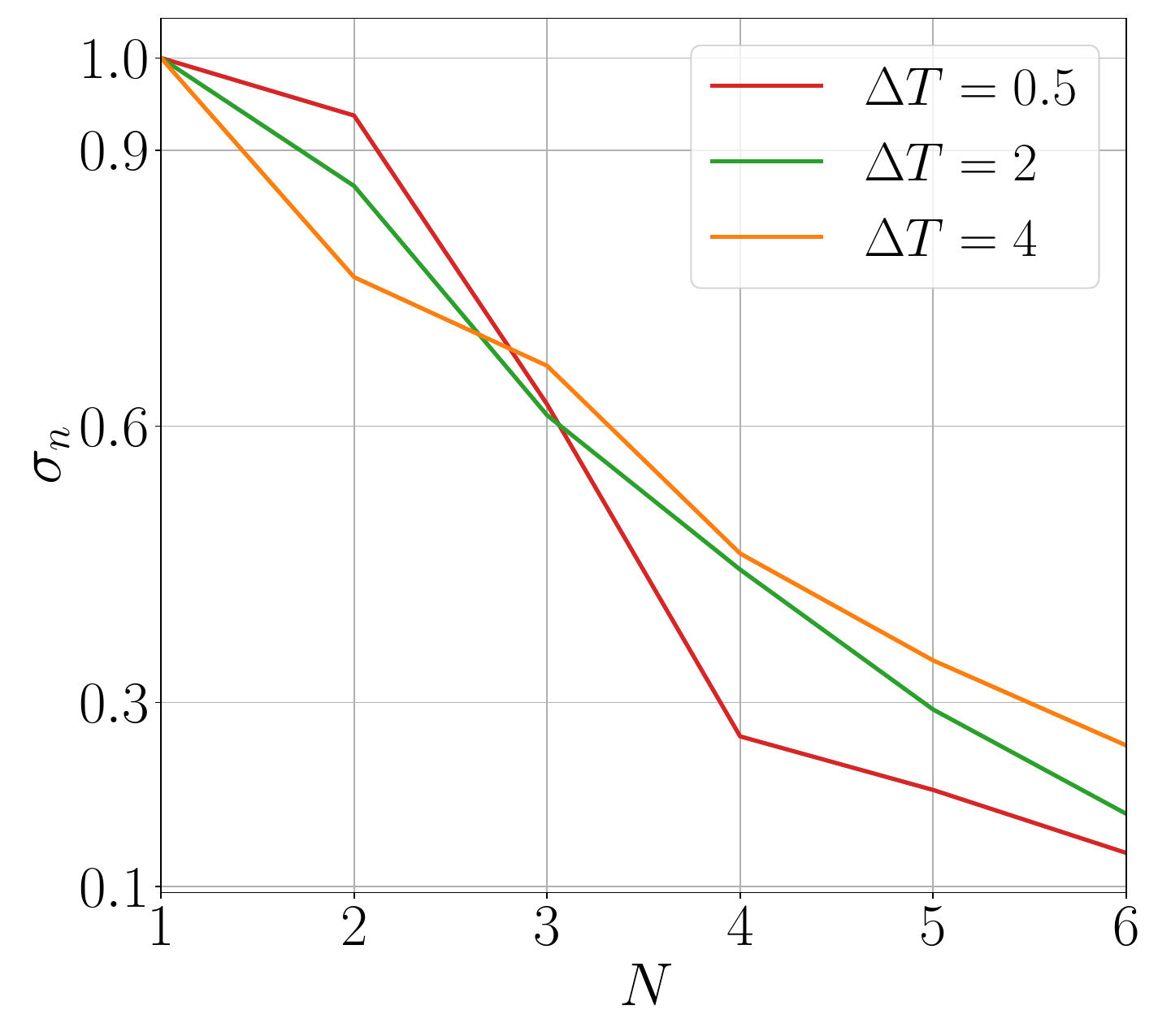}
     \caption{Averaged modal coefficients.}
     \label{fig:43p2_comp_dt:a}
 \end{subfigure}
 \begin{subfigure}{0.45\textwidth}
     \includegraphics[width=\textwidth]{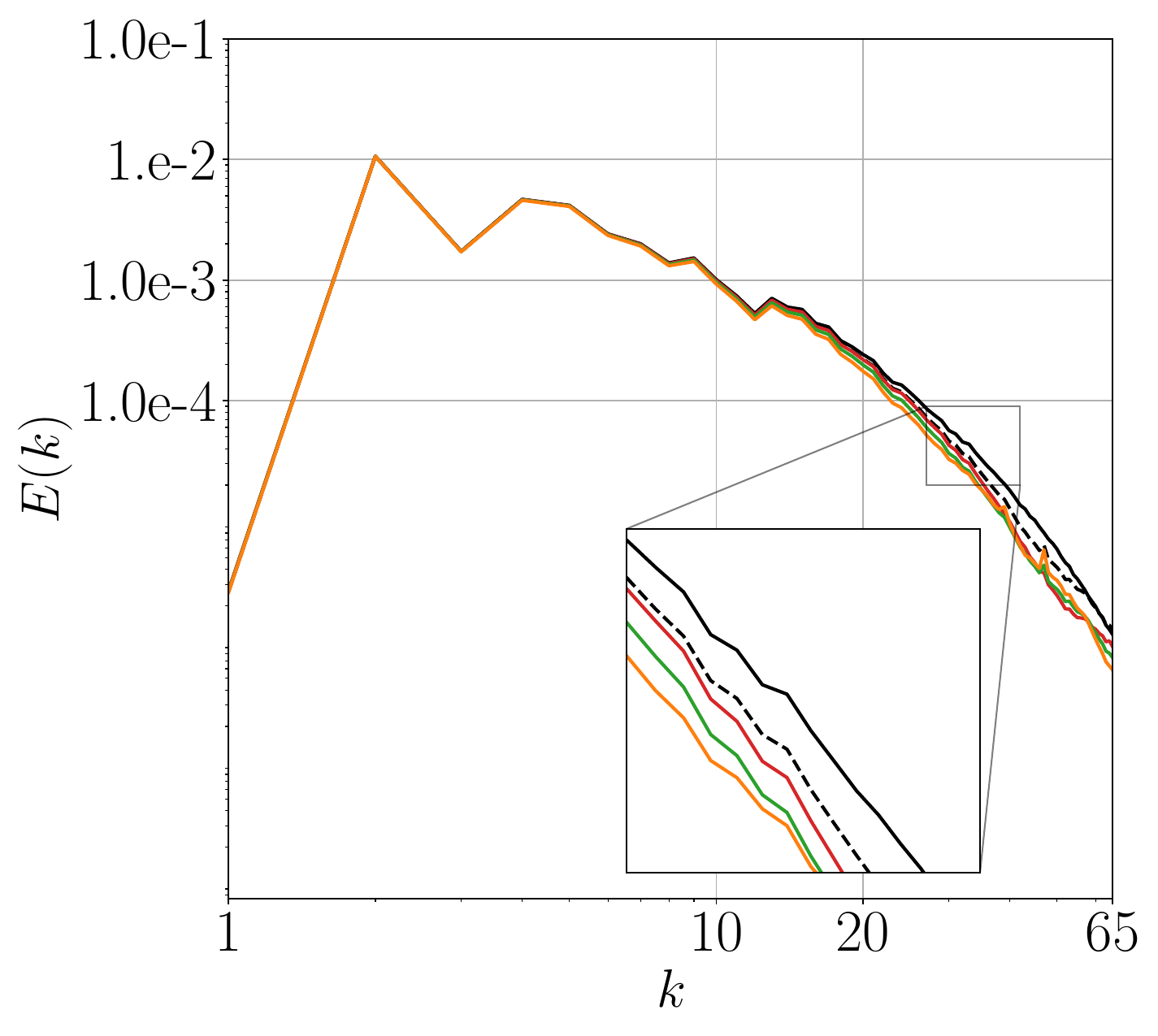}
     \caption{Kinetic energy spectrum.}
     \label{fig:43p2_comp_dt:b}
 \end{subfigure}
\caption{Averaged modal coefficients (left) and kinetic energy spectrum (right) for different time windows for $\mathrm{p}=2$. Solid black line, DNS; dashed line, filtered with $\Delta = 2h$.}
   \label{fig:43p2_comp_dt}
\end{figure}

On the contrary, regarding $\mathrm{p}=2$ (Figure~\ref{fig:43p2_comp_dt}), the variation of modal coefficients between different time windows is significantly more pronounced (compare Figure~\ref{fig:21p5_comp_dt:a} and Figure~\ref{fig:43p2_comp_dt:a}).
This plot supports the hypothesis that, as the time window increases, the qualitative behavior of the modal decay converges toward a certain trend, which is common among all the orders, as explained before in Figure~\ref{fig:CNN_DT_4}.
As a result, the kinetic energy spectra among the different time windows are quantitatively different as shown in Figure~\ref{fig:43p2_comp_dt:b}. Since spatial discretization errors are larger for $\mathrm{p}=2$ compared to $\mathrm{p}=5$, the errors in the former accumulate more substantially as the time window expands, resulting in an increased error propagation over time. Thus, the order of accuracy has a direct impact on the accumulation of errors, which subsequently affects the interpretability of the network concerning the modal coefficients.
\subsection{Fourier kernel analysis}\label{subsec:FOU}
In the context of the SD scheme (and more in general for DSEMs), the choice of the solution points is crucial for the overall stability and accuracy of the whole methodology~\cite{jameson2010proof}.
In this study we considered the classical choice of Gauss-Legendre quadrature points but different options are commonly employed. Once the appropriate locations of the solution points are defined, it is then possible to inspect the local spectral signature of the filter at the discrete level. Namely, the filter function in Fourier space can be defined as:
\begin{equation}
    \widehat{G}_\alpha(k) = \sum_{\beta=1}^{N}G_{\alpha \beta}e^{-\iota\left( \frac{x_\beta-x_\alpha}{\Delta} \right) k\Delta}. \label{eq:GFOU}
\end{equation}
Here, $G_{\alpha \beta}$ is the discrete weight matrix, $k$ is the wavenumber, $\iota$ is the imaginary unit and $x_\alpha$ is the position of the solution points inside the reference element, while $\Delta=1/N$ is the effective resolution inside each element.
The spectral signature offers valuable insights into the smoothing properties at various solution points~\cite{Lodato_DFOP_2012,SagautDFOP}.

\begin{figure}[h!]
 \centering  
 \begin{subfigure}{0.24\textwidth}
     \includegraphics[width=\textwidth]{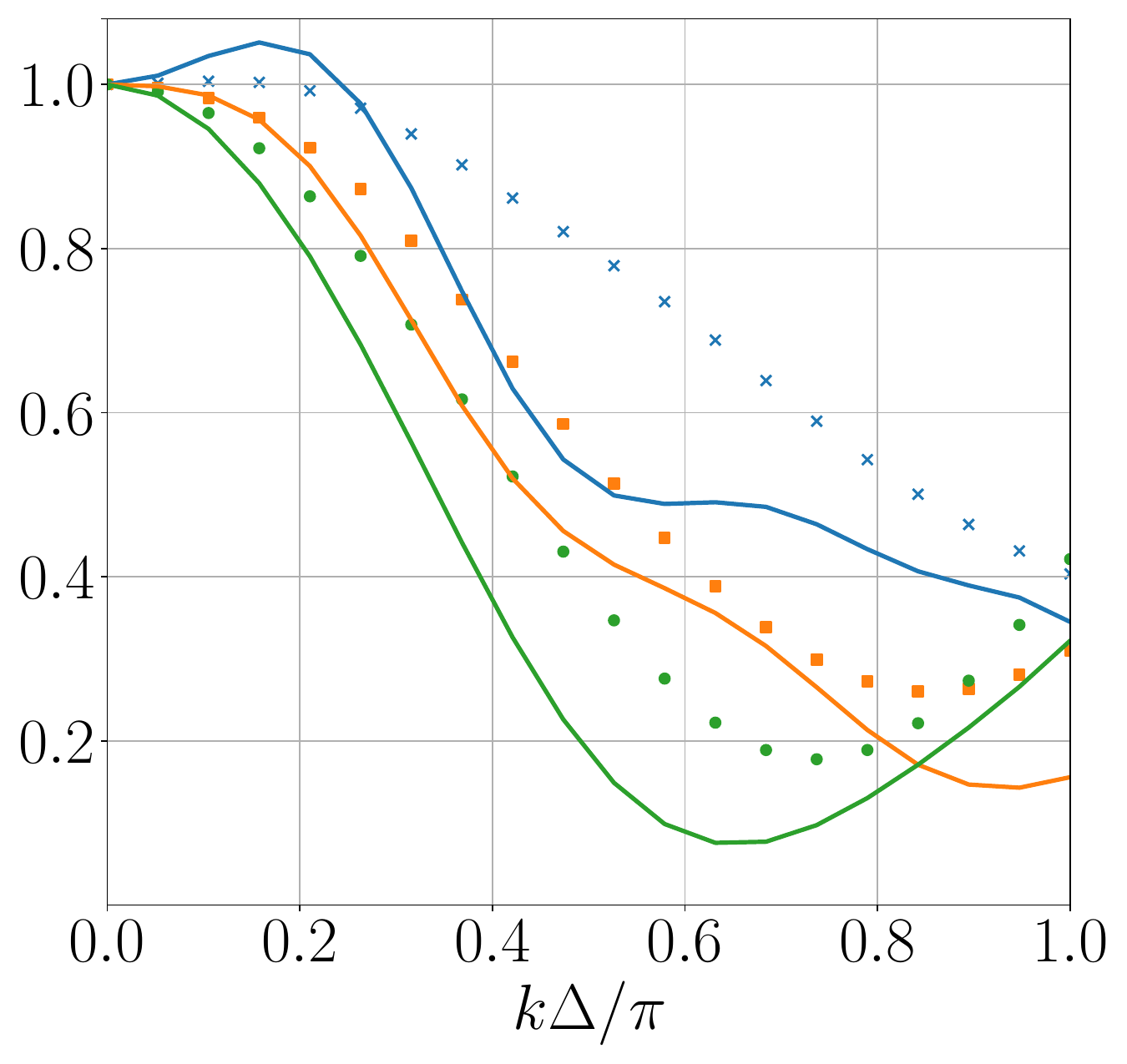}
     \caption{$\mathrm{p}=2$.}
     \label{fig:43p2_gfou:a}
 \end{subfigure}
 \begin{subfigure}{0.24\textwidth}
     \includegraphics[width=\textwidth]{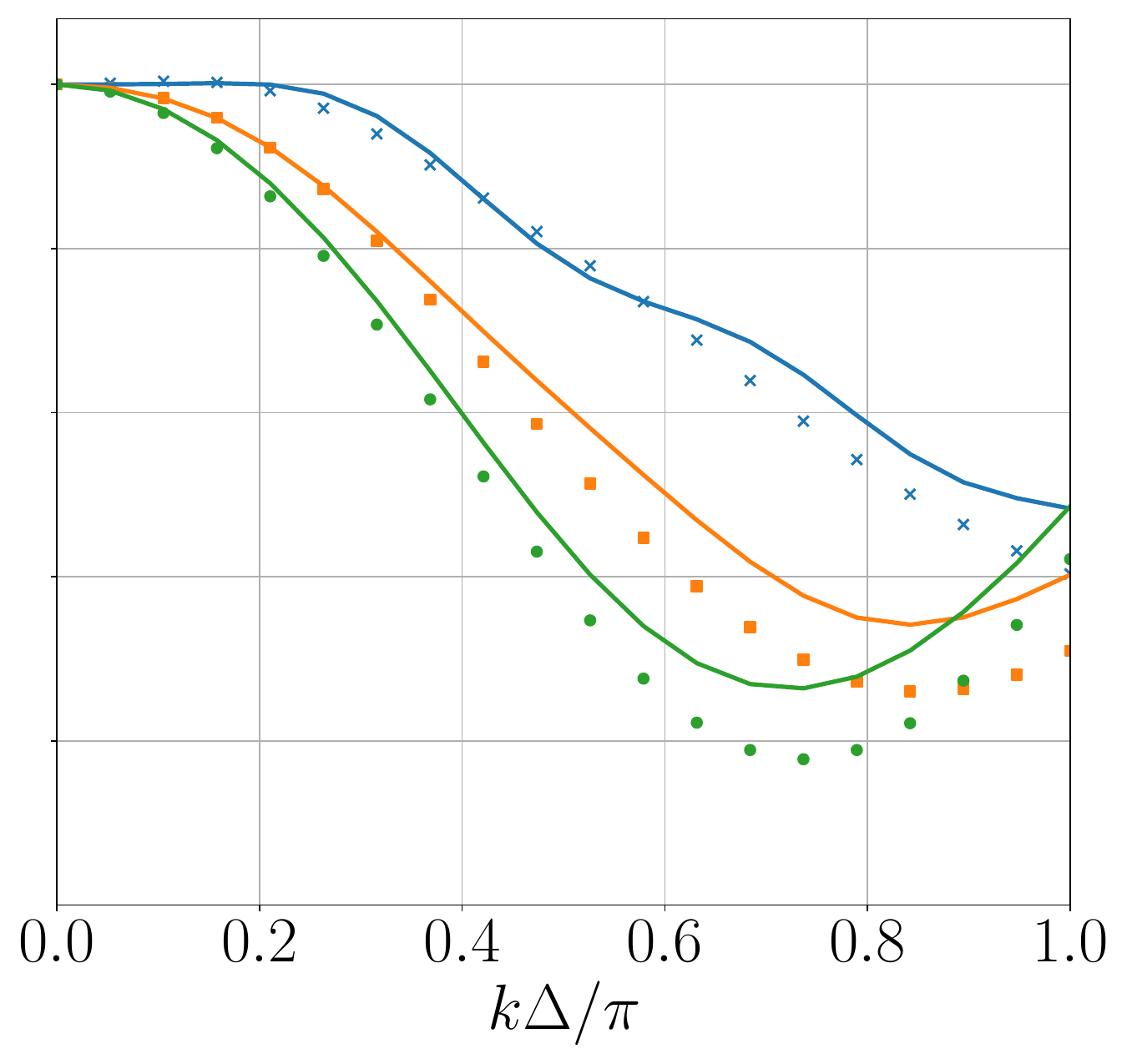}
     \caption{$\mathrm{p}=3$.}
     \label{fig:32p3_gfou:b}
 \end{subfigure}
 \begin{subfigure}{0.24\textwidth}
     \includegraphics[width=\textwidth]{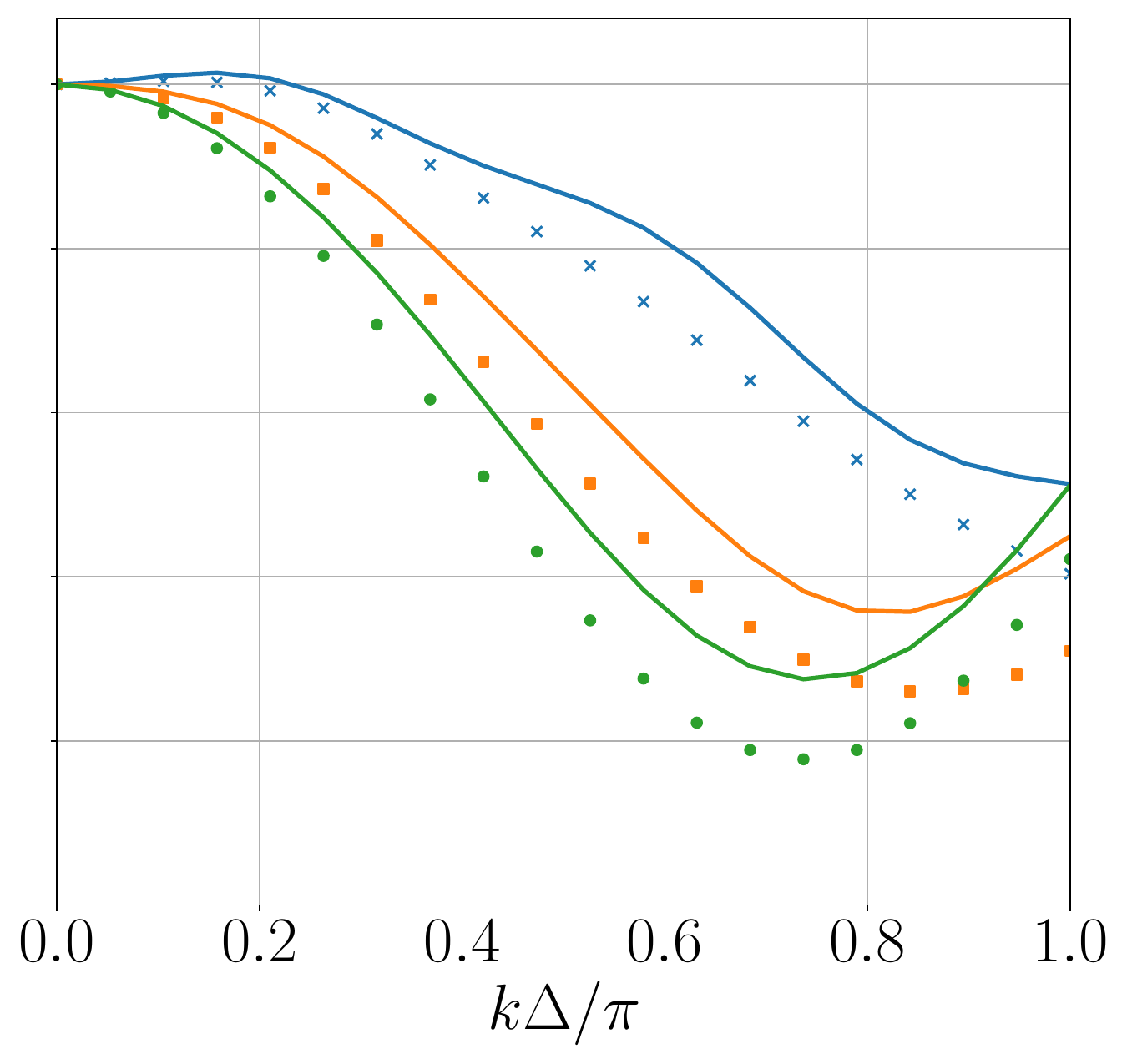}
     \caption{$\mathrm{p}=4$.}
     \label{fig:26p4_gfou:c}
 \end{subfigure}
 \begin{subfigure}{0.24\textwidth}
     \includegraphics[width=\textwidth]{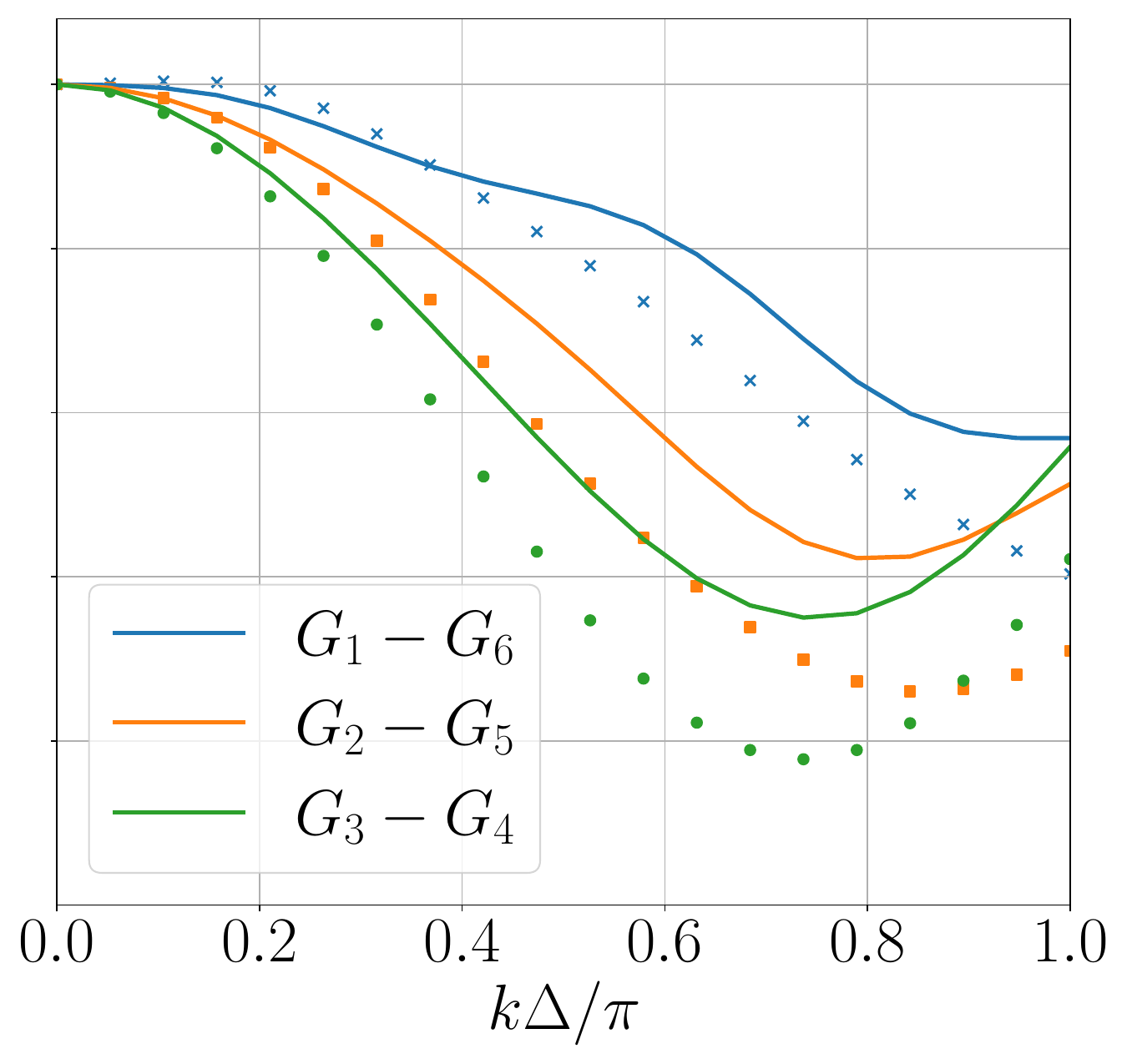}
     \caption{$\mathrm{p}=5$.}
     \label{fig:21p5_gfou:d}
 \end{subfigure} 
\caption{Real part $Re\left[\widehat{G}(\frac{k\Delta}{\pi})\right]$ of the Fourier transform for various polynomials at different locations of solution points. The symbols denote the reference exponential filter with cutoff of $\Delta=2h$.}
   \label{fig:FOUkernel_05}
\end{figure}

We report in Figure~\ref{fig:FOUkernel_05} the spectral signature of the filter at the discrete level for the smallest time integration window, namely $\Delta T=0.5$ along with a reference exponential filter with cutoff of $\Delta=2h$.

Firstly, we note that the curves for $\mathrm{p}=4$ (Figure \ref{fig:26p4_gfou:c}) and $\mathrm{p}=5$ (Figure \ref{fig:21p5_gfou:d}) exhibit the same qualitative trend for all the locations of the solution points, confirming that these two cases have almost the same smoothing properties. 
The values obtained for $\mathrm{p}=4$ in the low wavenumbers region (\ie, $k\Delta/\pi \lesssim 0.5$) for the outermost solution points are slightly larger with respect to $\mathrm{p}=5$. On the contrary, it appears that for the intermediate and the innermost solution points the trend is the opposite, as the curves for $\mathrm{p}=4$ are below the curves of $\mathrm{p}=5$.
For polynomials of degree $\mathrm{p}\leq 3$, there is a more pronounced decay for all the solution points with respect to $\mathrm{p}>3$, showing more pronounced smoothing properties for these orders, as expected.
These curves support the fact that lower polynomials tend to smooth out the solution more aggressively, which naturally results in a lower magnitude of the real part of the Fourier kernel. The filters based on low-order simulations are consequently more dissipative, particularly at lower frequencies. As the polynomial order increases, the filter becomes less dissipative, allowing the real part of the kernel to maintain higher values. Indeed, higher-order polynomials provide greater accuracy and resolution, preserving a larger percentage of the energy of the signal.

It is interesting to note that in all the cases, the filtering effect is stronger within the interior of the element. This finding is consistent with the recent work of Beck et al.~\cite{Beck23_RL0}, who developed a data-driven eddy viscosity model based on forced Homogeneous Isotropic turbulence data. They observed that the eddy viscosity of their model tended to increase near the center of each spectral element. As they suggested, this effect may be attributed to the implicit numerical dissipation introduced by the Riemann solver near the element interfaces. Similar concepts might be applied in the present framework, where the filter acts more in the interior of the cell rather than on the interface between elements.
In addition to this, we want to highlight that the proposed model is constructed as a linear combination of sharp modal filters. This assumption inherently constraints the kernel to exhibit certain predefined trends that are intrinsically related to the pre-assumed functional form of the filters and ultimately to the modal decay $\sigma_n$ observed. Indeed, we observe from Figure~\ref{fig:FOUkernel_05} that the curves for $\mathrm{p}\geq 3$ follow similar qualitative trends with respect to the reference exponential filter. Only the case of $\mathrm{p}=2$, deviates from the others and from the reference filter. This behavior is expected based on the previous discussions on the lowest orders. In this scenario, the network has more difficulties in replicating the correct behavior of the ILES.

We performed an additional analysis where we compared the Fourier kernels of the data-driven filter with exponential filters at different cutoff length scales. In particular, we manually calibrated the cutoff of the latter in order to match the curves of the data-driven filter as closely as possible. In this way, we can extrapolate the equivalent filter width, herein denoted as $\Delta_{eq}$.

\begin{figure}[h!]
 \centering  
 \begin{subfigure}{0.24\textwidth}
     \includegraphics[width=\textwidth]{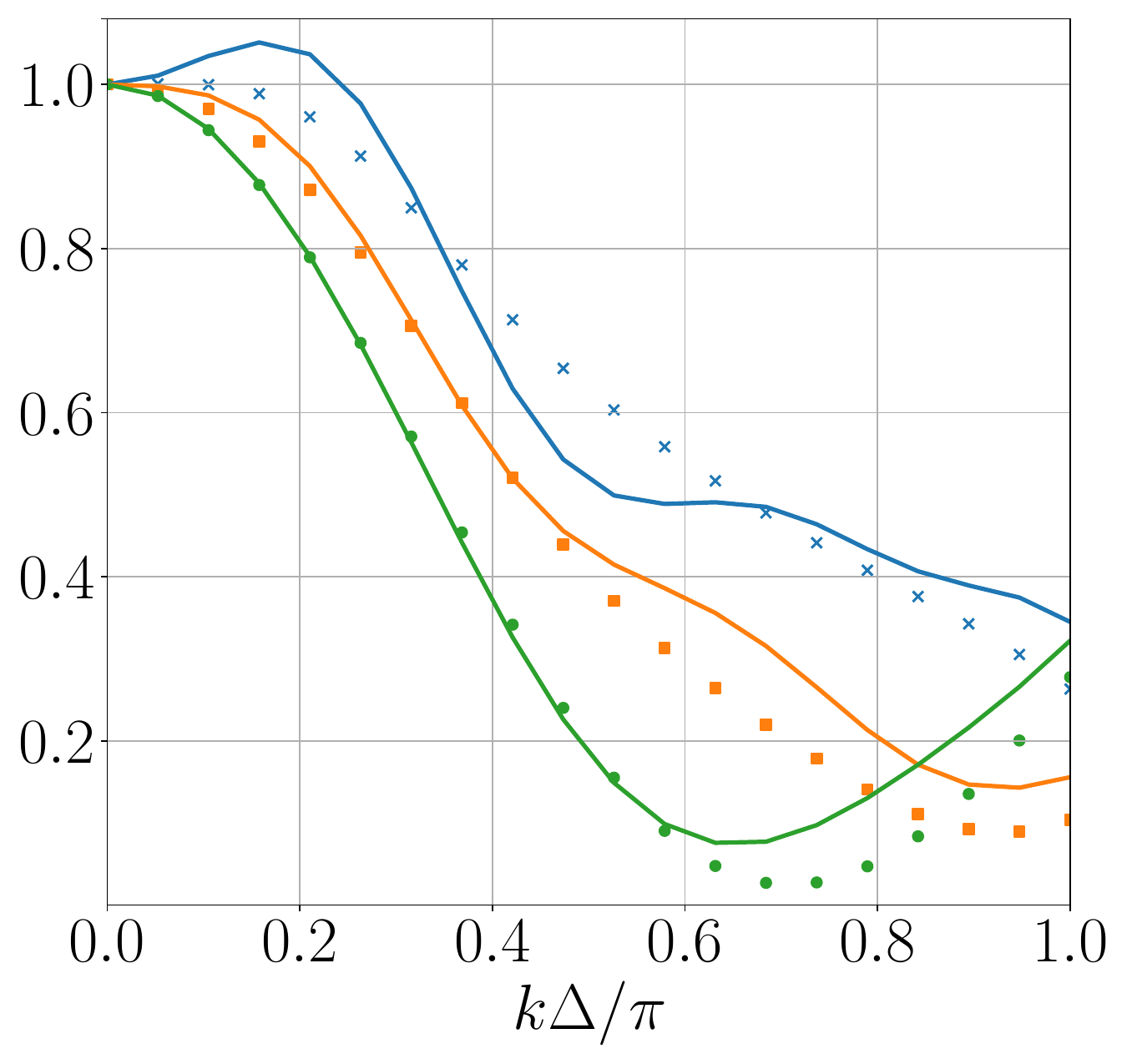}
     \caption{$\mathrm{p}=2$, $\Delta_{eq}=2.7h$.}
     \label{fig:43p2_gfou_flt:a}
 \end{subfigure}
 \begin{subfigure}{0.24\textwidth}
     \includegraphics[width=\textwidth]{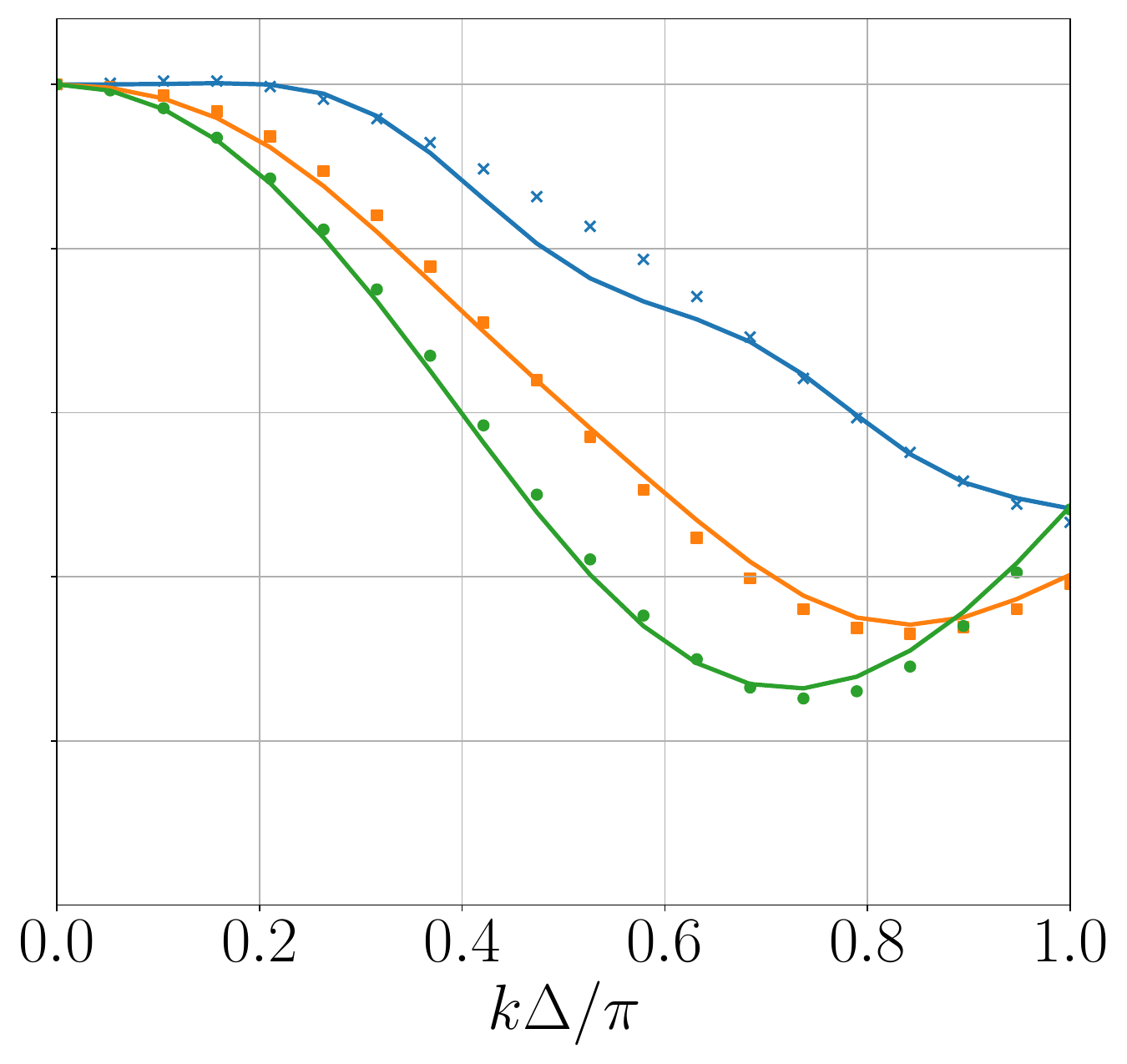}
     \caption{$\mathrm{p}=3$, $\Delta_{eq}=1.9h$.}
     \label{fig:32p3_gfou_flt:b}
 \end{subfigure}
 \begin{subfigure}{0.24\textwidth}
     \includegraphics[width=\textwidth]{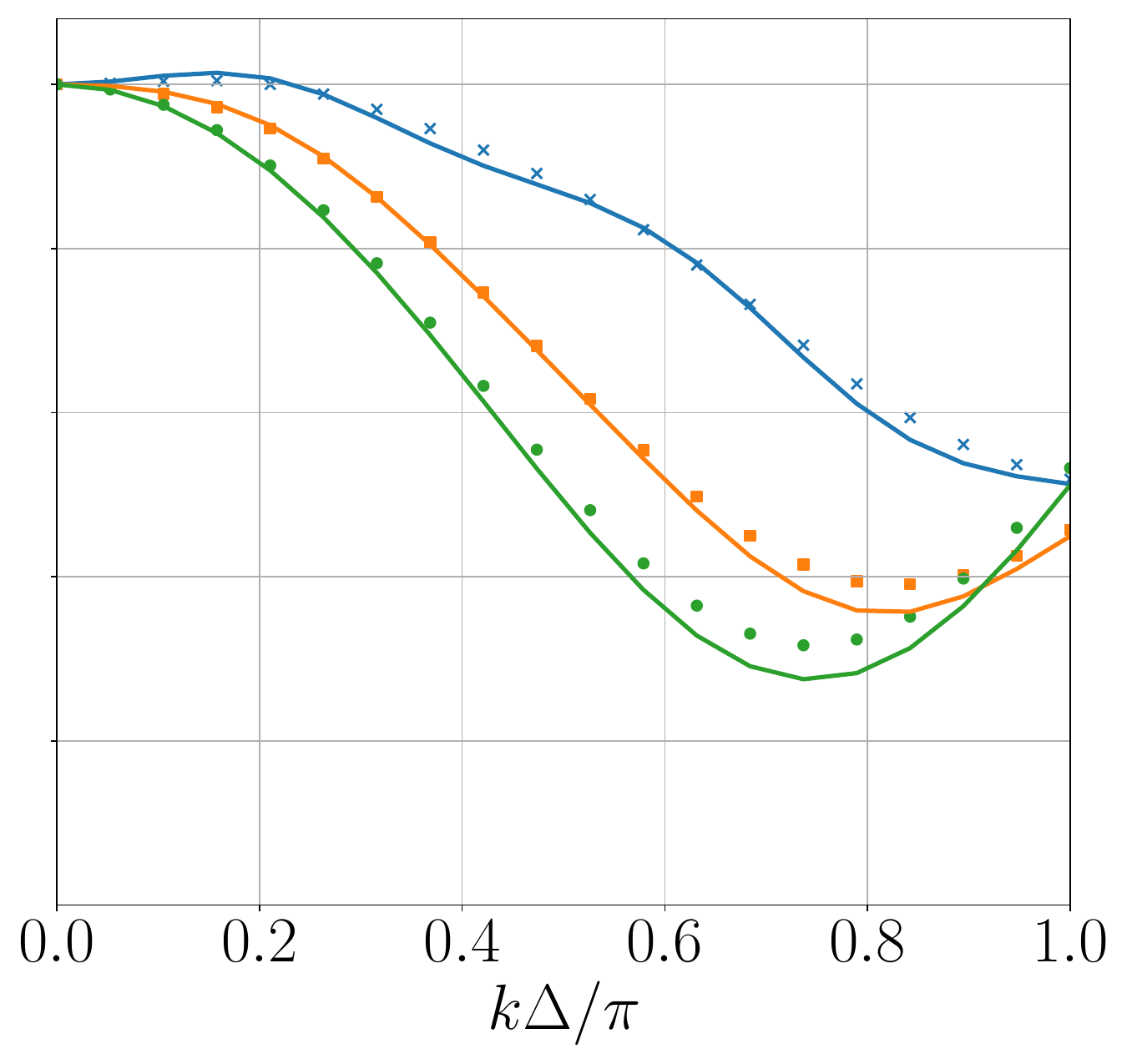}
     \caption{$\mathrm{p}=4$, $\Delta_{eq}=1.6h$.}
     \label{fig:26p4_gfou_flt:c}
 \end{subfigure}
 \begin{subfigure}{0.24\textwidth}
     \includegraphics[width=\textwidth]{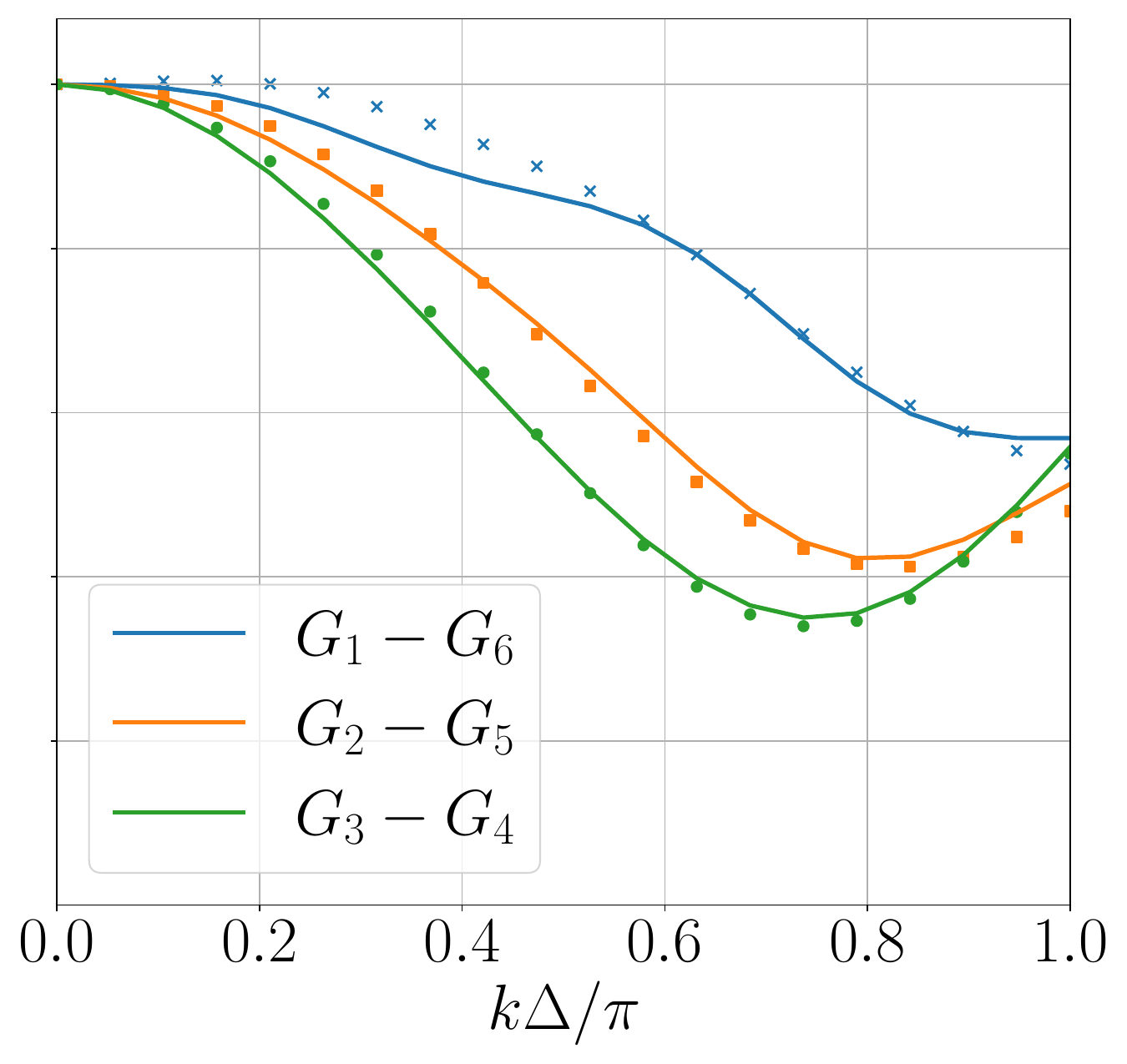}
     \caption{$\mathrm{p}=5$, $\Delta_{eq}=1.6h$.}
     \label{fig:21p5_gfou_flt:d}
 \end{subfigure} 
\caption{Real part $Re\left[\widehat{G}(\frac{k\Delta}{\pi})\right]$ of the Fourier transform for various polynomials at different locations of solution points. The symbols denote the equivalent exponential filter.}
   \label{fig:FOUkernel_05_fltwidth}
\end{figure}

Figure~\ref{fig:FOUkernel_05_fltwidth} displays the curves for various orders of accuracy, along with those obtained by manually adjusting an appropriate exponential filter.

For $\mathrm{p}=4$ (Figure \ref{fig:26p4_gfou_flt:c}) and $\mathrm{p}=5$ (Figure \ref{fig:21p5_gfou_flt:d}), the optimal filter width is $\Delta_{eq} \simeq 1.6h$, producing curves that most closely match the implicit filters. For $\mathrm{p}=3$, the equivalent filter width is slightly larger at $\Delta_{eq} \simeq 1.9h$. Lastly, for $\mathrm{p}=2$, the equivalent filter width is approximately $\simeq 2.7h$. 

These findings support the general assumption that for a fixed number of degrees of freedom, an higher polynomial degree provides less smoothing of the solution as they have an effective cutoff width which is smaller with respect to the one classically considered (\ie, $\Delta=2h$). Therefore, we can extrapolate that high orders provide an effective resolution which is higher compared to low orders. It is also interesting to notice that for $\mathrm{p}=3$ the equivalent filter width is close to the reference ($\Delta \simeq 2h$). Such finding is in agreement with what is commonly considered a classical choice for ILES based on DSEMs. In fact, third- and fourth-order DSEMs have been found suitable for ILES in many different configurations~\cite{FernandezSGSimplicit,chapelier2016spectral}.
\subsection{Kinetic energy transfers across scales}\label{subsec:FOU}
Up to this point we have considered a series of analyses to build a numerical characterization of the data-driven filter. For example, we focused on the modal decay for different orders of accuracy and on the shape of the filters in Fourier space within the spectral element. Such analyses are more related to the numerical nature of the filter itself rather than its effect on the physics of turbulent flows. Consequently, we herein consider the balance of the filtered kinetic energy to better understand the energy transfer mechanisms implied by the filtering operation. The transport equation for the filtered kinetic energy reads:
\begin{equation}
    \frac{\partial \frac{1}{2}\f{\rho} \ff{u_k}\ff{u_k}}{\partial t} = T_K + D_K - S_K + P_K - V_K, 
    \label{eq:Ktransport0}
\end{equation}
 with,
\begin{equation}
    T_K = \frac{\partial}{\partial x_j}\biggl\{ -\frac{1}{2} \f{\rho}\ff{u_k}\ff{u_k} \ff{u_j} -\f{p} \ff{u_j} + \left[2\mu \ff{S}^d_{ij}
    +\tau^{\mathrm{SGS}}_{ij} + \tau^v_{ij} \right]\ff{u_i} \biggr\}, \label{eq:Ktransport}
\end{equation}
\begin{equation}
    D_K = \f{p}\frac{\partial \ff{u_j}}{\partial x_j}, \;\;\; S_K = 2 \mu(T) \ff{S}^d_{ij}\frac{\partial \ff{u_i}}{\partial x_j}, \;\;\; P_K = \tau^{\mathrm{SGS}}_{ij} \frac{\partial \ff{u_i}}{\partial x_j}, \;\;\; V_K = \tau^v_{ij} \frac{\partial \ff{u_i}}{\partial x_j}, \label{eq:Kterms}
\end{equation}
where $\ff{S}^d_{ij}$ is the deviatoric part of the strain-rate tensor, namely
\begin{equation}    \ff{S}^d_{ij} = \frac{1}{2}\left( \frac{\partial \ff{u_i}}{\partial x_j} +\frac{ \partial \ff{u_j}}{\partial x_i} \right) -\frac{1}{3}\frac{\partial \ff{u_k}}{\partial x_k}\delta_{ij}, \label{eq:Sdev}
\end{equation}
and $\tau^v_{ij}$ is
\begin{equation}
     \tau^v_{ij} = \f{2\mu(T) S^d_{ij}} - 2\mu(T) \ff{S}^d_{ij}. \label{eq:TauV}
\end{equation}
In this study, $\tau^v_{ij}$ can be disregarded since the viscosity remains nearly constant within the temperature range of the test case considered here.

In equation \eqref{eq:Ktransport0}, the term $T_K$ serves for redistribution of energy, while other terms act as sources/sinks.
In this equation, the most important term for assessing the impact of the data-driven filters on kinetic energy dynamics is $P_K$, which is known as subgrid scale dissipation.
Indeed, this term appears in the unresolved part of the kinetic energy with an inverted sign and, as such, it represents the interactions between the resolved and unresolved scales in the LES formalism \cite{PIOMELLI_SGSBS_0}.
This information is crucial for a better understanding of the interactions between scales in turbulent flows.
Correctly representing such interactions is of fundamental importance for the overall accuracy and stability of \emph{a-posteriori} computations. 

\begin{figure}[h!]
 \centering  
 \begin{subfigure}{0.45\textwidth}
     \includegraphics[width=\textwidth]{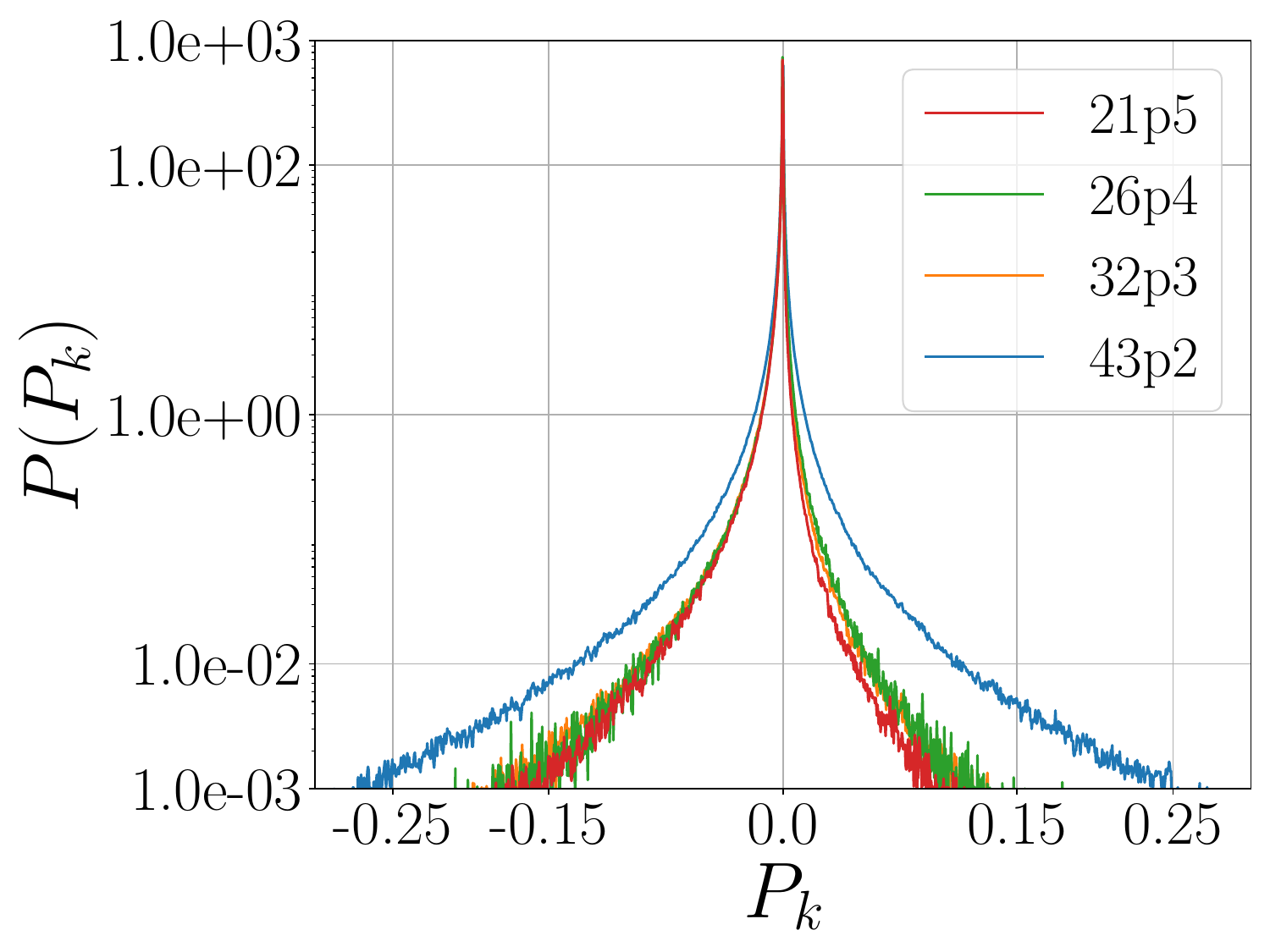}
     \caption{PDF of the dissipation term \eqref{eq:PK}.}
     \label{fig:BackScatter:a}
 \end{subfigure}
 \begin{subfigure}{0.45\textwidth}
     \includegraphics[width=\textwidth]{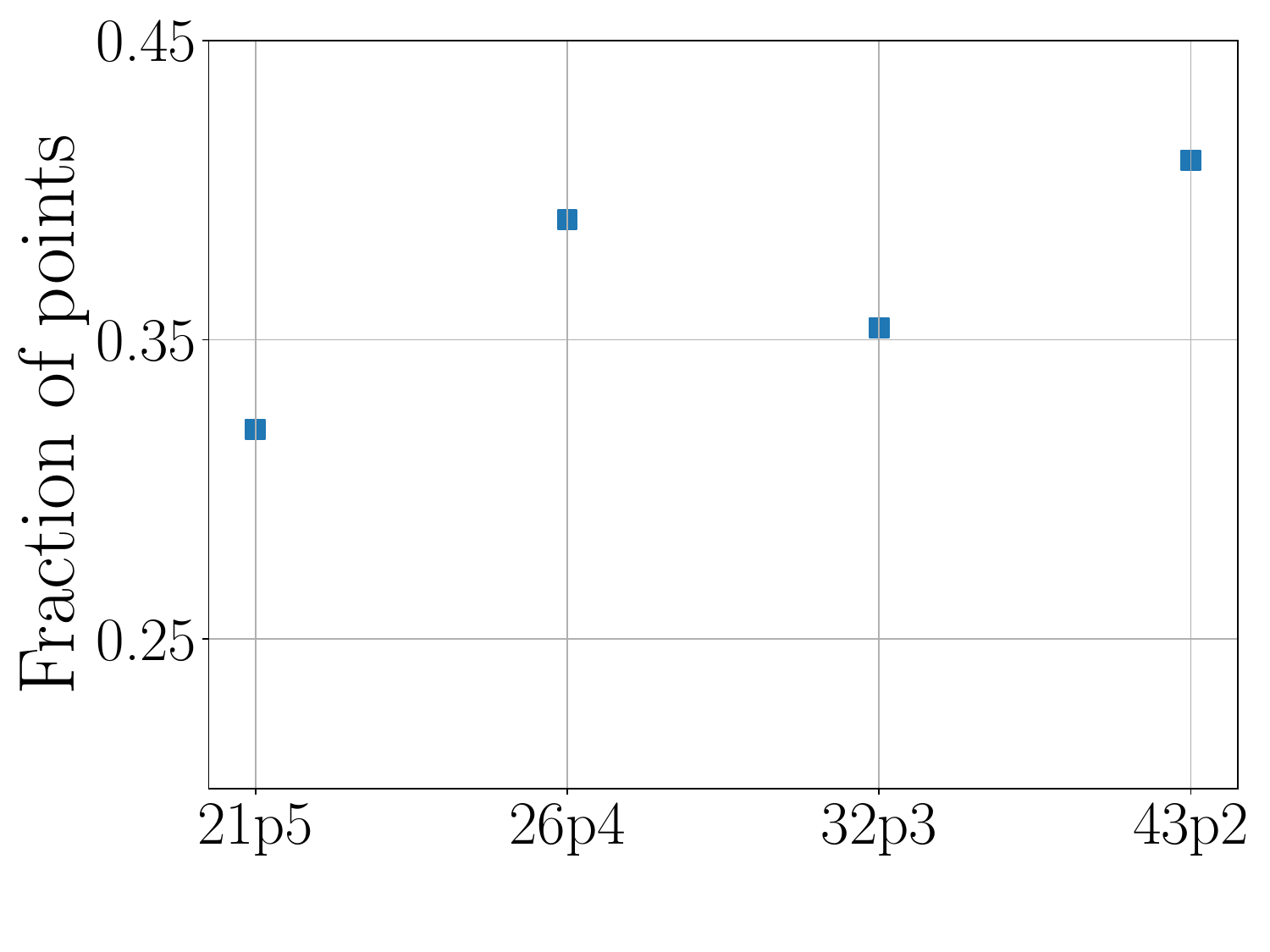}
     \caption{Fraction of Backscattered points.}
     \label{fig:BackScatter:b}
 \end{subfigure}
\caption{PDF of the dissipation term \eqref{eq:PK} (left) and fraction of points experiencing backscatter (right).}
   \label{fig:BackScatter}
\end{figure}

We consider the Probability Density Function (PDF) of the dissipation term $P_K$ obtained with the filtered velocities by the data-driven filters. In this way we can assess what types of energy transfer mechanisms are triggered by the present filtering operation. For clarity, we will denote the SGS stress tensor $\tau^{\mathrm{SGS}}_{ij}$ by $\tau^{DD}_{ij}$ to emphasize the dependence upon the data-driven filters:
\begin{equation}
    \tau^{\mathrm{SGS}}_{ij} = \tau^{DD}_{ij} = \f{\rho}(\ff{u_i u_j}^{DD} - \ff{u_i}^{DD} \ff{u_j}^{DD}), \label{eqn:tauDD}
\end{equation}
\begin{equation}
    P_K= \tau^{DD}_{ij} \frac{\partial \ff{u_i}}{\partial x_j}^{DD}. \label{eq:PK}
\end{equation}
The resulting PDF for the various data-driven models is shown in Figure~\ref{fig:BackScatter}. This has been computed by averaging instantaneous PDFs every two non-dimensional time units sampled in the Decaying Homogeneous Turbulence regime (\ie, $t^* \in [10,20]$). 

Firstly, the prevalence of negative values is indicative that all the data-driven filters are, on average, dissipative, confirming the general tendency of ILES in providing net forward energy transfers. However, the right portion ($P_k>0$) reveals also the existence of local backscatter phenomena. 

The curves for $\mathrm{p}\geq 3$ are almost overlapped. This  behavior is in agreement with the previous findings. As the modal decay is similar, the dissipation term is expected to be close among these orders.
Instead, $\mathrm{p}=2$ presents a PDF which is quantitatively different. Positioned above the other curves, it suggests that, greater local energy transfers are more likely to occur in both the forward and backward directions.
This trend confirms that, although a larger equivalent filter width ($\Delta_{eq}=2.7h$) is anticipated to decrease gradients, it may also cause a net increase in $\tau^{DD}_{ij}$, leading to an overall increase in $P_K$.
The presence of backscatter phenomena suggests that the effect of the numerical scheme is not merely dissipative, but also manifests local inverse kinetic energy cascade from the unresolved to the resolved scales. As a result, the numerical errors associated to the spectral difference scheme inherently provide both direct and inverse kinetic energy cascade.

To better characterize the backscatter implied by the data-driven filters, we will also consider the fraction of points which experience backscatter (\ie, $P_k>0$) with respect to the total. This is shown in Figure~\ref{fig:BackScatter:b}.
Although there is no clear trend, the fraction of points ranges from a minimum of $0.33$ to $0.42$ for all the orders in consideration. These values are in line with those obtained by Piomelli \cite{PIOMELLI_SGSBS_0} in the case of Compressible Isotropic Turbulence test case when a spectral cutoff filter was used.
%

\section{Conclusions and Perspectives}\label{sec:conclusion}
With the objective of characterizing the numerical footprint of Implicit Large-eddy Simulations in the context of the Spectral Difference scheme, we have proposed a data-driven filter based on CNNs to link the spatial features of the DNS with the ones from the ILES.
The pipeline to build a data-driven modal filter was presented. In particular, the proposed filter was formulated as a linear combination of sharp-modal filters that act at elementary cell level on different cutoff length scales. The coefficients of the linear combination are predicted using a convolutional neural network. The model has been trained on the data generated from the Taylor-Green vortex test case at $\Rey=1600$ in the low-Mach regime.
To separate the effects of spatial discretization from accumulation error, we generated ILES data by periodically restarting the simulations of the ILES from DNS data. With this approach, the curves of the quantities of interest for evaluating the numerical scheme (\eg, kinetic energy dissipation rate and numerical dissipation) are characterised by an easier interpretation, allowing us to better understand the connection with classical eigenanalysis for small time evolution.
A series of analyses regarding the data-driven filter were considered in order to better characterise different aspects of the present spatial discretization.
Among these, the modal decay of the resulting filter and kinetic energy spectra have been investigated. Overall, the proposed approach was found capable of accurately predicting the filtering operator for the smallest time window herein considered for all the orders of accuracy. Data-driven filters based on low-order simulations have shown slightly less satisfactory results, probably due to the omission of inter-element jumps within the modeling strategy. It was observed that larger time windows caused the network to reproduce less accurately the ILES results. We suspect that this trend is caused by the accumulation error which is allowed to grow for a larger time interval.

Additionally, an analysis of the spectral characteristics of the resulting filter was performed by examining the kernel in Fourier space. The findings confirmed that high polynomial orders are characterized by a smaller equivalent filter width in comparison to an equivalent exponential filter based on the classical cutoff scale $\Delta_{el}/N$. At the same time, data-driven filters based on low order simulations resemble exponential filters tuned on scales larger than $\Delta_{el}/N$. Polynomials orders of $\mathrm{p}=3-4$, instead, are particularly close to the reference exponential filter, further supporting this common choice for ILES using DSEMs.

Finally, the capacity of the data-driven filter in representing direct and inverse kinetic energy cascade was assessed. Both direct and inverse kinetic energy cascades were observed, with a prevalence of the former. The presence of backscatter suggests that Implicit Large-Eddy Simulations based on Discontinuous Spectral Element Methods might be equipped with an intrinsic mechanisms to transfer energy in both directions with a predominance of direct kinetic energy cascade.

Overall, it was found that the proposed data-driven filter was able to capture the main features of ILES in a reasonably accurate way, providing a more interpretable model to link DNS and ILES based on the spectral difference scheme. The trained model not only is able to reproduce ILES data, but it allows a large variety of additional analyses (\eg, spectral signature and local energy transfers) thanks to its interpretability. Such filter consequently represents a useful tool for diagnostics of numerical schemes which can be easily extended to different spatial discretizations and even to Explicit LES based on analytical SGS models. Although the present work focused only on \emph{a-priori} analyses, the trained filter can find many applications also in \emph{a-posteriori} settings. For example, a more detailed characterisation of the implicit filter induced by numerical schemes can lead to the development of \emph{a-posteriori} SGS models which are aware of the underlying spatial discretization. The proposed filter can then be easily implemented within scale-similarity models or dynamic procedures in classical SGS models (\eg, Smagorisnky, Vreman, WALE) in order to take into account the numerical scheme in the SGS modeling rationale. Also, designing filters which are aware of the underlying numerical scheme used in \emph{a-posteriori} computations can mitigate the instability issues related to offline data-driven SGS models (see \cite{BECK_19}).

\section*{Aknowledgements}

The use of the SD solver originally developed by Antony Jameson’s group at Stanford University is gratefully acknowledged.
We acknowledge the support provided by PRIN ``FaReX - Full and Reduced order modeling of coupled systems: focus on non-matching methods and automatic learning'' project, the European Research Council Executive Agency by the Consolidator Grant project AROMA-CFD ``Advanced Reduced Order Methods with Applications in Computational Fluid Dynamics'' - GA 681447, H2020-ERC CoG 2015 AROMA-CFD, PI G. Rozza, and INdAM-GNCS 2021-2023 projects. NT and GR acknowledge support from the European Union - NextGenerationEU, in the framework of the iNEST - Interconnected Nord-Est Innovation Ecosystem (iNEST ECS00000043 – CUP G93C22000610007).
We are grateful to Guido Lodato for the invaluable suggestions and guidance, which helped in improving the present work.

\appendix

\section{Results from the Decaying Hit Regime Training} \label{:train_II}

In this appendix, we present the  results obtained from our training experiments conducted under the Decaying Homogeneous Isotropic Turbulence regime (DHIT) of the Taylor-Green vortex test case, specifically from $t^{*}=10-20$.

\begin{figure}[h!]
        \centering
        \begin{subfigure}[t]{0.245\textwidth}
            \centering
            \includegraphics[width=\linewidth,height=50mm, keepaspectratio]{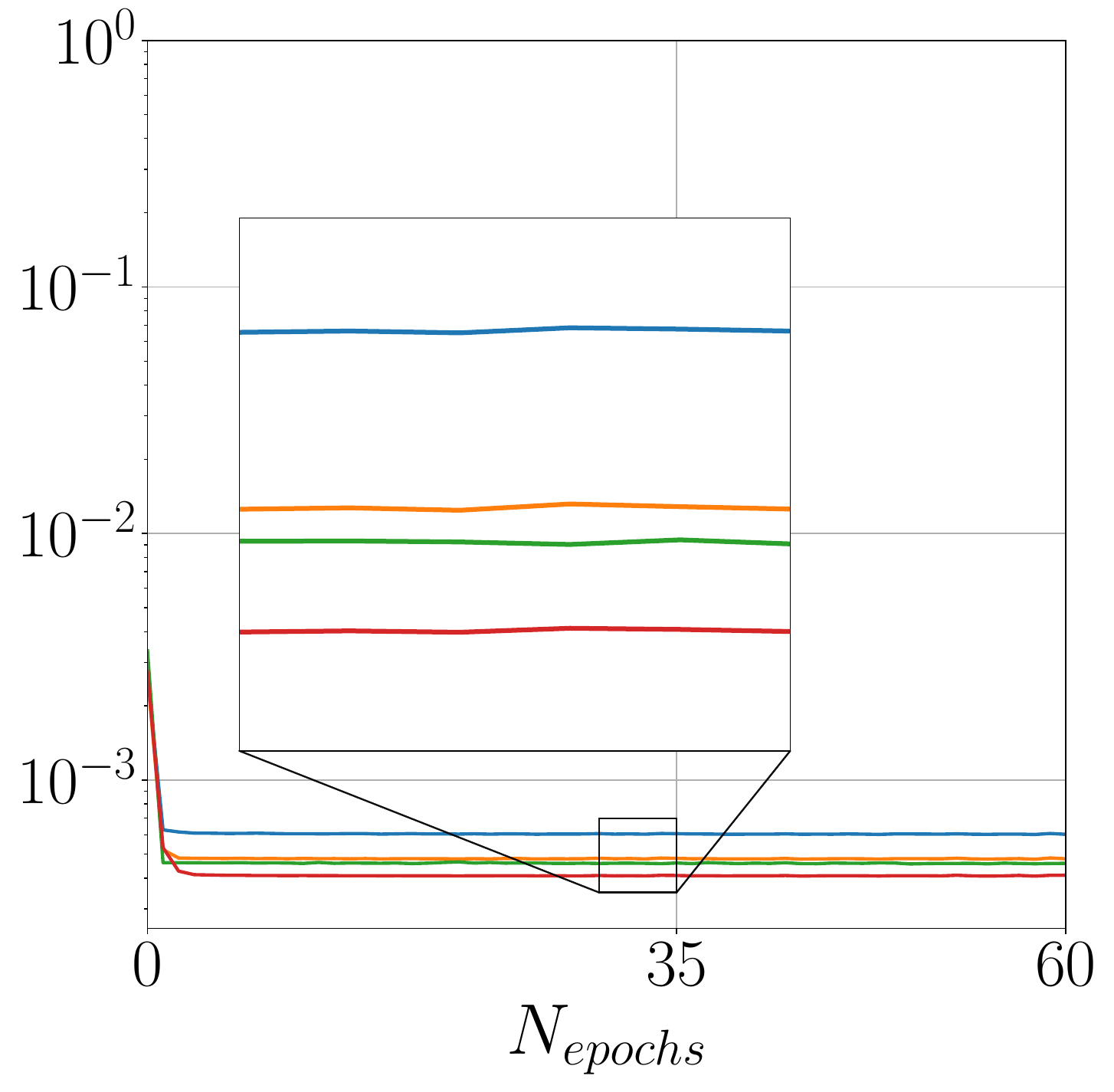}
            \caption{$\Delta T=0.5$.}
            \label{fig:Lt_05_2}
        \end{subfigure}
        \hfill
        \begin{subfigure}[t]{0.245\textwidth}
            \centering
            \includegraphics[width=\linewidth,height=40mm, keepaspectratio]{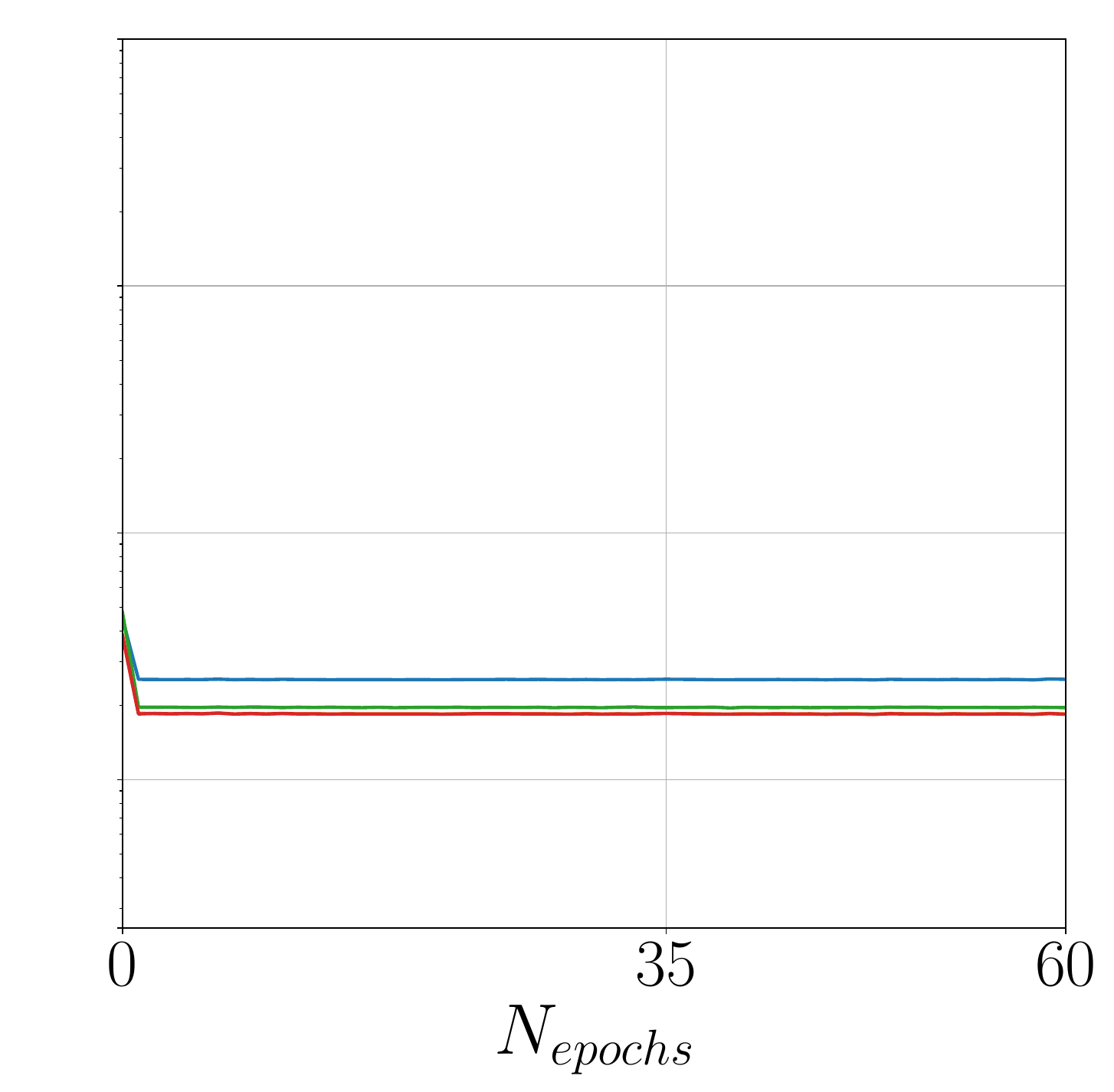}
            \caption{$\Delta T=2$.}
            \label{fig:Lt_2_2}
        \end{subfigure}
        \hfill
        \begin{subfigure}[t]{0.245\textwidth}
            \centering
            \includegraphics[width=\linewidth,height=50mm, keepaspectratio]{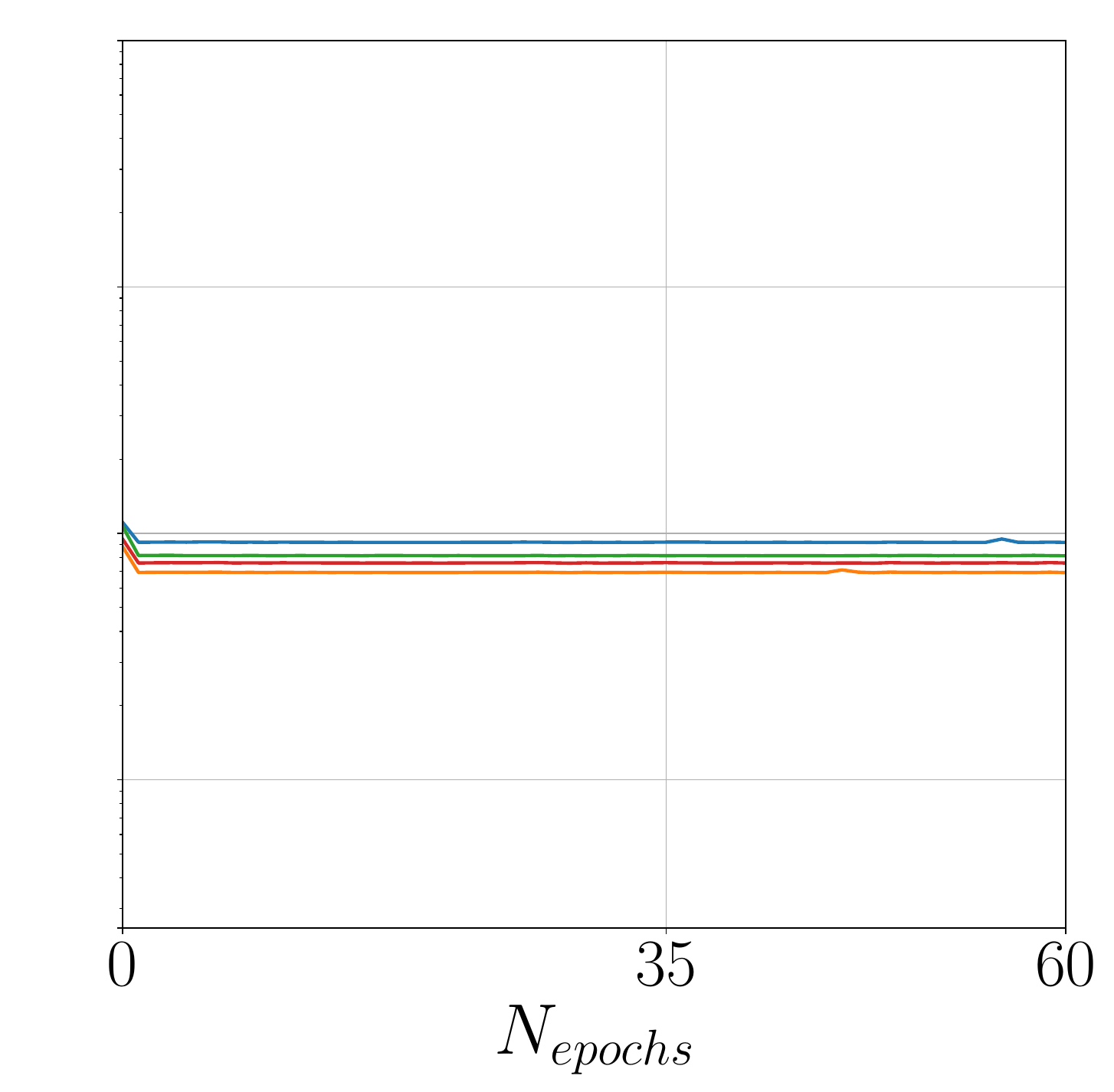}
            \caption{$\Delta T=4$.}
            \label{fig:Lt_4_2}
        \end{subfigure}
        \hfill
        \begin{subfigure}[t]{0.245\textwidth}
            \centering
            \includegraphics[width=\linewidth,height=40mm, keepaspectratio]{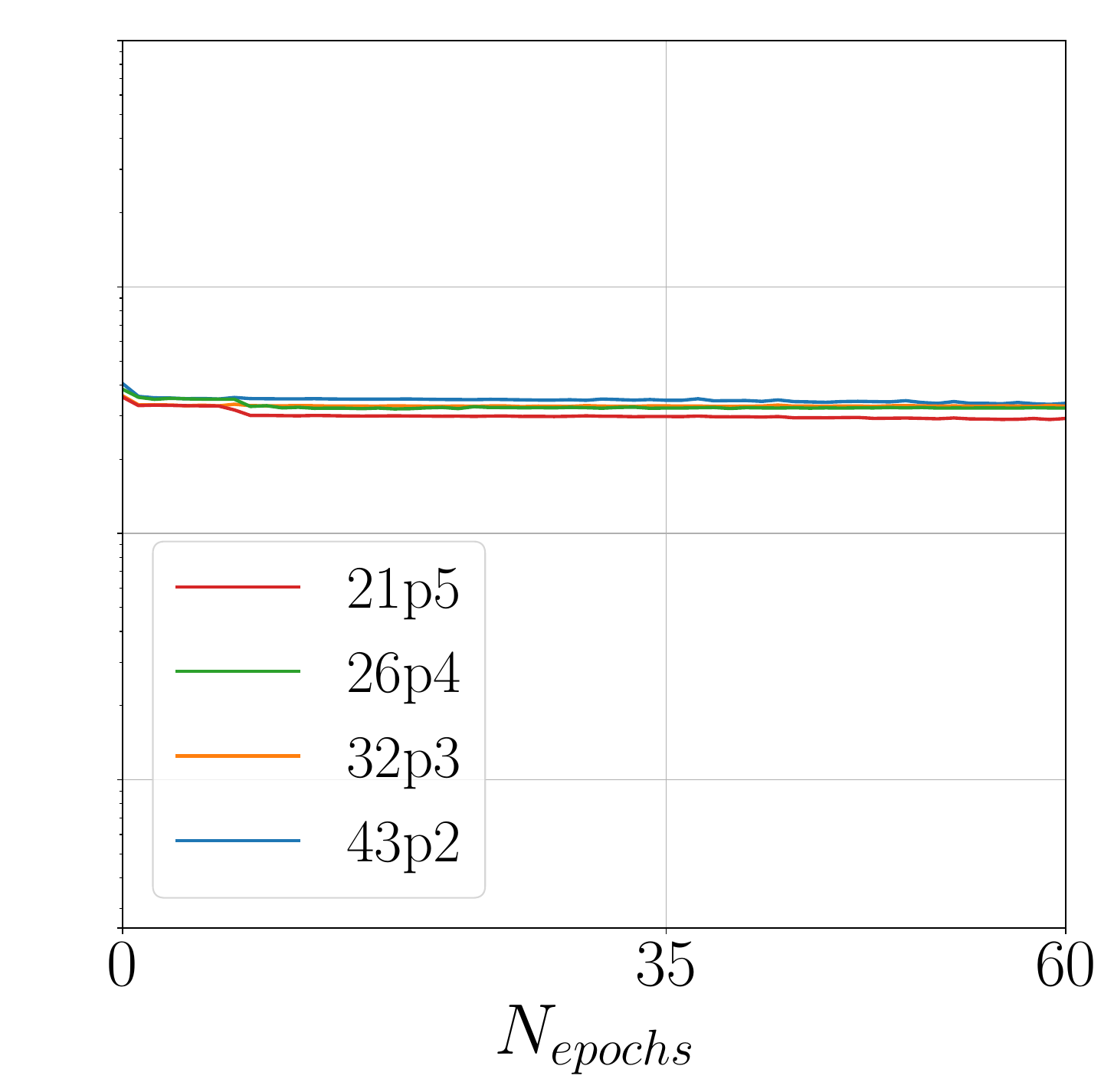}
            \caption{$\mathrm{ILES}$.}
            \label{fig:Lt_iles_2}
        \end{subfigure}
        \caption{Training MSE against epochs ($N_{epochs}$) for different polynomial orders for the model constructed with the DHIT data.. The integration time span increases from left to right.}
        \label{fig:TRAIN_LOSS_TIME_WINDOWS_2}
\end{figure}

\begin{figure}[h!]
        \centering
        \begin{subfigure}[t]{0.245\textwidth}
            \centering
            \includegraphics[width=\linewidth,height=50mm, keepaspectratio]{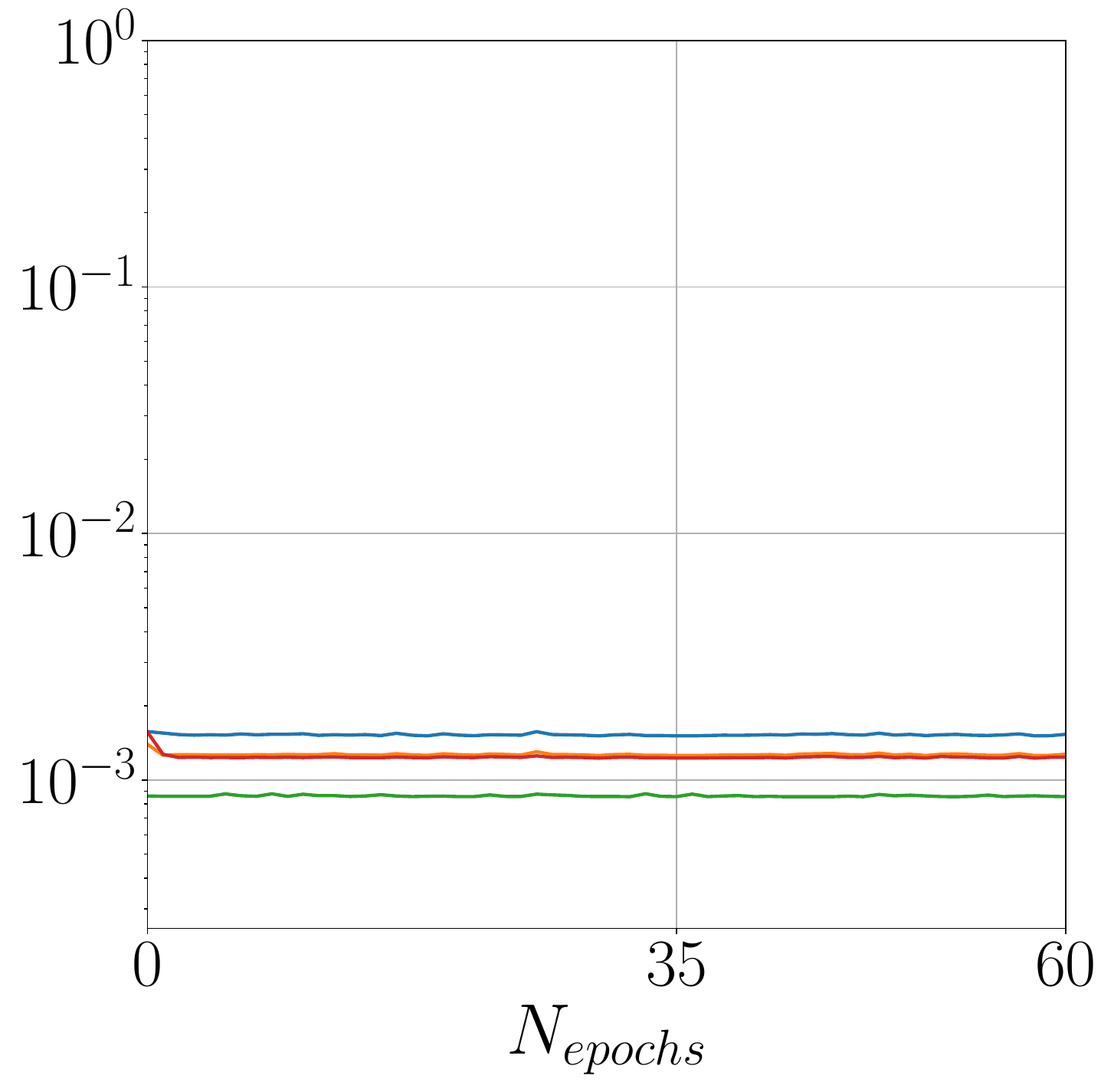}
            \caption{$\Delta T=0.5$.}
            \label{fig:Lv_05_2}
        \end{subfigure}
        \hfill
        \begin{subfigure}[t]{0.245\textwidth}
            \centering
            \includegraphics[width=\linewidth,height=40mm, keepaspectratio]{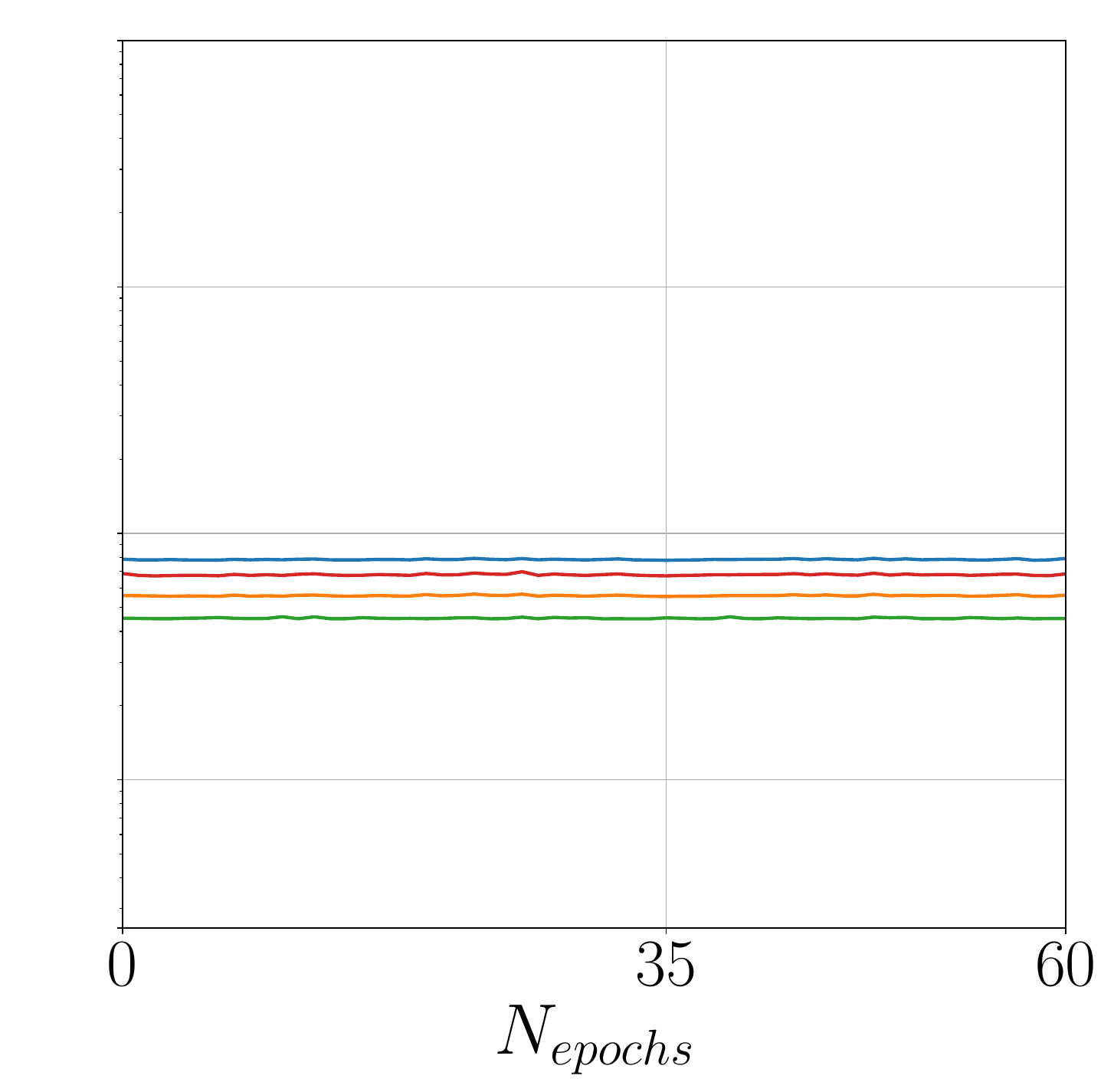}
            \caption{$\Delta T=2$.}
            \label{fig:Lv_2_2}
        \end{subfigure}
        \hfill
        \begin{subfigure}[t]{0.245\textwidth}
            \centering
            \includegraphics[width=\linewidth,height=50mm, keepaspectratio]{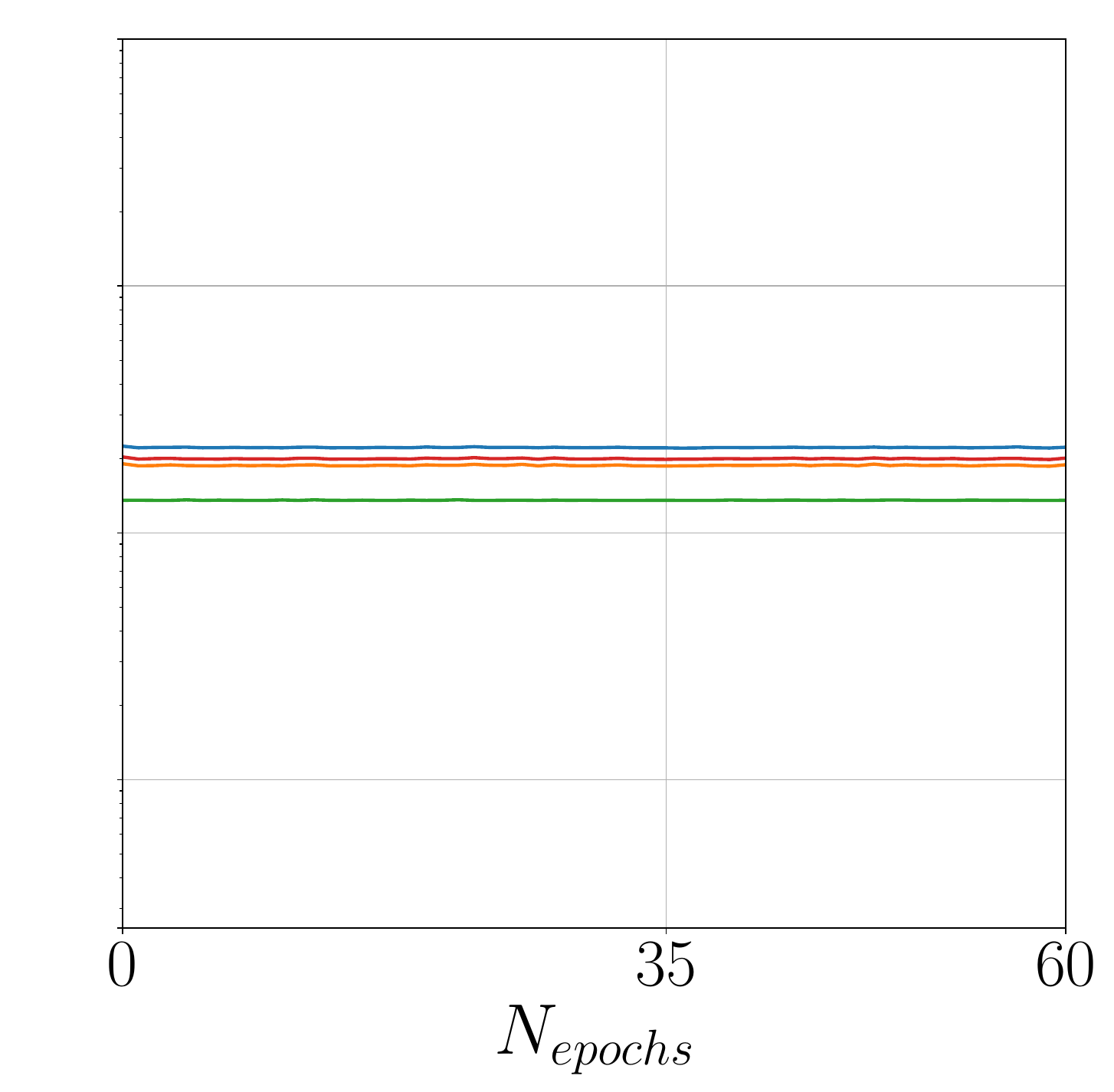}
            \caption{$\Delta T=4$.}
            \label{fig:Lv_4_2}
        \end{subfigure}
        \hfill
        \begin{subfigure}[t]{0.245\textwidth}
            \centering
            \includegraphics[width=\linewidth,height=40mm, keepaspectratio]{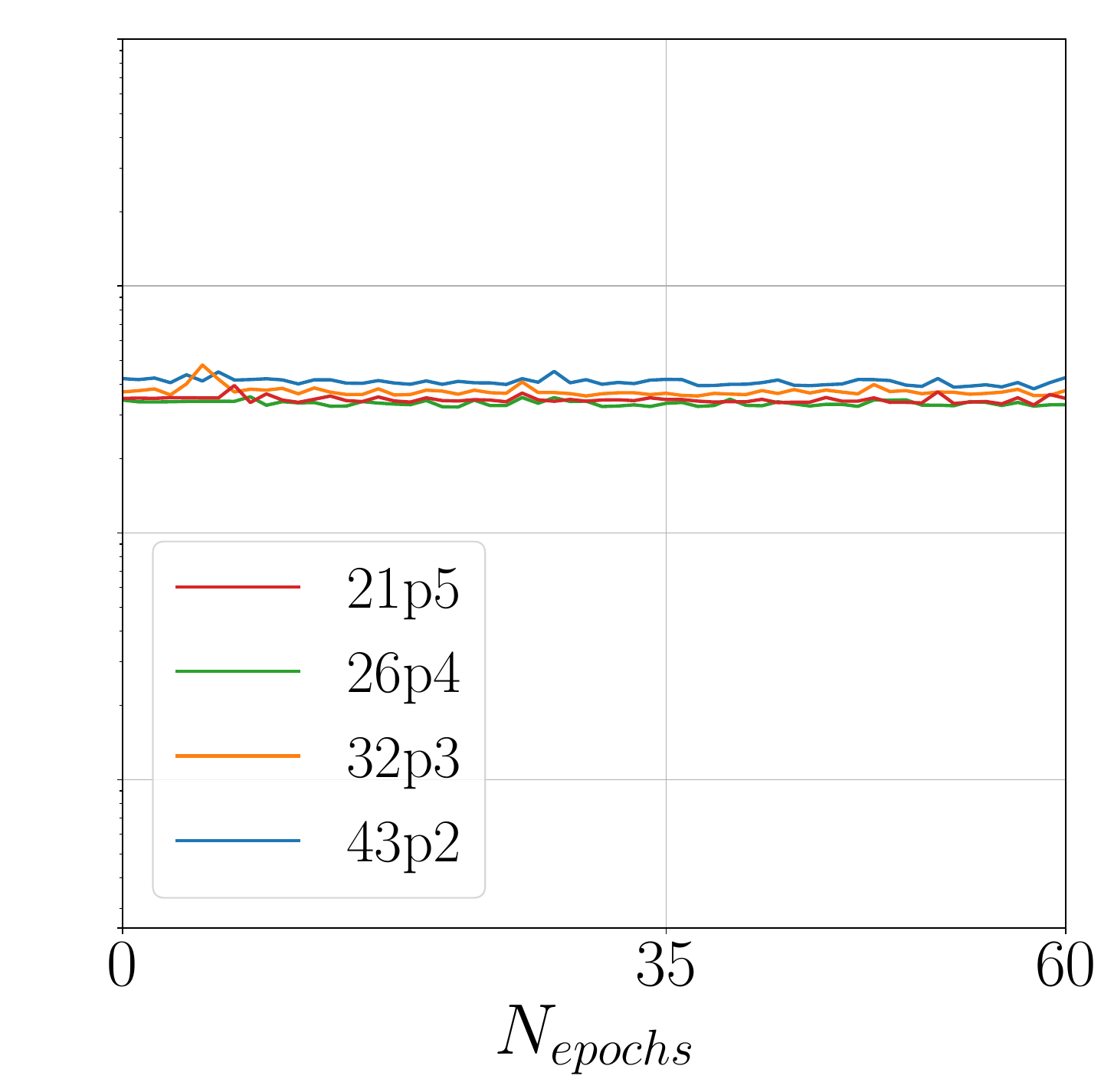}
            \caption{$\mathrm{ILES}$.}
            \label{fig:Lv_iles_2}
        \end{subfigure}
        \caption{Validation MSE against epochs ($N_{epochs}$) for different polynomial orders for the model constructed with the DHIT data.  The integration time span increases from left to right.}
        \label{fig:VAL_LOSS_TIME_WINDOWS_2}
\end{figure}

Figure~\ref{fig:TRAIN_LOSS_TIME_WINDOWS_2} compares the training MSE. obtained from different time windows for all the polynomial orders in consideration. 
The loss function increases as the time window lengthen. This is consistent with the training performed taking data over the entire time span.
Notably, the out-of-distribution peaks have disappeared in this scenario, indicating that parameter optimization is enhanced when training on this dataset.
Moreover, for $\Delta T=0.5$, even $\mathrm{p}=4$ and $\mathrm{p}=5$ are ordered. As a result, the network is capable, on average, of distinguishing features among higher orders in this case. For completeness, we have also reported the validation curves in Figure~\ref{fig:VAL_LOSS_TIME_WINDOWS_2} which follow similar trends with respect to training curves.

\begin{figure}[htb]
 \centering  
 \begin{subfigure}{0.32\textwidth}
     \includegraphics[width=\textwidth]{figures/PRIORI_DATA_DRIVEN/ILES_EXP_0.5_15.0_EK.pdf}
     \caption{Kinetic energy spectrum.}
     \label{fig:Res05_II:a}
 \end{subfigure}
 \begin{subfigure}{0.32\textwidth}
     \includegraphics[width=\textwidth]{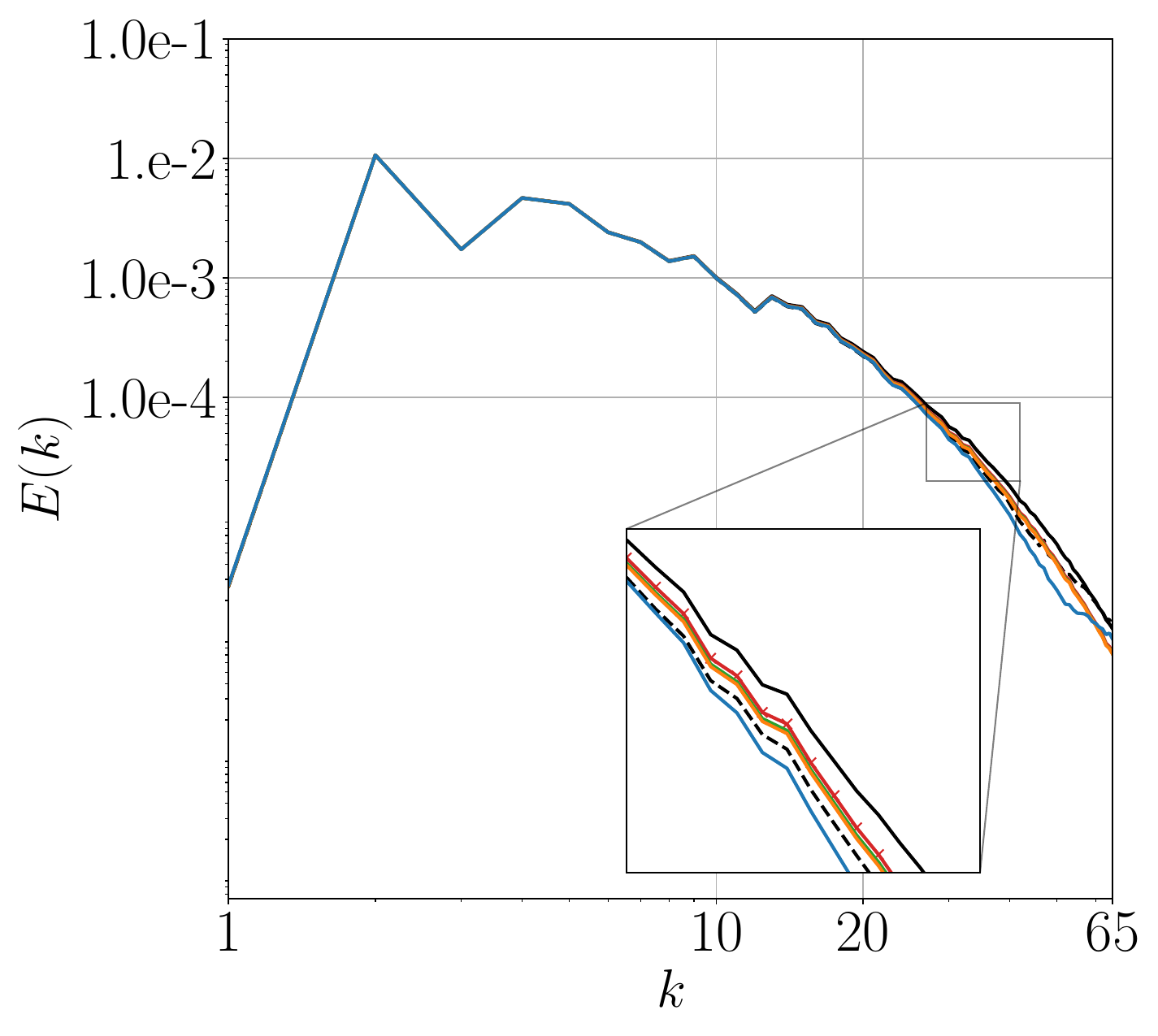}
     \caption{Kinetic energy spectrum.}
     \label{fig:Res05_II:b}
 \end{subfigure}
 \begin{subfigure}{0.32\textwidth}
     \includegraphics[width=\textwidth]{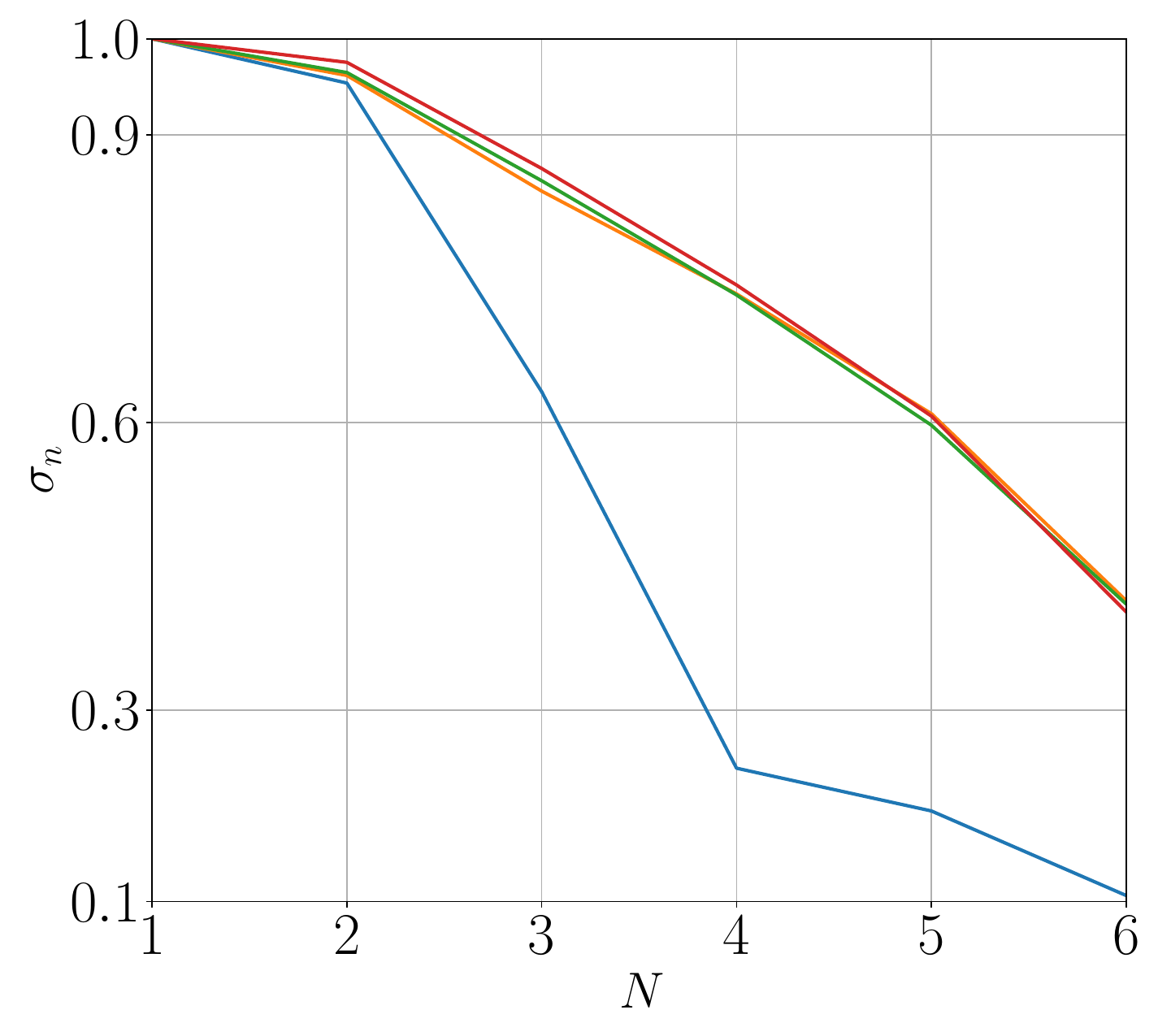}
     \caption{Averaged modal coefficients.}
     \label{fig:Res05_II:c}
 \end{subfigure}
 \vspace{-0.2cm}
\caption{Kinetic energy spectrum for different polynomial orders obtained at $t^*=15$ for the baseline ILES (left), from data-driven filtered DNS data with $\Delta T= 4.0$ (middle) and modal decay of the data-driven filters (right). Solid black line, DNS; dashed line, Filtered with $\Delta = 2h$.}
   \label{fig:Res05_II}
\end{figure}

Following similar analyses from section~\ref{sec:results}, in Figure~\ref{fig:Res05_II} we compare kinetic energy spectra of the DNS filtered with the data-driven filters and the ILES (with the same time window, $\Delta T$).
The data-driven models produce curves that follow the same ordering as in the ILES. This is evident from the modal decay in the right plot, where the curves for $\mathrm{p} \geq 3$ are ordered, though the variations between them are minor.

\newpage

\end{document}